\documentclass[a4paper,11pt]{article}
\pdfoutput=1
\usepackage{jheppub}
\usepackage[utf8x]{inputenc}
\usepackage{graphicx}
\usepackage{float}
\usepackage[T1]{fontenc}
\usepackage{epsfig}
\usepackage{color}
\usepackage{amsmath,mathtools}
\usepackage{subfloat}
\usepackage{amsfonts}
\usepackage{braket}
\usepackage{cleveref}
\usepackage{epstopdf}
\usepackage{caption}
\usepackage{etoolbox}
\usepackage{orcidlink}
\usepackage{subcaption}
\usepackage{enumitem}
\usepackage{comment}
\usepackage[titletoc,toc,title]{appendix}
\usepackage{hyperref}
\hypersetup{
	colorlinks=true,
	linkcolor=blue,
	filecolor=red,      
	urlcolor=blue,
	citecolor=red
} 
\DeclareMathOperator{\sech}{sech}
\DeclareMathOperator{\csch}{csch}

\title{\boldmath Holographic Reflected Entropy: Islands and Defect Phases}

\author{Ankur Dey\,\orcidlink{0009-0001-1077-0442}}
\author{and Gautam Sengupta\,\orcidlink{0000-0002-1118-6926}}
\affiliation{Department of Physics, Indian Institute of Technology, Kanpur 208016, India}

\emailAdd{ankurd21@iitk.ac.in}
\emailAdd{sengupta@iitk.ac.in}

\abstract{We investigate the mixed state entanglement structure through the reflected entropy for disjoint radiation subsystems coupled to a 2d eternal brane world black hole in a time dependent defect AdS$_3$/BCFT$_2$ scenario. Utilizing the island prescription and the defect extremal surface (DES) formula, we demonstrate a complex mixed state entanglement structure through the reflected entropy corresponding to distinct entanglement entropy phases. In each case we verify the holographic duality of the reflected entropy with the bulk entanglement wedge cross section (EWCS) and also obtain the Page curves for both the entanglement entropy and the associated reflected entropy phases. Furthermore, we extend our analysis to adjacent radiation subsystems and obtain consistent results using both the island and the DES prescription.}


\arxivnumber{2509.06551}

\begin{document} 
\maketitle
\flushbottom
\setcounter{tocdepth}{2}

\section{Introduction}\label{sec_intro}

For the past few decades the black hole information loss paradox has received intense research focus in the context of a viable quantum theory of gravity. Very recently a possible resolution of this paradox has been proposed for toy models of quantum field theories coupled to semi classical gravity which reproduces the elusive Page curve \cite{Page:1993wv, Page:1993df, Page:2013dx} for the Hawking radiation. This involved the emergence of the island formula for the fine-grained entanglement entropy of the Hawking radiation inspired by the construction of quantum extremal surface prescription due to \cite{Engelhardt:2014gca}. This framework suggests the appearance of spacetime regions denoted as islands at late times in the entanglement wedge of subsystems in radiation baths leading to the purification of the outgoing Hawking quanta \cite{Almheiri:2019hni, Almheiri:2019psf, Almheiri:2019qdq, Almheiri:2019psy, Almheiri:2019yqk, Almheiri:2020cfm}. 

An interesting perspective on the island formulation is offered by double holography \cite{Almheiri:2019hni, Sully:2020pza, Rozali:2019day, Chen:2020uac, Chen:2020hmv, Grimaldi:2022suv, Suzuki:2022xwv, Geng:2020qvw, Geng:2020fxl, Geng:2021iyq, Geng:2021mic, Geng:2021hlu} which describes a $d+1$-dimensional gravitational theory in a brane world as the holographic dual of a $d$-dimensional conformal field theory (CFT$_d$) coupled to semi-classical gravity on the brane (referred to as the brane description).\footnote{Since the brane description may be obtained via a partial reduction of the bulk AdS$_3$ geometry with the EOW brane, it is sometimes also described as the lower dimensional effective description \cite{Deng:2020ent, Chu:2021gdb, Li:2021dmf, Basu:2022reu, Shao:2022wrm, Basu:2024xjq}.} The gravity-matter system on the brane may be considered to be the holographic dual to a quantum mechanical system at the boundary of the CFT$_d$, known as the boundary description. The computation of the entanglement entropy in this setup using the island formula is equivalent to the Ryu-Takayanagi (RT) prescription in the bulk Anti-de Sitter (AdS$_{d+1}$) geometry. For recent works in double holography see \cite{Liu:2023ggg,Ling:2021vxe,Akal:2020wfl,Miao:2020oey,Myers:2024zhb}.

On a separate note a CFT defined on a domain with a boundary termed as a boundary conformal field theory (BCFT) was introduced in \cite{Cardy:2004hm}. Subsequently an AdS/BCFT correspondence was described in \cite {Takayanagi:2011zk} and later developed in  \cite { Fujita:2011fp, Sully:2020pza, Kastikainen:2021ybu,Izumi:2022opi,Cavalcanti:2018pta,Magan:2014dwa,Cavalcanti:2020rsp,Takayanagi:2020njm}. In the AdS$_3$/BCFT$_2$ framework the holographic dual for the BCFT$_2$ is described by an asymptotically AdS$_3$ geometry truncated by a codimension-1 end-of-the-world (EOW) brane with Neumann boundary conditions. In this framework the homology conditions for the Ryu-Takayanagi (RT) surfaces are modified and the holographic entanglement entropy formula will involve extremal surfaces which terminates on the EOW brane \cite{Takayanagi:2011zk, Fujita:2011fp}.

Interestingly this AdS$_3$/BCFT$_2$ scenario was further developed in \cite{Deng:2020ent} which considered the bulk as a defect spacetime with conformal matter on the EOW brane. In this case the gravity region relevant to the computation of the QES is localized on the EOW brane leading to the defect extremal surface (DES) formula as the holographic counterpart of the island formula. The $2d$ gravity on the brane emerges via a partial Randall-Sundrum reduction of the bulk geometry with transparent boundary conditions at the interface between the gravitational and the non-gravitational regions of the CFT$_2$ in the effective description. Computation of the fine grained entanglement entropy from the DES formula and the boundary quantum extremal surface prescription were shown to be in exact agreement with each other. Subsequently in \cite{Chu:2021gdb} the DES formula was further utilized to compute the entanglement entropy for subsystems in radiation baths associated with a $2d$ eternal black hole.

While the entanglement entropy effectively captures the entanglement structure of bipartite pure states it fails to describe mixed state entanglement due to the contributions from irrelevant correlations. Certain computable mixed state entanglement and correlation measures such as the entanglement negativity \cite{Vidal:2002zz, Plenio:2005cwa}, reflected entropy \cite{Dutta:2019gen, Akers:2021pvd}, the entanglement of purification \cite{Takayanagi:2017knl, Nguyen:2017yqw}, balanced partial entanglement entropy \cite{Wen:2021qgx} and the odd entanglement entropy \cite{Tamaoka:2018ned}
amongst others, has been introduced in the literature. 

In \cite{Li:2020ceg, Chandrasekaran:2020qtn} the authors proposed an island formula for the reflected entropy to characterize the mixed state entanglement structure of the Hawking radiation from black holes in models of quantum field theories coupled to semi-classical gravity. The reflected entropy computed from the island formula was shown to match exactly with the bulk computations involving the area of the entanglement wedge cross section (EWCS) \cite{Takayanagi:2017knl, Nguyen:2017yqw}. 

Subsequently the authors in \cite{Li:2021dmf} proposed a DES formula for the mixed state correlation measure of the reflected entropy in the modified AdS$_3$/BCFT$_2$ framework as described earlier in \cite{Deng:2020ent}. In this context they computed the reflected entropy for subsystems in both time-independent and time-dependent scenarios. The latter scenario corresponds to a $2d$ eternal black hole located on the EOW brane and the reflected entropy and their Page curves were obtained for subsystems in the black hole interior and the radiation baths for the left and the right copies of the thermofield double (TFD) state. In this context a corresponding DES formula for the entanglement negativity was proposed by the authors in \cite{Basu:2022reu,Shao:2022wrm}. In the latter work \cite{Shao:2022wrm} the authors computed the entanglement negativity between adjacent subsystems in radiation baths for the $2d$ eternal black hole using the island and the DES prescription and obtained an exact match.

As described above the reflected entropy for the 2d eternal black hole on the EOW brane has been obtained in the context of the DES formula and the modified AdS$_3$/BCFT$_2$ framework in \cite{Li:2021dmf} for TFD dual states corresponding to various configurations of single subsystems in the black hole interior and the radiation bath. However it is an important and significant exercise to extend their analysis for the reflected entropy to more complex mixed state configurations described by adjacent and disjoint subsystems in the radiation bath. This is expected to lead to further insights into the mixed state entanglement structure for the radiation baths in these braneworld models and serve as further significant verification of the proposed duality between the island and the DES prescriptions. In this article we address this interesting issue and explore such mixed state configurations utilizing both the island and the DES proposal for the reflected entropy leading to an interesting entanglement phase structure and their corresponding Page curves. The time evolution of the reflected entropy across various subsystem size and configuration illustrates an intricate mixed state entanglement pattern for the Hawking radiation.

The structure of this article is as follows. In \cref{sec_review}, we briefly review some of the necessary concepts, including the AdS$_3$/BCFT$_2$ framework, $2d$ eternal black holes, the DES formalism for the entanglement entropy, and the island and the DES formula for the reflected entropy. Following this, in \cref{sec_disjoint} we analyze the mixed state entanglement between two disjoint subsystems in the left and the right radiation baths and plot the Page curves for both the entanglement entropy and the reflected entropy. Next we extend our analysis to adjacent subsystems in the radiation baths in \cref{sec_adjacent}. Finally, in \cref{sec_summary} we conclude by summarizing our results and discuss their implications and open issues.

\section{Review of earlier literature}\label{sec_review}

This section provides a concise overview of several essential concepts required for our analysis. We begin with a brief review of the AdS$_3$/BCFT$_2$ correspondence, along with the incorporation of defect conformal matter on the EOW brane leading to the DES formula. Following this, we describe the emergence of a $2d$ eternal black hole on the EOW brane from the AdS$_3$/BCFT$_2$ framework. Subsequently we discuss the DES formula for the entanglement entropy and its relation to the black hole island framework. Finally, we review the definition of the mixed state correlation measure of reflected entropy and its description in the island and the DES formulations.

\subsection{AdS$_3$/BCFT$_2$ and brane defects}

The author in \cite{Cardy:2004hm} described boundary conformal field theories which are defined on a domain with a boundary. The corresponding holographic dual was proposed in \cite{Takayanagi:2011zk,Fujita:2011fp} as an  AdS$_3$ geometry truncated by a codimension-1 brane $Q$ with Neumann boundary conditions described by a bulk action as follows
\begin{align}
I=\frac{1}{16 \pi G_N} \int_N \sqrt{-g} (R-2 \Lambda)+\frac{1}{8 \pi G_N} \int_Q \sqrt{-h} (K-T),
\end{align}
where $N$ denotes the bulk AdS$_3$ spacetime and $Q$ as the brane with an induced metric $h_{ab}$, extrinsic curvature $K_{ab}$, and a constant tension $T$. Considering the variation of the above action with respect to $h_{ab}$, imposition of the Neumann boundary condition leads to the condition
\begin{align}
K_{ab}=(K-T)h_{ab}.
\end{align}

\begin{figure}[t]
\centering
\includegraphics[scale=1]{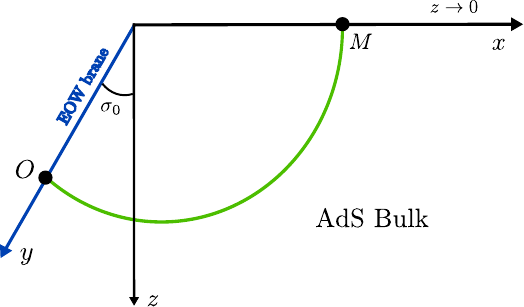}
\caption{Holographic dual of a BCFT$_2$ defined on a half plane $(x>0)$. The EOW brane is considered to be at a constant hyperbolic angle $\sigma_0$. Figure modified from \cite{Deng:2020ent}.}\label{fig_BCFT}
\end{figure}

On a separate note, the bulk AdS$_3$ geometry may be expressed in terms of AdS$_2$
foliations as follows
\begin{align}
ds^2 & =d\sigma^2+\ell ^2 \cosh^2 (\sigma / \ell) \left( \frac{-dt^2+dy^2}{y^2} \right),
\end{align}
where $y$ is a radial coordinate along the foliation, while $\sigma$ is the hyperbolic angle between the foliation and the normal to the asymptotic boundary (as illustrated in \cref{fig_BCFT}). It is fairly straightforward to retrieve the well known Poincar\'e AdS$_3$ metric using the following relations
\begin{align}
z=y \sech (\sigma / \ell), \qquad x=- y \tanh (\sigma / \ell),
\end{align}
with $\ell$ being the AdS radius. Assuming that the brane $Q$ is located at a positive constant hyperbolic angle $\sigma_0$, the induced metric on the brane may be obtained as
\begin{align}\label{eq_branemetric}
ds_{brane}^2=\frac{\ell ^2 }{y^2 \sech^2 (\sigma_0 / \ell)} \left( -dt^2+dy^2 \right),
\end{align}
where we see that the metric on the brane is simply a $2d$ flat metric with a conformal factor $\Omega_y=\frac{\ell}{y \sech (\sigma_0 / \ell)}$. It is also fairly simple to show that
\begin{align}\label{eq_bten}
K_{ab}=\frac{\tanh (\sigma_0 / \ell)}{\ell} h_{ab}, \qquad
T=\frac{\tanh (\sigma_0 / \ell)}{\ell}.
\end{align}
Thus we observe that the brane tension is dependent on the brane angle which in this scenario is a constant.

From \cref{eq_bten} it is evident that when the tension of the brane $Q$ is zero, then it is orthogonal to the asymptotic boundary. Addition of conformal matter on a tensionless brane induces a finite tension and displaces the brane to a constant hyperbolic angle $\sigma_0$ \cite{Fujita:2011fp,Deng:2020ent,Chu:2021gdb}. The resultant modified bulk action in this setup is given as
\begin{align}
I=\frac{1}{16 \pi G_N} \int_N \sqrt{-g} (R-2 \Lambda)+\frac{1}{8 \pi G_N} \int_Q \sqrt{-h} (K-T)+I_{CFT},
\end{align}
where $I_{CFT}$ corresponds to the actions of the conformal matter on the brane. Once again, considering the variation of the above action with respect to $h_{ab}$ and imposing the Neumann boundary condition leads us to
\begin{align}
K_{ab}-(K-T)h_{ab}=8 \pi G_N \chi h_{ab}.
\end{align}
Here $\chi$ is related to the vacuum 1-point function of the CFT stress tensor as
\begin{align}
\langle T_{ab} \rangle _{AdS_2}=\chi h_{ab}.
\end{align}
Using the above relations and the trace anomaly $\langle T_a^a \rangle = \frac{c'}{24 \pi} R$ on AdS$_2$ elaborated in \cite{DiFrancesco:639405}, it is possible to solve for $\chi$.

Once again assuming that the EOW brane is located at some constant $\sigma_0$, following the prescription described earlier the metric on the brane may be obtained as \cref{eq_branemetric}. The scalar curvature may also be determined as
\begin{align}
R=- \frac{2}{\ell ^2 \cosh ^2 (\sigma _0 / \ell)}.
\end{align}
Combining the 1-point function, the trace formula and the Neumann boundary condition, the central charge $c'$ on the brane may be obtained in terms of the brane tension $T$ as
\begin{align}
c' = \frac{3 \ell \cosh ^2 (\sigma_0 / \ell)}{G_N} (\tanh (\sigma_0 / \ell)-\ell T)= 2 c \cosh ^2 (\sigma_0 / \ell) (\tanh (\sigma_0 / \ell)-\ell T)
\end{align}
where the second equality comes from the fact that the CFT central charge on the asymptotic boundary $c$ is related to the bulk Newton constant $G_N$ as $c=\frac{3 \ell}{2 G_N}$, also known as the Brown-Henneaux relation \cite{Brown:1986nw}.

In this modified framework, the EOW brane is regarded as defect within the bulk geometry. Given that the EOW brane involves a maximally symmetric AdS$_2$ metric, the expectation value of the stress tensor in the defect CFT$_2$ is proportional to the induced metric $h_{ab}$. Consequently, only the brane tension $T$ is affected by the inclusion of conformal matter. Furthermore, the CFT$_2$ degrees of freedom confined on the EOW brane do not act as dynamical fields, but rather modify the brane's intrinsic properties without backreacting on the bulk geometry. This allows the EOW to be treated as a lower dimensional defect embedded in the AdS$_3$ spacetime.

\subsection{Eternal black hole from time dependent defect AdS$_3$/BCFT$_2$}\label{ssec_ebh}

\begin{figure}[t]
\centering
\includegraphics[scale=1.2]{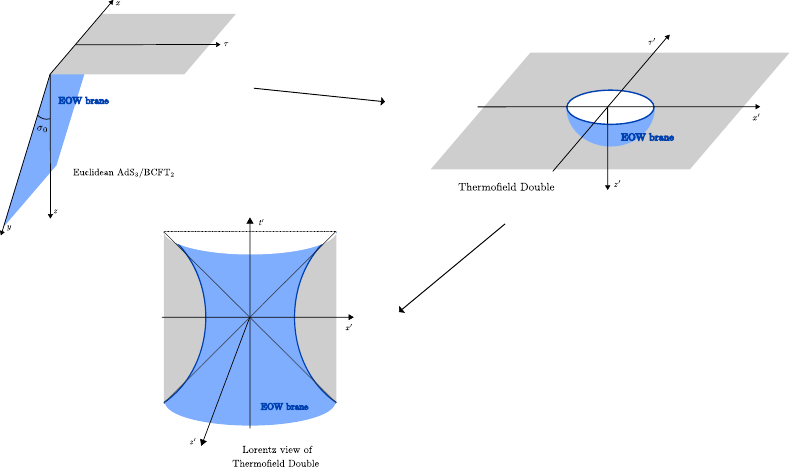}
\caption{A set of global conformal transformations takes us from a Euclidean AdS$_3$/BCFT$_2$ setup to a thermofield double state. Subsequently, by applying the analytic continuation $\tau' \to it'$ we obtain the Lorentzian solution. Figures modified from \cite{Chu:2021gdb}.}\label{fig_GCT}
\end{figure}

In this section we review how a 2d eternal black hole emerges from a time dependent defect AdS$_3$/BCFT$_2$ model, as elaborated in \cite{Fujita:2011fp,Li:2021dmf,Chu:2021gdb,Rozali:2019day}. As discussed earlier the holographic dual to BCFT$_2$ is an AdS$_3$ geometry truncated by a EOW brane. In this case the BCFT is defined on a half plane described by $(x,\tau>0)$, with the EOW brane being located at a constant hyperbolic angle $\sigma_0$ and described by the equation $\tau=-z \sinh (\sigma_0 / \ell)$.\footnote{It is important to note that the selection of $\tau$ or $x$ is purely a matter of computational convenience and is of no physical significance. Since there is no fundamental distinction between them in Euclidean spacetime, their roles are interchangeable within the given context.} The bulk AdS$_3$ is then described as
\begin{align}
ds^2 & =d\sigma^2+\ell ^2 \cosh ^2 (\sigma / \ell) \left( \frac{d\tau^2+dy^2}{y^2} \right), \notag \\
& = \frac{\ell ^2}{z^2} (d \tau ^2+dx^2+dz^2).
\end{align}
Using the set of global conformal transformations 
\begin{align}\label{eq_trans1}
\tau & =\frac{2({x'} ^2 + {\tau'} ^2 +{z'} ^2 -1)}{({\tau'} +1) ^2 + {x'} ^2 +{z'} ^2}, \notag \\
x & =\frac{4 {x'}}{({\tau'} +1) ^2 + {x'} ^2 +{z'} ^2},  \\
z & =\frac{4 {z'}}{({\tau'} +1) ^2 + {x'} ^2 +{z'} ^2} \notag
\end{align}
the boundary of the BCFT is mapped to the circle ${x'}^2+{\tau '}^2=1$, while the EOW brane is mapped to part of a sphere described by $(z'+\sinh (\sigma_0 / \ell))^2+{x'}^2+{\tau '}^2=\cosh ^2(\sigma_0 / \ell)$. Transformations in \cref{eq_trans1} are such that the form of the bulk metric is preserved. A Lorentzian solution may be obtained by analytically continuing $\tau' \to it'$, upon which the geometry of the EOW brane becomes part of a hyperboloid. A visual representations of the operations described above is provided in \cref{fig_GCT}.

To describe the black hole on the EOW brane, it is important to note that in the Lorentzian solution, $ x' = \pm \sqrt{ t'^2 +1} $ describe the left/right boundaries of the EOW brane. The light like curves at $ t' \to \pm \infty $ correspond to the horizons of this black hole. Consequently, the horizons may be described as
\begin{align}
x' = \pm t', \qquad \qquad z' = \cosh ^2(\sigma_0 / \ell) - \sinh (\sigma_0 / \ell).
\end{align}
The interior of the black hole is then identified with the region $ |x'| < t' $. Consequently, the two regions at the AdS$_3$ boundary are causally disconnected. However, they are still connected in the bulk AdS$_3$ geometry indicating that the two subregions are entangled with each other as also described in \cite{Maldacena:2001kr} for the case of general eternal AdS-Schwarzschild black holes.

Using a partial Randall-Sundrum reduction, the two-sided eternal black hole on the EOW brane is then a part of the effective $2d$ description and is coupled to the bath CFT. To describe the time evolution it is possible to introduce the Rindler coordinates\footnote{Note that the Rindler coordinates correspond to an observer moving with a constant acceleration in the Lorentzian geometry described above. A constant-accelerated observer in Rindler space corresponds to a static observer in Schwarzschild spacetime.} $(X,T)$ which captures the near horizon geometry of the black hole
\begin{align}\label{eq_trans21}
{x'}=e^X \cosh T, \qquad \qquad {t'}=e^X \sinh T.
\end{align}
This transformation may be applied to both the left and right patches, finally resulting in a $1+1$ dimensional two-sided eternal black hole coupled to the bath CFT.

In the course of the transformations in \cref{eq_trans1,eq_trans21}, the UV cut-off transforms as
\begin{align}\label{eq_trans22}
\epsilon=\frac{4 {\epsilon '}}{({\tau'} +1) ^2 + {x'} ^2}, \qquad {\epsilon '}=\epsilon _R e^X,
\end{align}
where $\epsilon _R$ in the Rindler coordinates is taken to be constant everywhere.

\subsection{DES formula and entanglement entropy }\label{ssec_eedes}

Consider a theory of semiclassical gravity coupled to a non-gravitating bath conformal field theory where the gravitational sector may also contain some conformal matter. In this framework, the fine grained entropy associated with an interval in the bath CFT may be computed using the island formula described in \cite{Almheiri:2019hni,Penington:2019npb, Almheiri:2019psf} as
\begin{align}\label{form_ee_is}
S_{Is}= \min_{X} \biggl[ \text{ext}_{X} \biggl\{ S_{eff}(A\cup I_S(A)) +S_{area} (X) \biggr\} \biggr], \qquad X= \partial I_S
\end{align}
where $A$ refers to the subsystem on the radiation bath, $I_S (A)$ is the island region on the brane corresponding to $A$, and $X$ denotes the boundary of the island $I_S(A)$.

From the holographic perspective, the Defect Extremal Surface (DES) formula \cite{Deng:2020ent} was introduced as a counterpart to the island formula. It was motivated by the observation that the gravity region relevant to determining the quantum extremal surface (QES) may be considered to be localized on the brane. The authors enhanced the standard AdS$_3$/BCFT$_2$ correspondence by incorporating contributions from the conformal matter located on the EOW brane. In this case the contribution from the bulk conformal matter located on the brane must also be accounted for while computing the entanglement entropy. In this setup, the fine grained entropy using the DES formula may be described as
\begin{align}\label{form_ee_des}
S_{DES}= \min_{\Gamma, X} \biggl[ \text{ext}_{\Gamma,X} \biggl\{ S_{RT}(\Gamma)+S_{defect}(D) \biggr\} \biggr], \qquad X= \Gamma \cap D
\end{align}
where $\Gamma$ is a co-dimension two RT surface homologous to $A$, and $D$ is the defect brane. $S_{defect}(D)$ represents the entanglement entropy contribution arising from the conformal matter localized on the brane.

\begin{figure}[t]
	\centering
	\begin{subfigure}{0.45\linewidth}
		\includegraphics[scale=0.75]{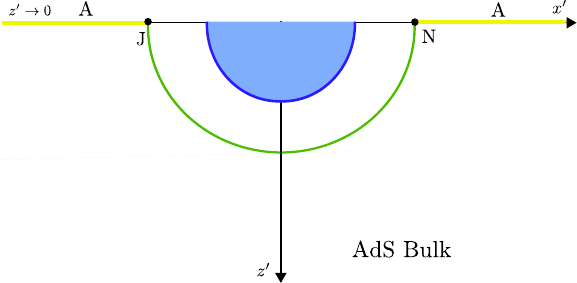}\caption{}
		\label{fig_eec}
	\end{subfigure}
	\hfill
	\begin{subfigure}{0.45\linewidth}
		\includegraphics[scale=0.75]{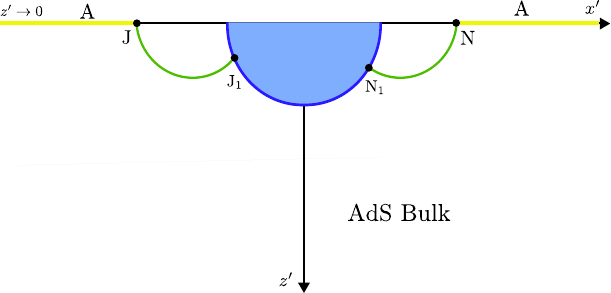}\caption{}
		\label{fig_eedc}
	\end{subfigure}
	\caption{(a) Connected and (b) disconnected phases of the extremal surfaces. Figures modified from \cite{Chu:2021gdb}.}
	\label{fig_e}
\end{figure}

Utilizing both the island and DES prescription, the computation of the entanglement entropy of a subsystem $A$ on the radiation bath connected to a $2d$ eternal black hole system described in the previous subsection is discussed in details in \cite{Chu:2021gdb}. As illustrated in \cref{fig_e}, the subsystem $A$ is defined as the interval $[-\infty , -x_1'] \cup [x_1',\infty]$ at some $\tau' = \tau' _1$ and $z' = \epsilon'$. For simplicity of computation, the unprimed coordinates are chosen as the starting point, where the set of transformations in \cref{eq_trans1} allow us to map the endpoints of the subsystem to $(\pm x_1,\tau_1,\epsilon_1)$.

Depending on the configuration of the interval, specifically, its size and location, different phases of the entanglement entropy are observed. Illustrated in \cref{fig_e} are the connected and disconnected phases of the entanglement entropy. A complete match between the results obtained by the island and DES prescription establishes the holographic duality proposed in \cite{Deng:2020ent}. The explicit computations (as provided in \cite{Chu:2021gdb}) are as follows.

\subsubsection*{Connected Phase}\label{conn}

In the connected phase illustrated in \cref{fig_eec}, the subsystem $A$ is located far away from the brane, ensuring that it does not possess an entanglement entropy island. Under these conditions the generalized entanglement entropy may be computed from the field theory perspective using the island formula described in \cref{form_ee_is} as
\begin{align}
S_{gen}=S_{eff}(A\cup I_S(A))+S_{area},
\end{align}
where the first term may be expressed in terms of the 2-point correlators of twist fields at endpoints $J$ and $N$ as
\begin{align}
S_{eff}=\lim_{n \to 1} \frac{1}{1-n} \log \langle \sigma_{g_A} (J) \sigma_{g_A^{-1}} (N)  \rangle,
\end{align}
where $\sigma_{g_A} (J)$ and $\sigma_{g_A^{-1}} (N)$ correspond to the twist and anti-twist fields positioned at $J$ and $N$ respectively. Since there is no island on the brane, the area term in \cref{form_ee_is} does not contribute to the entanglement entropy. Now using the standard form of the 2-point twist field correlators given as
\begin{align}
\langle \sigma (z_1) \sigma ^{-1} (z_2) \rangle = \frac{1}{(z_2-z_1)^{2 h_m}} \times h.c. \qquad h_m=\frac{c}{24} \left( m-\frac{1}{m} \right),
\end{align}
the entanglement entropy in the connected phase may be computed as
\begin{align}\label{eq_eec}
S_{conn}=\frac{c}{3} \log \left[ \frac{2 x_1}{\epsilon_1} \right]=\frac{c}{3} \log \left[ \frac{2 \cosh T}{\epsilon _R} \right],
\end{align}
where in the final step the set of transformations in  \cref{eq_trans1,eq_trans21,eq_trans22} were utilized to express the result in the near-horizon Rindler coordinates.

\vspace{0.5cm}

From the bulk perspective, the holographic entanglement entropy may be determined by the DES formula outlined in \cref{form_ee_des}. Here the minimal geodesic corresponds to the Hartman Maldacena (HM) surface denoted by the green line in \cref{fig_eec}. The generalized entanglement entropy in this scenario may then be derived as
\begin{align}
S_{gen}=S_{RT}+S_{defect}.
\end{align}
The first term directly corresponds to the geodesic length connecting the boundary points $(x_1,\tau_1,\epsilon_1)$ and $(-x_1,\tau_1,\epsilon_1)$. This geodesic length can be explicitly computed as using
\begin{align}\label{eq_geol}
L_{ji}=\cosh ^{-1} [-\xi_{ji}],
\end{align}
where in the Poincar\'e coordinates $\xi_{ji}$ is related to the inner product between the position vectors of the two endpoints $Y_j$ and $Y_i$ as
\begin{align}\label{eq_ip}
\xi_{ji}=Y_j \cdot Y_i=-\frac{(x_j-x_i)^2+(\tau_j-\tau_i)^2+z_j^2+z_i^2}{2 z_i z_i} .
\end{align}
In this specific setup, $Y_1$ is assigned to $J$ and $Y_2 \to N$. Since the HM surface does not intersect the EOW brane, contributions from the bulk conformal matter on the brane does not contribute. As a result the generalized entropy directly simplifies to the effective entanglement entropy, which agrees with the field theory result in \cref{eq_eec} on employing the Brown-Henneaux relation $c=\frac{3 \ell}{2 G_N}$ \cite{Brown:1986nw}.

\subsubsection*{Disconnected Phase}\label{disconn}

In the disconnected phase illustrated in \cref{fig_eedc}, subsystem $A$ is assumed to be located near the brane such that an entanglement entropy island corresponding to $A$ emerges on the brane. According to \cref{form_ee_is}, the semiclassical contribution may be computed in terms of the 4-point correlator of twist fields at the boundary points $J,N$ and brane points $J_1, N_1$ as follows
\begin{align}
S_{eff} =\lim_{n \to 1} \frac{1}{1-n} \log \Omega_{J_1}^{2 h_{g_A}} \Omega_{N_1}^{2 h_{g_A}}  \langle \sigma_{g_A} (J) \sigma_{g_A^{-1}} (J_1) \sigma_{g_A} (N_1) \sigma_{g_A^{-1}} (N)  \rangle  _{CFT}.
\end{align}
where the conformal factors $\Omega_{J_1}$ and $\Omega_{N_1}$ arise due to the presence of twist fields at bulk points $J_1$ and $N_1$ respectively. In the large $c$ approximation, this 4-point correlator factorizes into a product of 2-point correlators, simplifying the expression as
\begin{align}
S_{eff} = \lim_{n \to 1} \frac{1}{1-n} \log \Omega_{J_1}^{2 h_{g_A}} \Omega_{N_1}^{2 h_{g_A}}  \langle \sigma_{g_A} (J) \sigma_{g_A^{-1}} (J_1) \rangle _{CFT} \langle \sigma_{g_A} (N_1) \sigma_{g_A^{-1}} (N)  \rangle _{CFT}.
\end{align}
From symmetry of the configuration, we may assume $J_1 \equiv (-y,-x_y)$ and $N_1 \equiv (-y,x_y)$, where $y$ is the coordinate along the brane. Using the standard 2-point twist field correlator we obtain
\begin{align}
S_{eff}=\frac{c}{3} \left( \log \left[ \frac{(y+\tau_1 )^2+(x_1-x_y)^2}{\epsilon_1} \right] + \log \left[ \frac{ \ell}{\epsilon _y y \sech(\sigma_0 / \ell)} \right] \right) .
\end{align}
Note that $\epsilon_y$ here denotes the UV cut-off on the brane. For simplicity of computation, we also assume that the central charge $c'$ of the CFT on the brane is the same as that on the asymptotic boundary, i.e., $c' = c$. Meanwhile, the area term is given directly from \cite{Deng:2020ent} as
\begin{align}
S_{area}=2 \times \frac{1}{4 G_N} \frac{\sigma _0}{\ell} = 2 \times \frac{c}{6} \frac{\sigma _0}{\ell}
\end{align}
where the factor of 2 arises due to the two brane points. Also, in the second step we utilize the Brown-Henneaux relation. Combining the two separate contributions evaluated above we obtain the generalized entanglement entropy $S_{gen}$, which is extremized at $y=\tau_1$ and $x_y=x_1$. Consequently, the resultant entanglement entropy is this scenario is derived as 
\begin{align}\label{eq_eedc}
S_{disc} & = \frac{c}{3} \left( \frac{\sigma_0}{\ell} + \log \left[ \frac{2 \tau_1}{\epsilon_1} \right] + \log \left[ \frac{2 \ell}{\epsilon _y \sech (\sigma / \ell)} \right] \right) \notag \\
& =\frac{c}{3} \left( \frac{\sigma_0}{\ell} + \log \left[ \frac{2 \sinh X_1}{\epsilon _R} \right] + \log \left[ \frac{2 \ell}{\epsilon _y \sech(\sigma_0 / \ell)} \right] \right)
\end{align}
where once again the set of transformations in \cref{eq_trans1,eq_trans21,eq_trans22} were applied in the final step. 

\vspace{0.5cm}

From the bulk perspective, the first term in \cref{form_ee_des} corresponds to the combination of the RT surfaces  $JJ_1$ and $NN_1$, which are illustrated by the green lines in \cref{fig_eedc}. Here, $J_1$ and $N_1$ are arbitrary brane points, defined in unprimed coordinates as $J_1 \equiv (-x_y, -y \tanh (\sigma_0 / \ell), y \sech (\sigma_0 / \ell))$ and $N_1 \equiv (x_y, -y \tanh (\sigma_0 / \ell), y \sech (\sigma_0 / \ell))$. Using these definitions, the contribution $S_{RT}$ may be determined as
\begin{align}
S_{RT}= 2 \times \frac{1}{4 G_N}  \cosh ^{-1} \left[ \frac{y^2 \sech ^2(\sigma_0 / \ell )+(y \tanh (\sigma_0 / \ell )+\tau_1 )^2+(x_1-x_y)^2+\epsilon_1 ^2}{2 \epsilon_1 y \sech(\sigma_0 / \ell )} \right] .
\end{align}
Since the RT surfaces intersect the brane, contributions due to the bulk conformal matter (in the form of the second term in \cref{form_ee_des}) must be accounted for. This may be directly extracted from \cite{Deng:2020ent} as
\begin{align}
S_{defect}=2 \times \frac{c}{6} \log \left[ \frac{2 \ell}{\epsilon _y \sech (\sigma / \ell)} \right],
\end{align}
where once again the factor of 2 correspond to the two intersections points $J_1$ and $N_1$. The generalized entanglement entropy $S_{gen}$ is determined by adding the two separate contributions evaluated above. By extremizing $S_{gen}$ we obtain $y=\tau_1$ and $x_y=x_1$. This leads to the final expression of the entanglement entropy in the disconnected phase, which with the application of the Brown-Henneaux relation, matches with the results from the island computations in \cref{eq_eedc}.

\subsection{Reflected Entropy}\label{ssec_re}

We now briefly describe the definition of the reflected entropy which is a mixed state correlation measure and its computation in the framework of the AdS$_3$/CFT$_2$, as detailed in \cite{Dutta:2019gen}. Considering a bipartite mixed state $A \cup B$, a canonically purified state $| \sqrt{\rho_{AB}} \rangle _{ABA^*B^*}$ may be constructed by doubling of the Hilbert space involving the CPT conjugate states $ A^*, B^*$. The reflected entropy between the subsystems $A$ and $B$, denoted as $S_R(A:B)$, may then be defined as the von Neumann entropy of the reduced density matrix $\rho_{AA^*}$ as
\begin{align}
S_R(A:B)=S_{vN}(\rho _{AA^*})_{\sqrt{\rho_{AB}}}.
\end{align}
The reduced density matrix $\rho_{AA^*}$ may be obtained by tracing out the degrees of freedom associated with $B$ and $B^*$ from the full density matrix $|\sqrt{\rho_{AB}} \rangle \langle \sqrt{\rho_{AB}}| $ defined on the doubled Hilbert space. 

The replica technique to obtain the reflected entropy between two disjoint intervals $A \equiv [z_1,z_2]$ and $B \equiv [z_3,z_4]$ in a CFT$_2$ is elaborated in \cite{Dutta:2019gen}. Considering a 4-point correlator of twist fields inserted at the endpoints of the two intervals, the reflected entropy $S_R(A:B)$ is given as 
\begin{align}\label{eq_re}
S_R(A:B)=\lim_{m,n \to 1} S_n(AA*)_{\psi _m}=\lim_{m,n \to 1} \frac{1}{1-n} \log \frac{\langle \sigma_{g_A}(z_1) \sigma_{g_A^{-1}}(z_2) \sigma_{g_B}(z_3) \sigma_{g_B^{-1}}(z_4) \rangle _{CFT ^{\otimes mn}}}{\langle \sigma_{g_m}(z_1) \sigma_{g_m^{-1}}(z_2) \sigma_{g_m}(z_3) \sigma_{g_m^{-1}}(z_4) \rangle ^n _{CFT ^{\otimes m}}},
\end{align}
where $m$ and $n$ represent the replica indices.\footnote{Note that the two replica limits $n \to 1$ and $m \to 1$ are non-commutating. As suggested in \cite{Akers:2022max,Kusuki:2019evw,Akers:2021pvd}, in this article we first take $n \to 1$, followed by the replica limit $m \to 1$ to compute the reflected entropy.} The conformal dimensions of the field $\sigma_{g_A},\sigma_{g_B},\sigma_{g_m}$ and the composite field $\sigma_{g_A^{-1} g_B}$ (the dominant operator exchange between $\sigma_{g_A^{-1}}$ and $\sigma_{g_B}$) are given as
\begin{align}\label{eq_cdim}
h_A=h_B=\frac{nc}{24}\left( m-\frac{1}{m} \right), \qquad h_m=\frac{c}{24}\left( m-\frac{1}{m} \right), \qquad h_{AB}=\frac{2c}{24}\left( n-\frac{1}{n} \right).
\end{align}
According to \cite{Dutta:2019gen}, in the semi-classical approximation the holographic dual of the reflected entropy for a bipartite system in CFT$_d$ is proposed to be twice the entanglement wedge cross section (EWCS) \cite{Takayanagi:2017knl} in the bulk AdS$_{d+1}$ geometry and is defined as the minimal cross sectional area of the entanglement wedge of $A \cup B$ as
\begin{align}\label{eq_hre}
S_R=\frac{E_W}{2G_N}=\frac{Area [\Sigma_{AB}]}{2G_N},
\end{align}
where $\Sigma_{AB}$ denotes the EWCS which partitions the entanglement wedge of  $A \cup B$ into two different parts.

In the defect AdS$_3$/BCFT$_2$ framework, the reflected entropy between two subsystems $A$ and $B$ on the asymptotic boundary is computed using the island formula, detailed in \cite{Li:2020ceg, Chandrasekaran:2020qtn} as
\begin{equation}\label{eq_sris}
S^{(Is)}_{R}(A:B)= \text{min}~ \text{Ext}_{\Gamma} \Bigl\{S^{\text{eff}}_{R}(A\cup I_{S_{R}}(A):B \cup I_{S_{R}}(B)) + S_R ^{area} (\Gamma)\Bigl\}
\end{equation}
where the reflected entropy islands $I_{S_{R}}(A)$ and $I_{S_{R}}(B)$ partition the entanglement entropy island $I_{S}(A \cup B)$ into two parts, while  $\Gamma$ refers to the island cross section $I_{S_{R}}(A) \cap I_{S_{R}}(B)$. The first term is the effective reflected entropy between the subregions $A \cup I_{S_{R}}(A)$ and $B \cup I_{S_{R}}(B)$, which may be computed using \cref{eq_re}.\footnote{Note that though in \cref{eq_re} the reflected entropy is computed in terms of 4-point correlators in both the numerator and the denominator (corresponding to the choice of two disjoint intervals $A$ and $B$), in the course of this article we will observe that in the island formalism the correlator in both the numerator and the denominator may be different based on the presence (or absence) of reflected entropy islands on the brane corresponding to the subsystems $A$ and $B$. Moreover, the presence of the composite operator $\sigma_{g_A^{-1} g_B}$ on the effective $2d$ CFT also impacts the choice of correlators, which will be evident in \cref{sec_adjacent}. \label{ft_re}} The second term takes into account the contributions due to the area of the island cross section $\Gamma$.

In the bulk description, the reflected entropy is computed using the DES formula described in \cite{Li:2021dmf} as
\begin{equation}\label{eq_srdes}
S^{(DES)}_{R}(A:B)= \text{min}~ \text{Ext}_{\Sigma} \Bigl\{S_R(\Sigma_{AB}) + S^{defect}_{R}(\mathcal{A}:\mathcal{B})\Bigl\},
\end{equation}
where once again the EWCS $\Sigma_{AB}$ splits the entanglement wedge of  $A \cup B$ into two distinct regions $\mathcal{A}$ and $\mathcal{B}$. The first term corresponds the holographic reflected entropy related to the area of the $\Sigma_{AB}$ which may be computed using \cref{eq_hre}, while the second terms takes into account the reflected entropy between the bulk regions $\mathcal{A}$ and $\mathcal{B}$. However, since we assume bulk conformal matter to be localized only on the EOW brane, the second term simplifies to the reflected entropy between the reflected entropy islands $I_{S_{R}}(A)$ and $I_{S_{R}}(B)$ on the brane. This results in $S^{defect}_{R}(\mathcal{A}:\mathcal{B})=S^{defect}_{R}(I_{S_{R}}(A):I_{S_{R}}(B))$.

\section{Holographic Reflected Entropy for Two Disjoint Intervals}\label{sec_disjoint}

Having reviewed the essential concepts relevant to this article in the previous section, we now investigate the entanglement structure for the bipartite mixed state of two disjoint intervals within the radiation bath at the asymptotic boundary. Specifically, we compute the reflected entropy between the intervals $A$ and $B$ in the left and right copy of the TFD state for a $2d$ eternal black hole. These intervals are defined in the primed coordinates mentioned in \cref{ssec_ebh} as $A \equiv [J(- x_2', \tau_2', \epsilon_2'),I(- x_1', \tau_1', \epsilon_1')] \cup [M(x_1', \tau_1', \epsilon_1'),N(x_2', \tau_2', \epsilon_2')]$ and $B \equiv [L(- x_4', \tau_4', \epsilon_4'),K(- x_3', \tau_3', \epsilon_3')] \cup [O(x_3', \tau_3', \epsilon_3'),P(x_4', \tau_4', \epsilon_4')]$. For convenience of computation the unprimed coordinates are chosen as the initial reference frame, where the corresponding endpoints of the intervals may be determined using the transformations provided in \cref{eq_trans1}.\footnote{Though initially distinct UV cut-offs are assumed at different points, an underlying assumption is that $\epsilon_1, \epsilon_2, \epsilon_3, \epsilon_4 \to 0$ due to the fact that all subsystems reside on the asymptotic boundary.} For such a configuration of disjoint intervals multiple entanglement entropy phases may arise, and for each entanglement entropy phase multiple reflected entropy phases are observed depending on the relative size and location of the intervals.

In what follows we begin with the identification of the different entanglement entropy phases for the interval $A \cup B$, and subsequently compute the reflected entropy between the subsystems $A$ and $B$ utilizing both the island and the DES prescription (assuming $c' = c$ for simplicity of computations). It is observed that the reflected entropy computed from the two proposals match for the corresponding phases illustrating the validity of the duality described in \cref{ssec_re}. Finally we describe and analyze the Page curves for both the entanglement entropy and the reflected entropy phases for various subsystem sizes and locations while maintaining a fixed brane angle.

\subsection{Entanglement Entropy Phase 1}

\begin{figure}[t]
\centering
\includegraphics[scale=1]{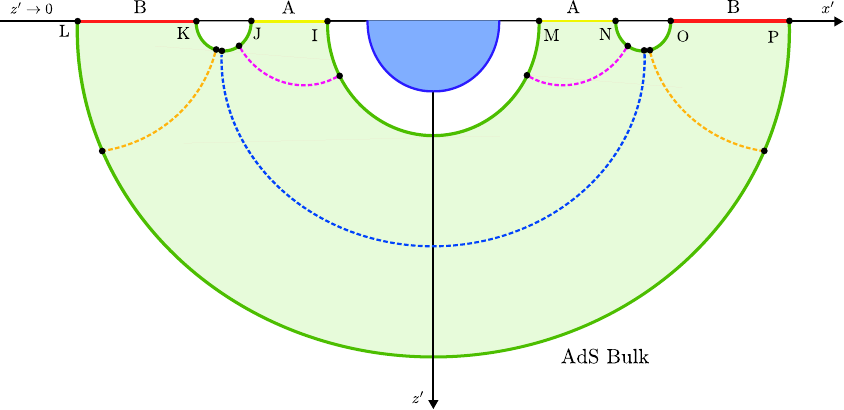}
\caption{EE Phase-1 for two disjoint intervals $A$ and $B$. The RT surfaces corresponding to $A \cup B$ are represented by the green lines, while the dashed lines depict the different configurations of the EWCS.}\label{fig_DE1}
\end{figure}

In this phase the two disjoint intervals $A$ and $B$ are considered to be large and located away from the brane. As outlined in \cref{ssec_eedes}, the entanglement entropy for this phase may be computed by combining the contributions from the two HM surfaces $LP$ and $IM$, along with RT surfaces $KJ$ and $NO$, all denoted by green lines in \cref{fig_DE1}. The lengths of these minimal surfaces can be determined using \cref{eq_geol}, and the entanglement entropy of $A \cup B$ in this phase may be obtained as
\begin{align}\label{eq_de1}
S(A \cup B)=\frac{2c}{3} \left( \log \left[ \frac{2 \cosh T}{\epsilon _R} \right] + \log \left[ \frac{2 \sinh \left( \frac{X_3-X_2}{2} \right) }{\epsilon _R} \right] \right).
\end{align}
For this entanglement entropy phase, there are three possible reflected entropy phases (depicted by the dashed lines in \cref{fig_DE1}) based on the relative sizes of $A$ and $B$ which are computed below along with the corresponding bulk EWCS as follows

\subsubsection*{Reflected Entropy Phase 1}\label{sssec_de1r1}

\textbf{Island prescription:}  Here, the size of the subsystem $A$ is assumed to be comparable to that of the subsystem $B$. The reflected entropy between $A$ and $B$ in this phase may be obtained using the island formula in \cref{eq_sris}, where the first term may be determined using \cref{eq_re}. The numerator on this case involves an
8-point correlator of twist fields corresponding to the boundary points $L,K,J,I,M,N,O$ and $P$ (refer to \cref{ft_re}), which in the large $c$ limit may be factorized as
\begin{align}\label{eq_8pf_de1r1}
& \langle \sigma_{g_B}(L) \sigma_{g_B^{-1}}(K) \sigma_{g_A}(J) \sigma_{g_A^{-1}}(I) \sigma_{g_A}(M) \sigma_{g_A^{-1}}(N) \sigma_{g_B}(O) \sigma_{g_B^{-1}}(P) \rangle _{CFT^{\otimes mn}} \notag \\
& \quad \approx \langle \sigma_{g_B}(L) \sigma_{g_B^{-1}}(P) \rangle  _{CFT^{\otimes mn}} \langle \sigma_{g_A^{-1}}(I) \sigma_{g_A}(M) \rangle  _{CFT^{\otimes mn}} \langle \sigma_{g_B^{-1}}(K) \sigma_{g_A}(J) \sigma_{g_A^{-1}}(N) \sigma_{g_B}(O) \rangle  _{CFT^{\otimes mn}}.
\end{align}
The 8-point twist correlator in the denominator (obtained by taking the $n \to 1$ limit of that in the numerator \cite{Dutta:2019gen}) undergoes a similar factorization in the large $c$ approximation as
\begin{align}
& \langle \sigma_{g_m}(L) \sigma_{g_m^{-1}}(K) \sigma_{g_m}(J) \sigma_{g_m^{-1}}(I) \sigma_{g_m}(M) \sigma_{g_m^{-1}}(N) \sigma_{g_m}(O) \sigma_{g_m^{-1}}(P) \rangle _{CFT^{\otimes m}} \notag \\
& \quad \approx \langle \sigma_{g_m}(L) \sigma_{g_m^{-1}}(P) \rangle  _{CFT^{\otimes m}} \langle \sigma_{g_m^{-1}}(I) \sigma_{g_m}(M) \rangle  _{CFT^{\otimes m}} \langle \sigma_{g_m^{-1}}(K) \sigma_{g_m}(J) \sigma_{g_m^{-1}}(N) \sigma_{g_m}(O) \rangle  _{CFT^{\otimes m}}.
\end{align}
In the limit $n \to 1$, the contributions arising from the 2-point correlators in the numerator cancels out with those in the denominator. As a result, the effective reflected entropy receives contribution only from the 4-point correlator and is given as follows
\begin{align}
S^{eff}_R (A:B)=\lim _{m,n \to 1} \frac{1}{1-n} \log  \frac{ \langle \sigma_{g_B^{-1}}(K) \sigma_{g_A}(J) \sigma_{g_A^{-1}}(N) \sigma_{g_B}(O) \rangle _{CFT^{\otimes mn}}}{\langle  \sigma_{g_m^{-1}}(K) \sigma_{g_m}(J) \sigma_{g_m^{-1}}(N) \sigma_{g_m}(O) \rangle ^n  _{CFT^{\otimes m}}}.
\end{align}
In this phase, the absence of twist fields on the brane leads to the vanishing of the area term in \cref{eq_sris}. Consequently, the generalized reflected entropy simplifies to the effective reflected entropy in this case, i.e. $S^{gen}_R (A:B)=S^{eff}_R (A:B)$. Utilizing the form of the 4-point function provided in \cite{Dutta:2019gen,Fitzpatrick:2014vua} the final reflected entropy for this phase may be derived as
\begin{align}\label{eq_de1r1}
S_R(A : B)=\frac{c}{6} \left( \log \left[ \frac{1+\sqrt{\eta}}{1-\sqrt{\eta}} \right]+\log \left[ \frac{1+\sqrt{\bar{\eta}}}{1-\sqrt{\bar{\eta}}} \right] \right),
\end{align}
where the cross ratios are expressed as
\begin{align}\label{eq_de1r1cross}
\eta =\bar{\eta}= \frac{4 x_2 x_3}{(x_3+x_2)^2+(\tau_3-\tau_2)^2} = \frac{1+\cosh (2T)}{\cosh (2T)+\cosh (X_3-X_2)}.
\end{align}
Note that in the final step of \cref{eq_de1r1cross} the set of transformations in \cref{eq_trans1,eq_trans21,eq_trans22} were utilized to express the result in the near horizon Rindler coordinates.

\vspace{0.5cm}

\textbf{Bulk prescription:} The holographic reflected entropy in this phase may be computed utilizing the DES formula given as \cref{eq_srdes}. The first term may be expressed in terms of the EWCS (depicted by the blue dashed line in \cref{fig_DE1}) involving the boundary points $K,J,N$ and $O$, which may be obtained using (check \cref{appD} for details) \cite{Kusuki:2019evw}
\begin{align}\label{eq_ewd}
E_W=\cosh ^{-1} \left[ \frac{\sqrt{\xi_{31}\xi_{42}}+\sqrt{\xi_{21}\xi_{43}}}{\sqrt{\xi_{41}\xi_{32}}} \right],
\end{align}
where we set $Y_1 \to J, Y_2 \to N, Y_3 \to O$ and $Y_4 \to K$. Here $\xi_{ji}$ is related to the inner product between the position vectors of the two points $Y_j$ and $Y_i$ and is given in \cref{eq_ip}. The holographic reflected entropy may now be computed from \cref {eq_ewd} using \cref{eq_hre}. Since this phase lacks an island region on the brane, the second term in \cref{eq_srdes} does not contribute. Consequently, the final expression for the holographic reflected entropy may be given by \cref{eq_de1r1,eq_de1r1cross} upon utilizing the Brown-Henneaux relation $c=\frac{3 \ell}{2 G_N}$ \cite{Brown:1986nw}.

\subsubsection*{Reflected Entropy Phase 2}

\textbf{Island prescription:} In this configuration, subsystem $A$ is assumed to be smaller than subsystem $B$. The reflected entropy is computed using the island formula \cref{eq_sris}, where the first term is once again determined using \cref{eq_re}. The numerator may be expressed in terms of the 8-point correlator of twist fields at the boundary points $L,K,J, I,M,N,O$ and $P$. In the large $c$ limit, the correlator may further be factorized as
\begin{align}\label{eq_8pf_de1r2}
& \langle \sigma_{g_B}(L) \sigma_{g_B^{-1}}(K) \sigma_{g_A}(J) \sigma_{g_A^{-1}}(I) \sigma_{g_A}(M) \sigma_{g_A^{-1}}(N) \sigma_{g_B}(O) \sigma_{g_B^{-1}}(P) \rangle _{CFT^{\otimes mn}} \notag \\
& \quad \approx \langle \sigma_{g_B}(L) \sigma_{g_B^{-1}}(P) \rangle  _{CFT^{\otimes mn}} \langle \sigma_{g_B^{-1}}(K) \sigma_{g_A}(J) \sigma_{g_A^{-1}}(I) \sigma_{g_A}(M) \sigma_{g_A^{-1}}(N) \sigma_{g_B}(O) \rangle _{CFT^{\otimes mn}}.
\end{align}
The 6-point function may be expanded in terms of the conformal block $\mathcal{F}_6$ which factorizes into a product of two 4-point conformal blocks $\mathcal{F}_4$ in the OPE channel. This specific factorization, referred to as the $\Omega_1$ channel in \cite{Banerjee:2016qca}, is given by
\begin{align}\label{eq_fact_de1r2}
\mathcal{F}_6 (K,J,I,M,N,O;h_A,h_B,h_{AB}) = & \mathcal{F}_4 (K,J,I,M;h_A,h_B,h_{AB}) \notag \\ & \qquad \times \mathcal{F}_4 (I,M,N,O;h_A,h_B,h_{AB}).
\end{align}
In the large $c$ limit the dominant contribution in the above 4-point conformal blocks arises from the composite field $\sigma_{g_A^{-1} g_B}$ characterized by the conformal dimension $h_{AB}$ as provided in \cref{eq_cdim}.

The denominator in \cref{eq_re}, which is obtained by taking the $n \to 1$ limit of that in the numerator \cite{Dutta:2019gen}, is also a 8-point correlator of twist fields $\sigma_{g_m}$ in this scenario and factorizes in a similar fashion. The effective reflected entropy may finally be derived as
\begin{align}\label{eq_de1r2_eff}
S^{eff}_R (A:B)=\lim _{m,n \to 1} \frac{1}{1-n} \log & \left( \frac{ \langle \sigma_{g_B^{-1}}(K) \sigma_{g_A}(J) \sigma_{g_A^{-1}}(I) \sigma_{g_A}(M) \rangle _{CFT^{\otimes mn}}}{\langle  \sigma_{g_m^{-1}}(K) \sigma_{g_m}(J) \sigma_{g_m^{-1}}(I) \sigma_{g_m}(M) \rangle ^n  _{CFT^{\otimes m}}} \right. \notag \\
& \qquad \left. \times \frac{ \langle \sigma_{g_A^{-1}}(I) \sigma_{g_A}(M) \sigma_{g_A^{-1}}(N) \sigma_{g_B}(O) \rangle _{CFT^{\otimes mn}}}{\langle  \sigma_{g_m^{-1}}(I) \sigma_{g_m}(M) \sigma_{g_m^{-1}}(N) \sigma_{g_m}(O) \rangle ^n  _{CFT^{\otimes m}}} \right).
\end{align}
where once again the contributions from to the 2-point correlators in both the numerator and the denominator cancel out in the limit $n \to 1$. As argued earlier the area term (the second term in \cref{eq_sris}) also vanishes in this phase leading to $S^{gen}_R (A:B)=S^{eff}_R (A:B)$. By utilizing the form of the 4-point function provided in \cite{Dutta:2019gen,Fitzpatrick:2014vua}, the final expression for the reflected entropy may be determined as
\begin{align}\label{eq_de1r2}
S_R(A : B)=\frac{c}{3} \left( \log \left[ \frac{1+\sqrt{\eta}}{1-\sqrt{\eta}} \right] +  \log \left[ \frac{1+\sqrt{\bar{\eta}}}{1-\sqrt{\bar{\eta}}} \right] \right),
\end{align}
where the cross ratios in this case have the form
\begin{align}\label{eq_de1r2cross}
\eta =  \frac{\cosh \left( \frac{X_3-X_1-2T}{2} \right) \sinh \left( \frac{X_2-X_1}{2} \right)}{\cosh \left( \frac{X_2-X_1-2T}{2} \right) \sinh \left( \frac{X_3-X_1}{2} \right)}, \qquad \bar{\eta} =  \frac{\cosh \left( \frac{X_3-X_1+2T}{2} \right) \sinh \left( \frac{X_2-X_1}{2} \right)}{\cosh \left( \frac{X_2-X_1+2T}{2} \right) \sinh \left( \frac{X_3-X_1}{2} \right)}.
\end{align}
As in the previous case, the final step involves applying the transformations outlined in \cref{eq_trans1,eq_trans21,eq_trans22} to express the result in near-horizon Rindler coordinates.

\vspace{0.5cm}

\textbf{Bulk prescription:} From the bulk perspective the holographic reflected entropy is computed using the DES formula outlined in \cref{eq_srdes}, where the first term is determined in terms of the two contributions to the EWCS as depicted by the magenta dashed lines in \cref{fig_DE1}. For the left TFD copy the contribution to the EWCS may be obtained using \cref{eq_ewd} with the assignment of $Y_1 \to J, Y_2 \to I, Y_3 \to M$ and $Y_4 \to K$. Similarly, for the right TFD copy, assigning $Y_1 \to O, Y_2 \to I, Y_3 \to M$ and $Y_4 \to N$ provides the second contribution to the EWCS. The bulk EWCS is then obtained as the sum of the left and the right contributions and the holographic reflected entropy may be computed using \cref{eq_hre}. Similar to the previous phase, this reflected entropy configuration does not receive contribution from the brane defect term in \cref{eq_srdes}. As a result, the holographic reflected entropy between the subsystems $A$ and $B$ in this phase may be obtained as \cref{eq_de1r2,eq_de1r2cross} upon utilizing the Brown-Henneaux relation.

\subsubsection*{Reflected Entropy Phase 3}\label{sssec_de1r3}

In this phase subsystem $A$ is considered to be larger than subsystem $B$. The computation for the reflected entropy in this phase using the island prescription is very similar to that of the previous phase and the final result may be obtained by the substitution $X_1 \to -X_4$ in \cref{eq_de1r2} as
\begin{align}\label{eq_de1r3}
S_R(A : B)=\frac{c}{6} \left( \log \left[ \frac{1+\sqrt{\eta}}{1-\sqrt{\eta}} \right]+\log \left[ \frac{1+\sqrt{\bar{\eta}}}{1-\sqrt{\bar{\eta}}} \right] \right),
\end{align}
with cross ratios
\begin{align}\label{eq_de1r3cross}
\eta =  \frac{\cosh \left( \frac{X_3+X_4-2T}{2} \right) \sinh \left( \frac{X_2+X_4}{2} \right)}{\cosh \left( \frac{X_2+X_4-2T}{2} \right) \sinh \left( \frac{X_3+X_4}{2} \right)}, \qquad \bar{\eta} =  \frac{\cosh \left( \frac{X_3+X_4+2T}{2} \right) \sinh \left( \frac{X_2+X_4}{2} \right)}{\cosh \left( \frac{X_2+X_4+2T}{2} \right) \sinh \left( \frac{X_3+X_4}{2} \right)}.
\end{align}

From the bulk perspective, the corresponding holographic reflected entropy in this phase may be computed using the DES prescription in terms of the contributions from the left and the right TFD copies to the EWCS, as depicted by the orange dashed lines in \cref{fig_DE1}. The holographic reflected entropy matches with the field theory result expressed in \cref{eq_de1r3,eq_de1r3cross}.

\subsection{Entanglement Entropy Phase 2}

\begin{figure}[t]
\centering
\includegraphics[scale=1]{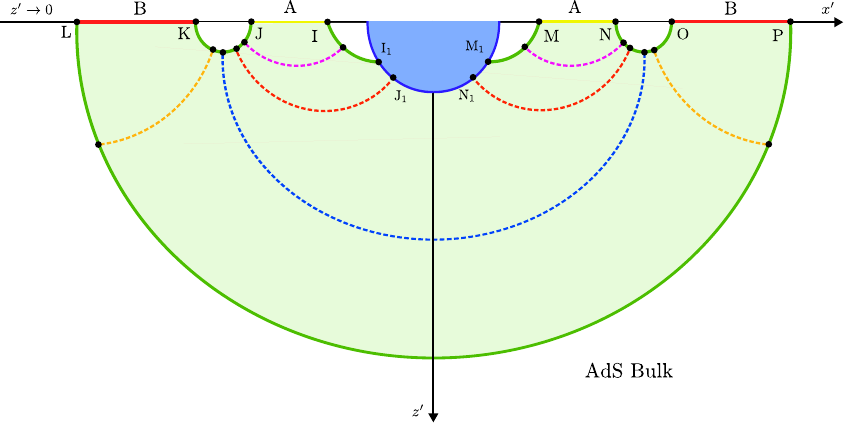}
\caption{EE Phase-2 for two disjoint intervals $A$ and $B$. The RT surfaces corresponding to $A \cup B$ are represented by the green lines, while the dashed lines depict the different configurations of the EWCS.}\label{fig_DE2}
\end{figure}

This phase corresponds to the scenario where the subsystem $A$ is close to the brane while $B$ is far away. The entanglement entropy in this configuration may be expressed in terms of the HM surface $LP$, along with the two RT surfaces $II_1$ and $MM_1$ which terminate on the brane, and the two RT surfaces $JK$ and $NO$, all represented by the green lines in \cref{fig_DE2}. Here $I_1 \equiv (- x_1,-\tau_1 \tanh (\sigma / \ell),\tau_1 \sech (\sigma / \ell))$ and $M_1 \equiv (x_1,-\tau_1 \tanh (\sigma / \ell),\tau_1 \sech (\sigma / \ell))$ are the points where the EOW brane and the RT surfaces $II_1$ and $MM_1$ intersect. In this case the entanglement entropy is obtained as 
\begin{align}\label{eq_de2}
S(A \cup B) & = \frac{c}{3} \left( \frac{\sigma}{\ell} + \log \left[ \frac{2 \cosh T}{\epsilon _R} \right] + \log \left[ \frac{2 \sinh X_1}{\epsilon _R} \right] \right. \notag \\
& \qquad \left. + 2 \log \left[ \frac{2 \sinh \left( \frac{X_3-X_2}{2} \right) }{\epsilon _R} \right] + \log \left[ \frac{2 \ell}{\epsilon _y \sech (\sigma / \ell)} \right] \right),
\end{align}
where it is observed that the entanglement entropy receives contribution from the bulk matter on the brane. Corresponding to this entanglement entropy phase, there are four distinct reflected entropy phases, each determined by the relative sizes of subsystems $A$ and $B$. The computations corresponding to these phases are outlined below.

\subsubsection*{Reflected Entropy Phase 1}

This phase corresponds to a configuration where the subsystems $A$ and $B$ are of comparable sizes. Using the island prescription in \cref{eq_sris}, the effective reflected entropy term may be determined using \cref{eq_re}, where in both the numerator and the denominator, we have 10-point correlators of twist fields at the boundary points $L,K,J,I,M,N,O,P$ and the brane points $I_1,M_1$. In the large $c$ limit, they factorize into products of three 2-point correlators and a 4-point correlator. Consequently, the effective reflected entropy in this phase is the same as \cref{eq_de1r2_eff} in the limit $n \to 1$. Due to absence of a reflected entropy island cross section on the brane, the second term in \cref{eq_sris} does not contribute. As a result, the final reflected entropy in this phase may directly be expressed as \cref{eq_de1r1,eq_de1r1cross}.

From the bulk perspective, the holographic reflected entropy is computed using the DES prescription in \cref{eq_srdes}, where the first term may be computed in terms of the EWCS depicted by the blue dashed line in \cref{fig_DE2}. This configuration of the EWCS closely resembles the first case in the previous entanglement entropy phase and thus the final holographic reflected entropy may be obtained directly from \cref{sssec_de1r1}.

\subsubsection*{Reflected Entropy Phase 2}\label{sssec_de2r2}

\textbf{Island perspective:} In this phase the subsystem $A$ is considered to be significantly smaller than the subsystem $B$, such that the entire entanglement entropy island of $A \cup B$ corresponds to the reflected entropy island of the subsystem $B$, while the subsystem $A$ lacks a reflected entropy island on the brane. Using the island prescription \cref{eq_sris} the effective reflected entropy may be described using \cref{eq_re}, where the numerator in this case is a 10-point correlator of twist fields at the boundary points $L,K,J,I,M,N,O,P$ and the brane points $I_1,M_1$. In the large $c$ limit, this correlator undergoes factorization into the product of a 2-point correlator and two 4-point correlators, which may be expressed as
\begin{align}\label{eq_8pf_de2r2}
& \langle \sigma_{g_B}(L) \sigma_{g_B^{-1}}(K) \sigma_{g_A}(J) \sigma_{g_A^{-1}}(I) \sigma_{g_A}(I_1) \sigma_{g_A^{-1}}(M_1) \sigma_{g_A}(M) \sigma_{g_A^{-1}}(N) \sigma_{g_B}(O) \sigma_{g_B^{-1}}(P) \rangle _{CFT^{\otimes mn}} \notag \\
& \quad \approx \langle \sigma_{g_B}(L) \sigma_{g_B^{-1}}(P) \rangle  _{CFT^{\otimes mn}} \langle \sigma_{g_B^{-1}}(K) \sigma_{g_A}(J) \sigma_{g_A^{-1}}(I) \sigma_{g_A}(I_1) \rangle _{CFT^{\otimes mn}} \notag \\
& \qquad \qquad \times \langle \sigma_{g_A^{-1}}(M_1) \sigma_{g_A}(M) \sigma_{g_A^{-1}}(N) \sigma_{g_B}(O) \rangle _{CFT^{\otimes mn}}.
\end{align}
The denominator is also a 10-point correlator, which undergoes a similar factorization as
\begin{align}
& \langle \sigma_{g_m}(L) \sigma_{g_m^{-1}}(K) \sigma_{g_m}(J) \sigma_{g_m^{-1}}(I) \sigma_{g_m}(I_1) \sigma_{g_m^{-1}}(M_1) \sigma_{g_m}(M) \sigma_{g_m^{-1}}(N) \sigma_{g_m}(O) \sigma_{g_m^{-1}}(P) \rangle _{CFT^{\otimes m}} \notag \\
& \quad \approx \langle \sigma_{g_m}(L) \sigma_{g_m^{-1}}(P) \rangle  _{CFT^{\otimes m}} \langle \sigma_{g_m^{-1}}(K) \sigma_{g_m}(J) \sigma_{g_m^{-1}}(I) \sigma_{g_m}(I_1) \rangle _{CFT^{\otimes m}} \notag \\
& \qquad \qquad \times \langle \sigma_{g_m^{-1}}(M_1) \sigma_{g_m}(M) \sigma_{g_m^{-1}}(N) \sigma_{g_m}(O) \rangle _{CFT^{\otimes m}}.
\end{align}
The effective reflected entropy may finally be expressed as 
\begin{align}\label{eq_de2r2_eff}
S^{eff}_R (A:B)=\lim _{m,n \to 1} \frac{1}{1-n} \log & \left( \Omega_{I_1}^{2 h_{g_{A}}} \frac{ \langle \sigma_{g_B^{-1}}(K) \sigma_{g_A}(J) \sigma_{g_A^{-1}}(I) \sigma_{g_A}(I_1) \rangle _{CFT^{\otimes mn}}}{\langle \sigma_{g_m^{-1}}(K) \sigma_{g_m}(J) \sigma_{g_m^{-1}}(I) \sigma_{g_m}(I_1) \rangle ^n  _{CFT^{\otimes m}}} \right. \notag \\
& \quad \left. \times \Omega_{M_1}^{2 h_{g_{A}}} \frac{ \langle \sigma_{g_A^{-1}}(M_1) \sigma_{g_A}(M) \sigma_{g_A^{-1}}(N) \sigma_{g_B}(O) \rangle _{CFT^{\otimes mn}}}{\langle \sigma_{g_m^{-1}}(M_1) \sigma_{g_m}(M) \sigma_{g_m^{-1}}(N) \sigma_{g_m}(O) \rangle ^n  _{CFT^{\otimes m}}} \right).
\end{align}
where the contributions due to the 2-point correlators in both the numerator and denominator cancel out in the limit $n \to 1$. Also recall that $\Omega _{I_1}$ and $\Omega _{M_1}$ are the conformal factors corresponding to the brane points $I_1$ and $M_1$. 

It is important to note that, while the entanglement entropy in this phase includes contributions from conformal matter on the brane, the area term (the second term in \cref{eq_sris}) does not contribute to the reflected entropy. This is due to the absence of a reflected entropy island corresponding to subsystem $A$, which in turn eliminates any reflected entropy island cross section within this phase. Consequently, we have $S^{gen}_R (A:B)=S^{eff}_R (A:B)$. Once again utilizing the form of the 4-point twist field correlator provided in \cite{Dutta:2019gen,Fitzpatrick:2014vua}, the final reflected entropy may be determined as
\begin{align}\label{eq_de2r2}
S_R(A : B)=\frac{c}{3} \left( \log \left[ \frac{1+\sqrt{\eta}}{1-\sqrt{\eta}} \right] +  \log \left[ \frac{1+\sqrt{\bar{\eta}}}{1-\sqrt{\bar{\eta}}} \right] \right),
\end{align}
with cross ratios
\begin{align}\label{eq_de2r2cross}
\eta =\bar{\eta}= \frac{ \sinh \left( \frac{X_2-X_1}{2} \right)  \sinh \left( \frac{X_3+X_1}{2} \right)}{\sinh \left( \frac{X_2+X_1}{2} \right)  \sinh \left( \frac{X_3-X_1}{2} \right)}.
\end{align}
As earlier the final step involves utilizing the transformations in \cref{eq_trans1,eq_trans21,eq_trans22} to reformulate the result in the near-horizon Rindler coordinates.

\vspace{0.5cm}

\textbf{Bulk perspective:} The holographic reflected entropy from the bulk perspective can be determined using the DES formula detailed in \cref{eq_srdes}, where the first term is expressed by the contributions of the left and the right TFD copies to the bulk EWCS, depicted by the magenta dashed lines in \cref{fig_DE2}. The contribution from the left copy of the TFD may be computed in terms of the three boundary points and a bulk point as (check \cite{Basu:2023jtf} and  \cref{appD} for details)
\begin{align}\label{eq_ewdm1}
E_W=\cosh ^{-1} \left[ \frac{\sqrt{(2 \xi_{32} \xi_{31}+\xi_{21})(2 \xi_{32} \xi_{42}+\xi_{43})}+\sqrt{(2 \xi_{21} \xi_{32}+\xi_{31})(2 \xi_{32} \xi_{43}+\xi_{42})}}{2\sqrt{\xi_{41}\xi_{32}^3}} \right],
\end{align}
where we assign $Y_1 \to J, Y_2 \to I, Y_3 \to I_1$ and $Y_4 \to K$.  Similarly the contribution from the right TFD copy to the EWCS is computed by assigning $Y_1 \to N, Y_2 \to M, Y_3 \to M_1$ and $Y_4 \to O$. The bulk EWCS is then obtained from the sum of the left and the right contributions and the holographic reflected entropy may then be obtained using \cref{eq_hre}.

Since the subsystem $A$ lacks a corresponding reflected entropy island on the brane, the brane defect term (the second term in \cref{eq_srdes}) does not contribute to the total holographic reflected entropy in this phase. By applying the Brown-Henneaux relation, the final expression for the holographic reflected entropy may thus be described as \cref{eq_de2r2,eq_de2r2cross}.

\subsubsection*{Reflected Entropy Phase 3}

In this phase the subsystem $A$ is significantly larger than the subsystem $B$, as a result of which the subsystem $B$ lacks a corresponding reflected entropy island on the brane. Consequently, the entire entanglement entropy island of $A \cup B$ corresponds to the reflected entropy island of $A$. Using the island formula the effective reflected entropy term in the limit $n \to 1$ is obtained to be the same as that described in the third phase of \cref{sssec_de1r3}. Once again the final reflected entropy does not receive contribution from the area term, and thus may be expressed directly as \cref{eq_de1r3,eq_de1r3cross}.

From the bulk perspective, the holographic reflected entropy may be computed using the DES formula, where the first term may be expressed in terms of the contributions of the two TFD copies to the bulk EWCS, which are denoted by the orange dashed lines in \cref{fig_DE2}. This configuration of the bulk EWCS is similar to that in the third case in the previous entanglement entropy phase, as a result of which the holographic reflected entropy may once again be determined directly from \cref{sssec_de1r3}.

\subsubsection*{Reflected Entropy Phase 4}\label{sssec_de2r4}

\textbf{Island prescription:} In this phase both the subsystems $A$ and $B$ possess corresponding reflected entropy islands on the brane. Using the island prescription in \cref{eq_sris}, the effective reflected entropy may be formulated using \cref{eq_re}, where in the numerator we have a 12-point correlator of twist fields located at the boundary points $L,K,J,I,M,N,O,P$ and the brane points $I_1,J_1,N_1,M_1$. In the large $c$ limit, this correlator factorizes into
\begin{align}\label{eq_8pf_de2r4}
& \langle \sigma_{g_B}(L) \sigma_{g_B^{-1}}(P) \rangle  _{CFT^{\otimes mn}} \langle \sigma_{g_A^{-1}}(I) \sigma_{g_A}(I_1) \rangle _{CFT^{\otimes mn}} \langle \sigma_{g_A^{-1}}(M_1) \sigma_{g_A}(M) \rangle _{CFT^{\otimes mn}} \notag \\
& \qquad \times \langle \sigma_{g_B^{-1}}(K) \sigma_{g_A}(J) \sigma_{g_A^{-1} g_B}(J_1) \rangle _{CFT^{\otimes mn}} \langle \sigma_{g_A g_B^{-1}}(N_1) \sigma_{g_A^{-1}}(N) \sigma_{g_B}(O) \rangle _{CFT^{\otimes mn}}.
\end{align}
The 12-point correlator in the denominator, determined by the $n \to 1$ limit of that in the numerator, undergoes a similar factorization in the large $c$ approximation. The effective reflected entropy may then be computed as
\begin{align}\label{eq_de2r4_eff}
S^{eff}_R (A:B)=\lim _{m,n \to 1} \frac{1}{1-n} \log & \left(\Omega_{J_1}^{2 h_{g_{AB}}} \frac{ \langle \sigma_{g_B^{-1}}(K) \sigma_{g_A}(J) \sigma_{g_A^{-1} g_B}(J_1) \rangle _{CFT^{\otimes mn}}}{\langle  \sigma_{g_m}(K) \sigma_{g_m}(J) \rangle ^n  _{CFT^{\otimes m}}} \right. \notag \\
& \qquad \left. \times \Omega_{N_1}^{2 h_{g_{AB}}} \frac{ \langle \sigma_{g_A g_B^{-1}}(N_1) \sigma_{g_A^{-1}}(N) \sigma_{g_B}(O) \rangle _{CFT^{\otimes mn}}}{\langle  \sigma_{g_m}(N) \sigma_{g_m}(O) \rangle ^n  _{CFT^{\otimes m}}} \right),
\end{align}
where in the large $c$ limit, the contributions of the first three terms in \cref{eq_8pf_de2r4} cancel out against those in the denominator. The conformal factors $\Omega _{J_1}$ and $\Omega _{N_1}$ correspond to the brane points $J_1$ and $N_1$ respectively. From symmetry of the configuration, the brane points may be parametrized as $I_1 \equiv (x_y, -y)$ and $N_1 \equiv (-x_y, -y)$, where $y$ represents the brane coordinate. Using the form of the 2-point twist field correlators described in \cref{ssec_eedes} and 3-point twist field correlators for reflected entropy provided in \cite{Dutta:2019gen}, the effective reflected entropy between $A$ and $B$ may be expressed as 
\begin{align}\label{eq_de2r4_sreff}
S_R^{eff}= \frac{c}{3} \left(\log \left[ \frac{((x_2-x_y)^2+(y+\tau_2)^2)  ((x_3-x_y)^2+(y+\tau_3)^2)}{y^2 ((x_3-x_2)^2+(\tau_3-\tau_2)^2) } \right] + 2 \log \left[ \frac{2 \ell }{\epsilon_y \sech (\sigma_0 / \ell)} \right] \right).
\end{align}
Since points $J_1$ and $N_1$ divide the entanglement entropy island into reflected entropy islands of $A$ and $B$, the contributions of the area terms (second term in \cref{eq_sris}) must be incorporated. This term may be directly obtained from \cite{Li:2021dmf} as
\begin{align}\label{eq_area}
S^{Area}_{R}=2 \times \frac{1}{2 G_N} \frac{\sigma _0}{\ell}=2 \times \frac{c}{3} \frac{\sigma _0}{\ell},
\end{align}
where in the second step the Brown-Henneaux relation was utilized. The generalized reflected entropy $S_R^{gen}$ in this phase may then be expressed as the sum of \cref{eq_de2r4_sreff,eq_area}. Extremizing $S_R^{gen}$ for $x_y$ and $y$ gives
\begin{align}\label{eq_yxy}
x_y=\frac{x_3 \tau_2+x_2\tau_3}{\tau_3+\tau_2}, \qquad y=\sqrt{\frac{\tau_2 \tau_2 ((x_3-x_2)^2+(\tau_3+\tau_2)^2)}{(\tau_3+\tau_2)^2}}.
\end{align}
The final reflected entropy between $A$ and $B$ may then be obtained by substituting \cref{eq_yxy} in $S_R^{gen}$ as follows 
\begin{align}\label{eq_de2r4}
S_R(A : B) & =\frac{c}{3} \left( 2 \frac{\sigma}{\ell} + \log \left[ 1-\frac{4 \left( \sinh X_2 \sinh X_3 + \sqrt{\sinh X_2 \sinh X_3} \sinh \left( \frac{X_3+X_2}{2} \right) \right)}{1-\cosh (X_3-X_2)} \right] \right. \notag \\
& \qquad \qquad \left. + 2 \log \left[ \frac{2 \ell}{\epsilon _y \sech (\sigma / \ell)} \right] \right).
\end{align}
Following the approach used in previous cases, the transformations in \cref{eq_trans1,eq_trans21,eq_trans22} were applied and the final result is expressed in near-horizon Rindler coordinates.

\vspace{0.5cm}

\textbf{Bulk perspective:} The holographic reflected entropy for this phase may be computed using the DES formula given as \cref{eq_srdes}, where the first term corresponds to the sum of the left and the right contributions to the EWCS represented by the red dashed lines in \cref{fig_DE2}. For the left copy of the TFD, the EWCS has one of the endpoints on an RT surface $KJ$, while the other lies on the brane at $J_1$ and is 
determined using the formulation described in \cref{appA} \cite{Basu:2023jtf} as
\begin{align}\label{eq_ewa}
E_W=\cosh ^{-1} \left[ \sqrt{-\frac{2 \xi_{21} \xi_{32}}{\xi_{31}}} \right],
\end{align}
where we have assigned $Y_1 \to K, Y_2 \to J_1$ and $Y_3 \to J$. The contribution from the right TFD copy to the EWCS may similarly be determined by assigning $Y_1 \to O, Y_2 \to N_1$ and $Y_3 \to N$. From symmetry of the configuration, the brane points $I_1$ and $M_1$ may be parametrized as $I_1 \equiv (x_y, -y \tanh (\sigma / \ell),y \sech (\sigma / \ell))$ and $N_1 \equiv (-x_y, -y \tanh (\sigma / \ell),y \sech (\sigma / \ell))$.  The effective holographic reflected entropy for this phase may be computed from the sum of the two contributions to the EWCS using \cref{eq_hre} as follows
\begin{align}\label{eq_de2r4_eweff}
S_R^{eff} & =2 \times \frac{1}{2G_N} \notag \\ & \times  \cosh ^{-1} \left[ \sqrt{\frac{((x_2-x_y)^2+y^2+\tau_2^2+2y\tau_2 \tanh (\sigma / \ell))((x_3-x_y)^2+y^2+\tau_3^2+2y\tau_3 \tanh (\sigma / \ell))}{\sech (\sigma / \ell)^2 y^2 ((x_3-x_2)^2+(\tau_3-\tau_2)^2)}} \right].
\end{align}
Since the EWCS partitions the entanglement entropy island into two reflected entropy islands, it is crucial to incorporate contributions from the second term in \cref{eq_srdes}. The contribution to the holographic reflected entropy arising from the brane defects may be obtained directly from \cite{Li:2021dmf} and is given as
\begin{align}\label{eq_defect}
S^{defect}_R=2 \times \frac{c}{3} \log \left[ \frac{2 \ell}{\epsilon _y \sech (\sigma / \ell)} \right],
\end{align}
where the factor of 2 arises due to the two points of intersection between the brane and the EWCS. The generalized reflected entropy $S_R^{gen}$ for this phase is therefore determined by adding \cref{eq_de2r4_eweff,eq_defect}, which is extremized at the values of $x_y$ and $y$ provided in \cref{eq_yxy}. Finally, using the Brown-Henneaux relation, the reflected entropy using the bulk prescription can be expressed as \cref{eq_de2r4}.

\subsection{Entanglement Entropy Phase 3}

\begin{figure}[t]
\centering
\includegraphics[scale=1]{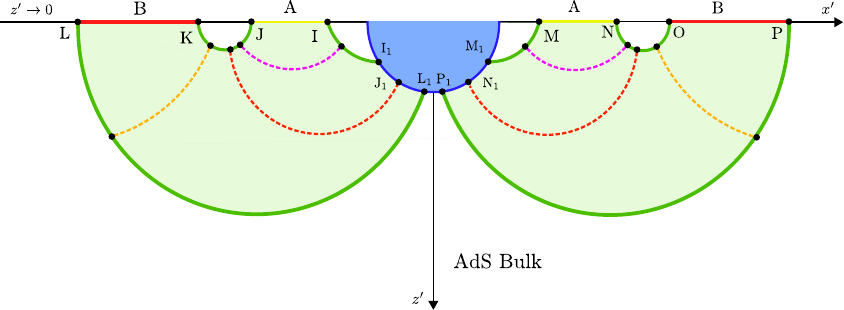}
\caption{EE Phase-3 for two disjoint intervals $A$ and $B$. The RT surfaces corresponding to $A \cup B$ are represented by the green lines, while the dashed lines depict the different configurations of the EWCS.}\label{fig_DE3}
\end{figure}

In this phase, both the subsystems $A$ and $B$ are positioned close to the brane, and the separation between them is also considered to be small. The entanglement entropy may be determined in terms of the combinations of RT surfaces $LL_1, PP_1,MM_1, II_1, JK$ and $NO$, all represented by the green lines in \cref{fig_DE3}. Utilizing \cref{eq_geol}, the entanglement entropy may be computed as
\begin{align}\label{eq_de3}
S(A \cup B)= \frac{c}{3} \left( 2 \frac{\sigma}{\ell} + \log \left[ \frac{4 \sinh X_4 \sinh X_1}{\epsilon _R ^2} \right] +  2 \log \left[ \frac{2 \sinh \left( \frac{X_3-X_2}{2} \right) }{\epsilon _R} \right] + 2 \log \left[ \frac{2 \ell}{\epsilon _y \sech (\sigma / \ell)} \right] \right).
\end{align}
In this entanglement entropy phase, three distinct reflected entropy phases are observed depending on the relative sizes of the subsystems $A$ and $B$.

\subsubsection*{Reflected Entropy Phase 1}

This phase corresponds to the scenario where the subsystem $A$ is smaller that the subsystem $B$, as a result of which $A$ has no corresponding reflected entropy island on the brane and the reflected entropy island of $B$ spans the entanglement entropy island of $A \cup B$. The reflected entropy may be computed using the island prescription, where the effective reflected entropy term is given by \cref{eq_de2r2_eff} in the $n \to 1$ limit. The final reflected entropy in this phase may be directly described by \cref{eq_de2r2,eq_de2r2cross}.

Subsequently, the holographic reflected entropy may be determined using the DES formula, where the first term receives contributions from the bulk EWCS represented by the magenta dashed lines in \cref{fig_DE3}. Given the similarity in bulk EWCS structures with the second phase in \cref{sssec_de2r2}, the holographic reflected entropy may be expressed directly, and it matches with the field theory results above.

\subsubsection*{Reflected Entropy Phase 2}

In this scenario the subsystem $A$ is considered to be larger than the subsystem $B$. As a result, only $A$ has a corresponding reflected entropy island on the brane. Applying the island formula in \cref{eq_sris}, the effective reflected entropy may be determined using \cref{eq_re}, and in the limit $n \to 1$ it can be obtained by replacing $X_1 \to X_4$ and swapping $X_2 \leftrightarrow X_3$ in \cref{eq_de2r2_eff}. Consequently, the final reflected entropy in this phase is described as 
\begin{align}\label{eq_de3r2}
S_R(A : B)=\frac{c}{3} \left( \log \left[ \frac{1+\sqrt{\eta}}{1-\sqrt{\eta}} \right] +  \log \left[ \frac{1+\sqrt{\bar{\eta}}}{1-\sqrt{\bar{\eta}}} \right] \right),
\end{align}
with cross ratios
\begin{align}\label{eq_de3r2cross}
\eta =\bar{\eta}= \frac{ \sinh \left( \frac{X_4-X_3}{2} \right)  \sinh \left( \frac{X_4+X_2}{2} \right)}{\sinh \left( \frac{X_4+X_3}{2} \right)  \sinh \left( \frac{X_4-X_2}{2} \right)}.
\end{align}

From the bulk perspective, the holographic reflected entropy may be computed using the DES formula in terms of the bulk EWCS, depicted by the orange dashed lines in \cref{fig_DE3}. Once again the EWCS structure has similarities with those described in the second phase in \cref{sssec_de2r2} with the substitution of $X_1 \to X_4$ and the interchange of $X_2 \leftrightarrow X_3$. Utilizing these substitutions the holographic reflected entropy may then be directly obtained, and it matches with the field theory result described above.

\subsubsection*{Reflected Entropy Phase 3}

In this phase both the subsystems $A$ and $B$ are of comparable sizes, and have corresponding reflected entropy islands on the brane. From the island formula the effective reflected entropy may be expressed as \cref{eq_de2r4_eff}. Combined with the contributions from the area term, the final reflected entropy in this phase may be directly expressed as \cref{eq_de2r4}.

The holographic reflected entropy is computed using the DES formula where the first term may be expressed in terms of the two contributions to the EWCS represented by the red dashed lines in \cref{fig_DE3}, which resemble the EWCS structures described in the fourth phase in \cref{sssec_de2r4}. Including the contributions from the defect term the final holographic reflected entropy thus obtained matches with the field theory results described above.

\subsection{Entanglement Entropy Phase 4}

\begin{figure}[t]
\centering
\includegraphics[scale=1]{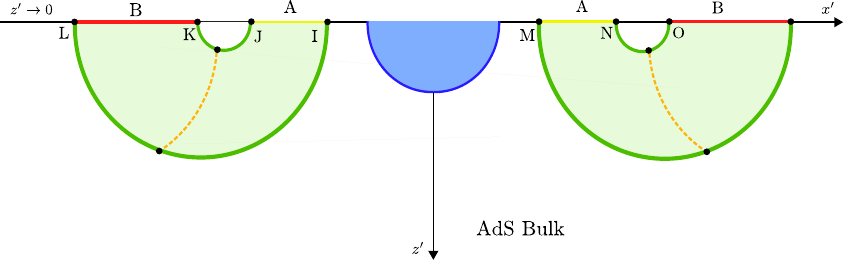}
\caption{EE Phase-4 for two disjoint intervals $A$ and $B$. The RT surfaces corresponding to $A \cup B$ are represented by the green lines, while the dashed lines depict the EWCS.}\label{fig_DE4}
\end{figure}

In this entanglement entropy phase the subsystems $A$ and $B$ are considered to be very small and close to each other. The entanglement entropy can be determined in terms of the RT surfaces $IL, JK, MO$ and $NO$, represented by the green lines in \cref{fig_DE4}. Utilizing \cref{eq_geol}, the entanglement entropy is computed as
\begin{align}\label{eq_de4}
S(A \cup B)=\frac{2c}{3} \left( \log \left[ \frac{2 \sinh \left( \frac{X_4-X_1}{2} \right) }{\epsilon _R} \right]+ \log \left[ \frac{2 \sinh \left( \frac{X_3-X_2}{2} \right) }{\epsilon _R} \right] \right).
\end{align}
In this phase, only a single reflected entropy phase is observed. The computations of the reflected entropy from both the island and the bulk perspective are as follows.

\subsubsection*{Reflected Entropy Phase 1}

\textbf{Island prescription:} The reflected entropy may be computed using the island formula given as \cref{eq_sris}, where the effective reflected entropy in this phase can be determined using \cref{eq_re}. In the numerator we have an 8-point correlator of twist fields at the boundary point $L,K,J,I,M,N,O,P$, which in the large $c$ limit factorizes into a product of two 4-point correlators as
\begin{align}\label{eq_8pf_de4r1}
\langle \sigma_{g_B}(L) \sigma_{g_B^{-1}}(K) \sigma_{g_A}(J) \sigma_{g_A^{-1}}(I) \rangle _{CFT^{\otimes mn}} \times \langle \sigma_{g_A}(M) \sigma_{g_A^{-1}}(N) \sigma_{g_B}(O) \sigma_{g_B^{-1}}(P) \rangle _{CFT^{\otimes mn}}
\end{align}
The 8-point correlator in the denominator undergoes a similar factorization in the large $c$ approximation. The effective reflected entropy may then be evaluated as 
\begin{align}
S^{eff}_R (A:B)=\lim _{m,n \to 1} \frac{1}{1-n} \log & \left( \frac{ \langle \sigma_{g_B}(L) \sigma_{g_B^{-1}}(K) \sigma_{g_A}(J) \sigma_{g_A^{-1}}(I) \rangle _{CFT^{\otimes mn}}}{\langle \sigma_{g_m}(L) \sigma_{g_m^{-1}}(K) \sigma_{g_m}(J) \sigma_{g_m^{-1}}(I) \rangle ^n  _{CFT^{\otimes m}}} \right. \notag \\
& \qquad \left. \times \frac{ \langle \sigma_{g_A}(M) \sigma_{g_A^{-1}}(N) \sigma_{g_B}(O) \sigma_{g_B^{-1}}(P) \rangle _{CFT^{\otimes mn}}}{\langle  \sigma_{g_m}(M) \sigma_{g_m^{-1}}(N) \sigma_{g_m}(O) \sigma_{g_m^{-1}}(P) \rangle ^n  _{CFT^{\otimes m}}} \right).
\end{align}
Since there are no reflected entropy islands in this phase the contribution from the area term vanishes, as a result we have $S^{gen}_R=S^{eff}_R$. By utilizing the form of the 4-point correlator given in \cite{Dutta:2019gen,Fitzpatrick:2014vua}, the final expression for the reflected entropy is then determined as
\begin{align}\label{eq_de4r1}
S_R(A : B)=\frac{c}{3} \left( \log \left[ \frac{1+\sqrt{\eta}}{1-\sqrt{\eta}} \right] +  \log \left[ \frac{1+\sqrt{\bar{\eta}}}{1-\sqrt{\bar{\eta}}} \right] \right),
\end{align}
with cross ratios
\begin{align}\label{eq_de4r1cross}
\eta =\bar{\eta}= \frac{ \sinh \left( \frac{X_4-X_3}{2} \right)  \sinh \left( \frac{X_2-X_1}{2} \right)}{\sinh \left( \frac{X_4-X_2}{2} \right)  \sinh \left( \frac{X_3-X_1}{2} \right)}.
\end{align}

\vspace{0.5cm}

\textbf{Bulk prescription:} The holographic reflected entropy may be computed using the DES formula given in \cref{eq_srdes}, where the first term is determined in terms of the two contributions to the bulk EWCS, depicted by the orange dashed lines in \cref{fig_DE4}. The contribution from the left copy of the TFD may be computed using \cref{eq_ewd} by assigning $Y_1 \to I, Y_2 \to J, Y_3 \to K$ and $Y_4 \to L$, while that from the right TFD copy is determined by setting $Y_1 \to M, Y_2 \to N, Y_3 \to O$ and $Y_4 \to P$. The bulk EWCS is obtained by combining the two contribution, and the first term in \cref{eq_srdes} may be computed using \cref{eq_hre}.

Since there are no reflected entropy islands on the brane, the defect term (the second term in \cref{eq_srdes}) does not contribute. The final holographic reflected entropy between $A$ and $B$ matches with the field theory results provided in \cref{eq_de4r1,eq_de4r1cross} on application of the Brown-Henneaux relation.

\subsection{Page Curves}\label{ssec_pcdisjoint}

\begin{figure}[t]
\centering
\begin{subfigure}{0.45\linewidth}
\includegraphics[scale=0.53]{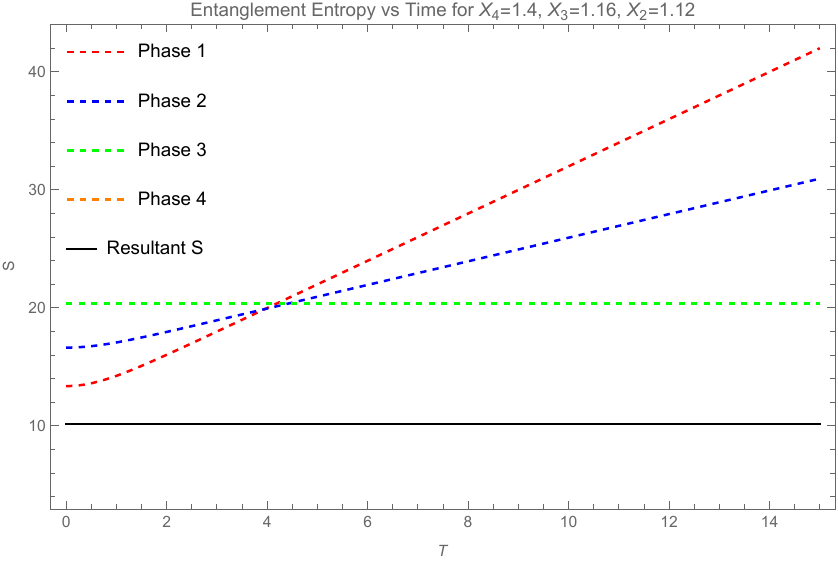}\caption{}
\label{fig_de1}
\end{subfigure}\hfill
\begin{subfigure}{0.45\linewidth}
\includegraphics[scale=0.53]{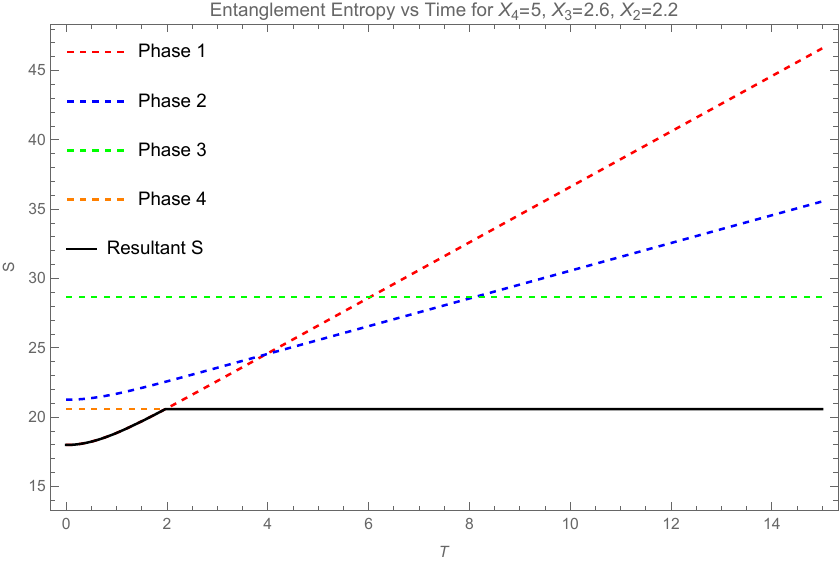}\caption{}
\label{fig_de2}
\end{subfigure}\hfill
\begin{subfigure}{0.45\linewidth}
\includegraphics[scale=0.53]{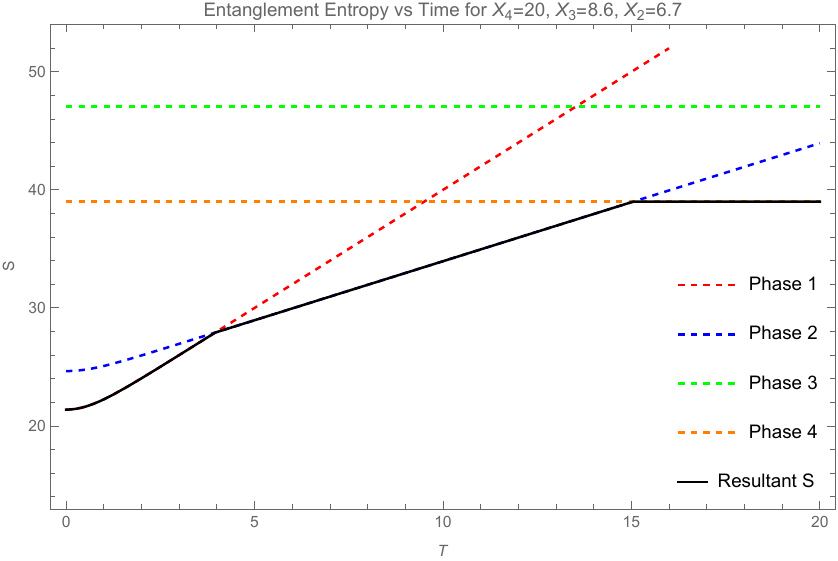}\caption{}
\label{fig_de3}
\end{subfigure}\hfill
\caption{Entanglement Entropy vs time plots for two disjoint radiation subsystems for different values of $X_4,X_3$ and $X_2$. The dashed curves represent the time evolution of the different EE phases, while the black solid curve denotes the minimum EE at any given time. We fix $c=3,\ell =1,\sigma_0 =1, X_1 = 1, \epsilon_R=0.01$ and $\epsilon_y=0.1$ to obtain the plots.}
\label{fig_de}
\end{figure}

In this subsection we plot and analyze the Page curves for the entanglement entropy and the reflected entropy involving the two disjoint subsystems $A$ and $B$ considered in the radiation bath. The Page curves for the entanglement entropy of the subsystem $A \cup B$, corresponding to varying subsystem sizes (specified by $X_4, X_3, X_2$ and $X_1$), are illustrated in \cref{fig_de}. This is followed by the Page curves for the reflected entropy between $A$ and $B$ for varying subsystem sizes. To this end, the entanglement entropy and the reflected entropy were evaluated under the assumption $c=3,\ell =1,\sigma_0 =1, X_1 = 1, \epsilon_R=0.01$ and $\epsilon_y=0.1$.

Note that a comparable analysis may be performed to examine how the variation of the brane angle $\sigma_0$ may effect the evolution of the entanglement entropy and the reflected entropy with time. Apart from a general shift in the transition times between phases and adjustments in the entanglement measure values, the overall structure of the curves remain unchanged as the brane angle is varied. 

\subsubsection{Entanglement Entropy Page Curves}

The Page curves for the entanglement entropy of subsystem $A \cup B$ are provided in \cref{fig_de}. We examine three different subsystem configurations specified by different values of $X_4, X_3$ and $X_2$ (with $X_1=1$). In these plots the dashed curves represent the time evolution of the entanglement entropy of $A \cup B$ for different phases, while the solid black curve represents the dominant entanglement entropy phase (the phase for which the entanglement entropy is minimum) as a function of time. The transition times relevant to the plots in \cref{fig_de} are as follows
\begin{align}\label{eq_dtt}
\text{Phase 1 to Phase 4 :} \quad T_{1 \to 4} ^{\text{Disj}} & = \cosh ^{-1} \left[ \sinh \left( \frac{X_4-X_1}{2} \right) \right], \notag \\
\text{Phase 1 to Phase 2 :} \quad T_{1 \to 2} ^{\text{Disj}} & =\cosh ^{-1} \left[ \frac{(1+e^{2 \sigma_0 /\ell}) \ell \sinh X_1}{\epsilon _y} \right], \\
\text{Phase 2 to Phase 4 :} \quad T_{2 \to 4} ^{\text{Disj}} & =\sech ^{-1} \left[ \frac{(1+e^{2 \sigma_0 /\ell}) \ell \sinh X_1}{\epsilon _y \sinh ^2 \left( \frac{X_4-X_1}{2} \right)} \right]. \notag
\end{align}
The above transition times are determined by equating the entanglement entropies of the initial and the final phase and solving for the time $T$. Subsequently, the transition time between the different reflected entropy phases are determined in a similar fashion.

\subsubsection{Reflected Entropy Page Curves}

\subsubsection*{Case 1: $X_4=1.4, X_3=1.16, X_2=1.12$}

\begin{figure}[t]
\centering
\begin{subfigure}{0.45\textwidth}
\includegraphics[scale=0.5]{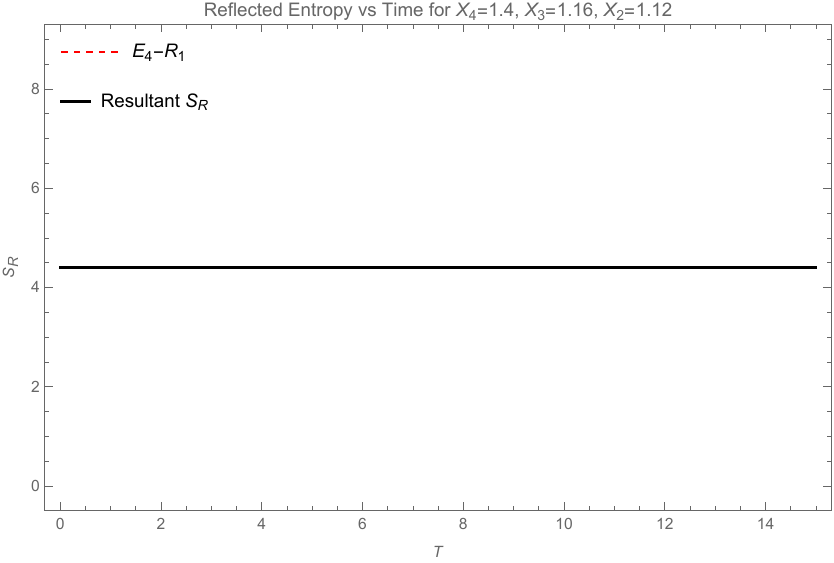}
\caption{}\label{fig_dr1}
\end{subfigure}\hfill
\begin{subfigure}{0.45\textwidth}
\includegraphics[scale=0.5]{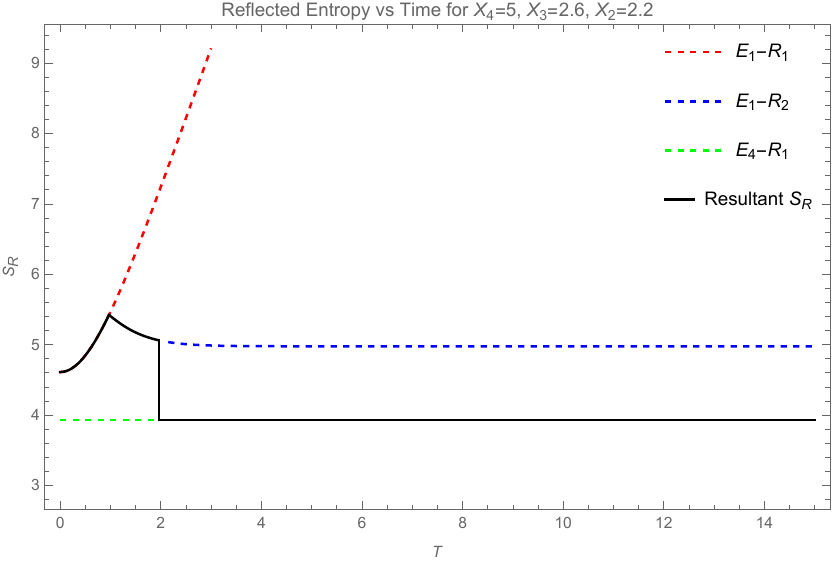}
\caption{}\label{fig_dr2}
\end{subfigure}\hfill \vspace{0.5cm}
\begin{subfigure}{0.45\textwidth}
\includegraphics[scale=0.5]{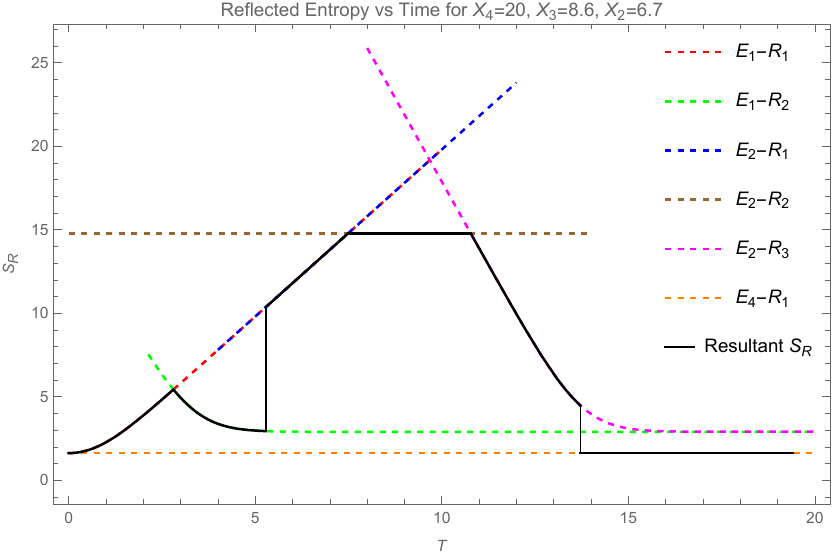}
\caption{}\label{fig_dr3}
\end{subfigure}\hfill
\caption{Analogous reflected entropy Page curves for (a) $X_4=1.4, X_3=1.16, X_2=1.12$, (b) $X_4=5, X_3=2.6, X_2=2.2$, and (c) $X_4=20, X_3=8.6, X_2=6.7$.}
\end{figure}

This case corresponds to a configuration where the subsystems $A$ and $B$ are very small and are very close to each other. The entanglement entropy Page curve is illustrated in \cref{fig_de1}, where we observe that \hyperref[eq_de4]{phase-4} of the EE dominates across all times. This EE phase contains only a single reflected entropy phase between $A$ and $B$, depicted by the dashed curve in \cref{fig_dr1}, which remains unchanged with time in this case. The solid black curve, which overlaps with the dashed curve in this case, represents the minimum of the reflected entropy.

\subsubsection*{Case 2: $X_4=5, X_3=2.6, X_2=2.2$}

In this scenario, where the subsystem $B$ is assumed to be larger and positioned further from both the EOW brane and the subsystem $A$, the entanglement entropy Page curve is illustrated in \cref{fig_de2}. The entanglement entropy initially rises with time due to the dominance of \hyperref[eq_de1]{phase-1}, before eventually stabilizing at a constant value determined by \hyperref[eq_de4]{phase-4}. The reflected entropy Page curve is presented in \cref{fig_dr2}, where once again the dashed curves represent the time evolution of the different reflected entropy phases involved, while the solid black curve is the minimum of the reflected entropy between $A$ and $B$ at any given time. As evident from \cref{fig_dr2}, the reflected entropy initially increases rapidly with time before transiting to a decreasing phase in EE \hyperref[eq_de1]{phase-1}. As the EE shifts to \hyperref[eq_de4]{phase-4}, the reflected entropy undergoes a transition to a lower, stable value.

\subsubsection*{Case 3: $X_4=20, X_3=8.6, X_2=6.7$}

In this scenario the subsystem $B$ is considered to be significantly larger and positioned very far from both the EOW brane and the subsystem $A$. \Cref{fig_de3} illustrates the Page curve of the entanglement entropy, where it shifts from a rapidly increasing \hyperref[eq_de1]{phase-1} to a much slowly rising \hyperref[eq_de2]{phase-2}, before finally stabilizing at some constant value in \hyperref[eq_de4]{phase-4}. The Page curves for the reflected entropy between $A$ and $B$ is illustrated in \cref{fig_dr3}, where once again the dashed curves and the black solid curve have their usual meaning. In this configuration the reflected entropy initially rises with time and then transitions to a decreasing phase within the EE \hyperref[eq_de1]{phase-1}. As the EE shifts to \hyperref[eq_de2]{phase-2}, the reflected entropy first increases, next it transits to a constant phase, and eventually starts decreasing. When the EE moves to \hyperref[eq_de4]{phase-4}, the reflected entropy undergoes a sharp transition to a lower constant value.

\section{Holographic Reflected Entropy for Two Adjacent Intervals}\label{sec_adjacent}

Having discussed the reflected entropy between two disjoint intervals, we now examine the reflected entropy between two adjacent subsystems $A$ and $B$ in the left and right copy of the TFD state for a $2d$ eternal black hole, defined as $A \equiv [J(- x_2', \tau_2', \epsilon_2'),I(- x_1', \tau_1', \epsilon_1')] \cup [M(x_1', \tau_1', \epsilon_1'),N(x_2', \tau_2', \epsilon_2')]$ and $B \equiv [K(- x_3', \tau_3', \epsilon_3'),J(- x_2', \tau_2', \epsilon_2')] \cup [N(x_2', \tau_2', \epsilon_2'),O(x_3', \tau_3', \epsilon_3')]$ in the primed coordinates. The unprimed coordinates are once again chosen as the initial reference frame for simplicity of the computations, and the corresponding endpoint of the intervals are obtained via the transformations provided in \cref{eq_trans1}. In this setup, multiple entanglement entropy phases can emerge. Within each of these phase, several distinct reflected entropy phases are observed, depending on the relative size of the subsystems and their location in relation to the brane.

We begin by identifying the different possible entanglement entropy phases of $A \cup B$. For each EE phase we compute the reflected entropy between $A$ and $B$ employing both the island and the DES formalism, assuming $c' = c$ for simplicity of the computations. It is observed that for each phase the reflected entropy computed using the two prescriptions are in agreement, illustrating the validity of the proposed duality described in \cref{ssec_re}.\footnote{The same results may also be reproduced from the disjoint interval computations described in \cref{sec_disjoint} by taking the adjacent limit where the separation between the two subsystems $A$ and $B$ is of the order of the UV cut-off. This illustrates the validity of our analysis and also provides additional substantiation for the proposed duality.} Finally we describe and analyze the Page curves for both the entanglement entropy and the reflected entropy for various subsystem sizes and locations relative to the brane, while maintaining a constant brane angle.

\subsection{Entanglement Entropy Phase 1}

\begin{figure}[t]
\centering
\includegraphics[scale=1]{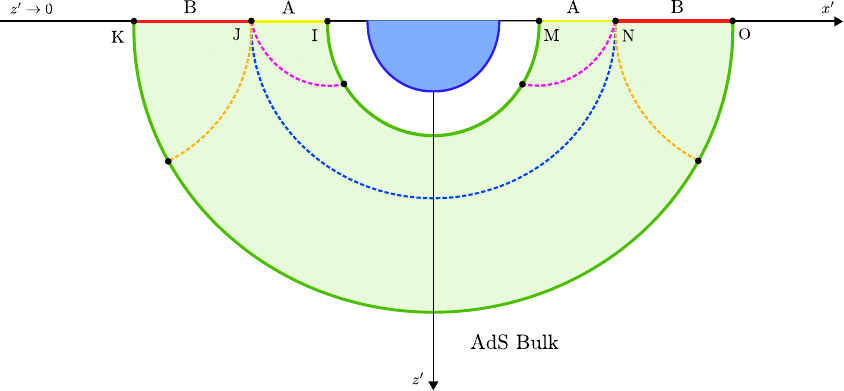}
\caption{EE Phase-1 for two adjacent intervals $A$ and $B$. The RT surfaces corresponding to $A \cup B$ are represented by the green lines, while the dashed lines depict the different configurations of the EWCS.}\label{fig_AE1}
\end{figure}

In this phase, the subsystems $A$ and $B$ are located far away from the brane. Consequently, the entanglement entropy of $A \cup B$ is determined by the combination of the two HM surfaces $KO$ and $IM$, represented by the green lines in \cref{fig_AE1}. The lengths of these geodesics may be computed using \cref{eq_geol}, and the entanglement entropy of $A \cup B$ in this phase may be obtained as 
\begin{align}\label{eq_ae1}
S(A \cup B)=\frac{2c}{3} \log \left[ \frac{2 \cosh T}{\epsilon _R} \right].
\end{align}
In this entanglement entropy phase three distinct reflected entropy phases are observed depending on the relative sizes of the subsystems $A$ and $B$, and are represented by the dashed lines in \cref{fig_AE1}. The explicit computations of the reflected entropy and the corresponding bulk EWCS are presented as follows.

\subsubsection*{Reflected Entropy Phase 1}\label{sssec_ae1r1}

\textbf{Island prescription:} In this phase the subsystems $A$ and $B$ are assumed to be approximately equal in size. The reflected entropy between $A$ and $B$ can be computed using the island prescription outlined in \cref{eq_sris}, where the effective reflected entropy may be determined using \cref{eq_re}. In the numerator we have a 6-point correlator of twist fields at the boundary points $K,J,I,M,N$ and $O$, which in the large $c$ limit may be factorized as
\begin{align}
& \langle \sigma_{g_B}(K) \sigma_{g_A g_B^{-1}}(J) \sigma_{g_A^{-1}}(I) \sigma_{g_A}(M) \sigma_{g_A^{-1} g_B}(N) \sigma_{g_B^{-1}}(O) \rangle _{CFT^{\otimes mn}} \notag \\
& \quad \approx \langle \sigma_{g_B}(K) \sigma_{g_B^{-1}}(O) \rangle _{CFT^{\otimes mn}} \langle \sigma_{g_A^{-1}}(I) \sigma_{g_A}(M) \rangle _{CFT^{\otimes mn}} \langle \sigma_{g_A g_B^{-1}}(J) \sigma_{g_A^{-1} g_B}(N) \rangle _{CFT^{\otimes mn}}.
\end{align}
In the denominator we have a 4-point correlator of twist fields $\sigma_{g_m}$ at the boundary points $K, I, M$ and $O$ determined by taking the $n \to 1$ limit of that in the numerator (as demonstrated in \cite{Dutta:2019gen}), which factorizes in a similar fashion. The effective reflected entropy may then be expressed as 
\begin{align}
S^{eff}_R (A:B)=\lim_{m,n \to 1} \frac{1}{1-n} \log \langle \sigma_{g_A g_B^{-1}}(J) \sigma_{g_A^{-1} g_B}(N) \rangle _{CFT^{\otimes mn}},
\end{align}
where we observe that the first two correlators in the numerator cancel out with those in the denominator in the limit $n \to 1$. Since neither of the subsystems have an island on the brane, the second term in \cref{eq_sris} does not contribute, leading to the simplification $S^{gen}_R=S^{eff}_R$. Applying the standard formula of the 2-point correlator described in \cref{ssec_eedes}, the final reflected entropy derived from the island prescription may be expressed as
\begin{align}\label{eq_ae1r1}
S_R(A : B)=\frac{2 c}{3} \log \left[ \frac{2 x_2}{\epsilon _2} \right]=\frac{2c}{3} \log \left[ \frac{2 \cosh T}{\epsilon _R} \right],
\end{align}
where in the second step the set of transformations outlined in \cref{eq_trans1,eq_trans21,eq_trans22} are applied and the final result is expressed in terms of the near horizon Rindler coordinates.

\vspace{0.5cm}

\textbf{Bulk prescription:} In this phase, the holographic reflected entropy between $A$ and $B$ may be calculated using the DES formula outlined in \cref{eq_srdes}. The first term is computed in terms the EWCS bound by the two boundary points $J$ and $N$ as depicted by the blue dashed line in \cref{fig_AE1}. The bulk EWCS can be computed using \cref{eq_geol} and its contribution to the holographic reflect entropy may be determined using \cref{eq_hre}.

Since the EWCS does not terminate on the brane in this scenario, the brane defect term (the second term in \cref{eq_srdes}) is zero. Subsequently applying the Brown-Henneaux relation, the final holographic reflected entropy may be obtained via the DES prescription, which matches with the field theory results in \cref{eq_ae1r1}.

\subsubsection*{Reflected Entropy Phase 2}

\textbf{Island prescription:} In this phase, the subsystem $A$ is assumed to be significantly smaller than the subsystem $B$. Using the island prescription detailed in \cref{eq_sris}, the effective reflected entropy between $A$ and $B$ may be obtained using \cref{eq_re} where in the numerator we have a 6-point correlator of twist fields at the boundary points $K,J,I,M,N$ and $O$. In the large $c$ limit it factorizes into a product of a 2-point and a 4-point correlator as
\begin{align}
& \langle \sigma_{g_B}(K) \sigma_{g_A g_B^{-1}}(J) \sigma_{g_A^{-1}}(I) \sigma_{g_A}(M) \sigma_{g_A^{-1} g_B}(N) \sigma_{g_B^{-1}}(O) \rangle _{CFT^{\otimes mn}} \notag \\
& \quad \approx \langle \sigma_{g_B}(K) \sigma_{g_B^{-1}}(O) \rangle _{CFT^{\otimes mn}} \langle \sigma_{g_A g_B^{-1}}(J) \sigma_{g_A^{-1}}(I) \sigma_{g_A}(M) \sigma_{g_A^{-1} g_B}(N) \rangle _{CFT^{\otimes mn}}.
\end{align}
In the 4-point correlator on the RHS of the above expression, the composite twist fields positioned at $J$ and $N$ may be expanded in terms of the individual twist fields $\sigma_{g_A}$ and $\sigma_{g_B}$. Upon this expansion, the 4-point correlator is then expressed as a 6-point correlator as
\begin{align}
& \langle \sigma_{g_A g_B^{-1}}(J) \sigma_{g_A^{-1}}(I) \sigma_{g_A}(M) \sigma_{g_A^{-1} g_B}(N) \rangle _{CFT^{\otimes mn}} \notag \\
& \quad \approx \langle \sigma_{g_B^{-1}}(J') \sigma_{g_A }(J) \sigma_{g_A^{-1}}(I) \sigma_{g_A}(M) \sigma_{g_A^{-1}}(N) \sigma_{g_B}(N') \rangle _{CFT^{\otimes mn}}
\end{align}
Under the assumption that $J'$ and $N'$ are located very close to $J$ and $N$ respectively, this 6-point correlator may be factorized in terms of two 4-point correlators, as described in \cite{Banerjee:2016qca}. This factorization is explicitly demonstrated as
\begin{align}
& \langle \sigma_{g_B^{-1}}(J') \sigma_{g_A }(J) \sigma_{g_A^{-1}}(I) \sigma_{g_A}(M) \sigma_{g_A^{-1}}(N) \sigma_{g_B}(N') \rangle _{CFT^{\otimes mn}} \notag \\
& \quad \approx \langle \sigma_{g_B^{-1}}(J') \sigma_{g_A }(J) \sigma_{g_A^{-1}}(I) \sigma_{g_A}(M) \rangle _{CFT^{\otimes mn}} \langle \sigma_{g_A^{-1}}(I) \sigma_{g_A}(M) \sigma_{g_A^{-1}}(N) \sigma_{g_B}(N') \rangle _{CFT^{\otimes mn}}
\end{align}
By applying the OPE limit to the twist fields positioned at $J,J'$ and $N,N'$, the above expression may be rewritten as products of two 3-point correlators as
\begin{align}
& \langle \sigma_{g_A g_B^{-1}}(J) \sigma_{g_A^{-1}}(I) \sigma_{g_A}(M) \sigma_{g_A^{-1} g_B}(N) \rangle _{CFT^{\otimes mn}} \notag \\
& \quad \approx \langle \sigma_{g_A g_B^{-1}}(J) \sigma_{g_A^{-1}}(I) \sigma_{g_A}(M) \rangle _{CFT^{\otimes mn}} \langle \sigma_{g_A^{-1}}(I) \sigma_{g_A}(M) \sigma_{g_A^{-1} g_B}(N) \rangle _{CFT^{\otimes mn}}.
\end{align}
The correlator in the denominator can be determined as the $n \to 1$ limit of that in the numerator (as elaborated in \cite{Dutta:2019gen}) and it factorizes in a similar fashion. The effective reflected entropy may then be obtained as
\begin{align}\label{eq_ae1r2_eff}
S^{eff}_R (A:B)=\lim _{m,n \to 1} \frac{1}{1-n} \log & \left( \frac{ \langle \sigma_{g_A g_B^{-1}}(J) \sigma_{g_A^{-1}}(I) \sigma_{g_A}(M)\rangle _{CFT^{\otimes mn}}}{ \langle \sigma_{g_m ^{-1}}(I) \sigma_{g_m}(M)\rangle ^n _{CFT^{\otimes m}}} \right. \notag \\
& \quad \times \left. \frac{ \langle \sigma_{g_A}(M) \sigma_{g_A^{-1}}(I) \sigma_{g_A^{-1} g_B}(N) \rangle _{CFT^{\otimes mn}}}{ \langle \sigma_{g_m}(M) \sigma_{g_m ^{-1}}(I)  \rangle ^n _{CFT^{\otimes m}}} \right).
\end{align}
where the contribution of the 2-point correlator in the numerator cancels out with those in the denominator as $n \to 1$, effectively simplifying the expression. Additionally, the area term does not contribute to the generalized reflected entropy since neither $A$ not $B$ have any reflected entropy island on the brane. Utilizing the standard form of 2-point correlator \cref{ssec_eedes} and 3-point correlators from \cite{Dutta:2019gen}, the reflected entropy may be computed as
\begin{align}\label{eq_ae1r2}
S_R(A : B) & =\frac{2 c}{3} \log \left[ \frac{\sqrt{((x_2-x_1)^2+(\tau_2-\tau_1)^2)((x_2+x_1)^2+(\tau_2-\tau_1)^2)}}{\epsilon _2 x_1} \right], \notag \\
& =\frac{2c}{3} \log \left[ \frac{2 \sqrt{2 (\cosh(2T)+\cosh(X_2-X_1))}\sech(T)\sinh \left( \frac{X_2-X_1}{2} \right)}{\epsilon _R} \right].
\end{align}
Once again the transformations outlined in \cref{eq_trans1,eq_trans21,eq_trans22} are applied in the second step to express the final result in terms of the Rindler coordinates.

\vspace{0.5cm}

\textbf{Bulk prescription:} The holographic reflected entropy in this phase may be determined using the DES formula outlined in \cref{eq_srdes}. The first term may be expressed in terms of the two contributions to the bulk EWCS, represented by the magenta dashed lines in \cref{fig_AE1}. The contribution from the left TFD copy, which is defined in terms of the three boundary points $J,I$ and $M$, may be determined using \cref{eq_ewa}
\begin{align}
E_W=\cosh ^{-1} \left[ \sqrt{-\frac{2 \xi_{21} \xi_{32}}{\xi_{31}}} \right],
\end{align}
where we assign $Y_1 \to I$, $Y_2 \to J$ and $Y_3 \to M$. Similarly the contribution from the right TFD copy may be determined by the assignment $Y_1 \to I$, $Y_2 \to N$ and $Y_3 \to M$. Combining the two contributions determined above, and first term in \cref{eq_srdes} is then computed using \cref{eq_hre}.

Similar to the previous reflected entropy phase, the second term in \cref{eq_srdes} does not contribute. The final holographic reflected entropy between $A$ and $B$ thus obtained is in agreement with the field theory computation results provided in \cref{eq_ae1r2} on implementation of the Brown-Henneaux relation.

\subsubsection*{Reflected Entropy Phase 3}\label{sssec_ae1r3}

In this phase the subsystem $A$ is assumed to be larger than the subsystem $B$. The computation for the reflected entropy in this phase using the island prescription is very similar to that of the previous phase and the final result may be obtained by the substitution $X_1 \to X_3$ in \cref{eq_ae1r2} as
\begin{align}\label{eq_ae1r3}
S_R(A : B)=\frac{2c}{3} \log \left[ \frac{2 \sqrt{2 (\cosh(2T)+\cosh(X_3-X_2))}\sech(T)\sinh \left( \frac{X_3-X_2}{2} \right)}{\epsilon _R} \right].
\end{align}

From the bulk perspective, the corresponding holographic reflected entropy may be computed using the DES prescription in terms of the two contributions to the EWCS from the TFD state, as depicted by the orange dashed lines in \cref{fig_AE1}. The holographic reflected entropy in this phase matches with the fields theory result expressed as \cref{eq_ae1r3}.

\subsection{Entanglement Entropy Phase 2}

\begin{figure}[t]
\centering
\includegraphics[scale=1]{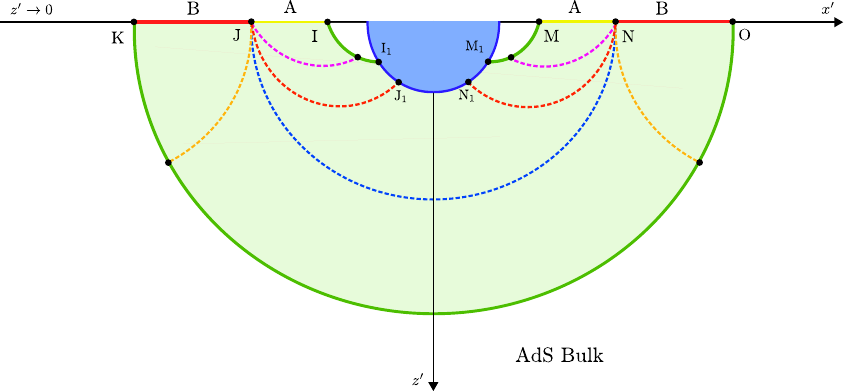}
\caption{EE Phase-2 for two adjacent intervals $A$ and $B$. The RT surfaces corresponding to $A \cup B$ are represented by the green lines, while the dashed lines depict the different configurations of the EWCS.}\label{fig_AE2}
\end{figure}

In this configuration, the subsystem $A$ is positioned relatively closer to the brane, while the subsystem $B$ is located further away. In this scenario the entanglement entropy can be expressed in terms of the HM surface $KO$, and RT surfaces $II_1$ and $MM_1$ ending on the brane, illustrated by the green lines in \cref{fig_AE2}. Here $I_1 \equiv (- x_1,-\tau_1 \tanh (\sigma / \ell),\tau_1 \sech (\sigma / \ell))$ and $M_1 \equiv (x_1,-\tau_1 \tanh (\sigma / \ell),\tau_1 \sech (\sigma / \ell))$ represent the locations where the two RT surfaces $II_1$ and $MM_1$ terminate on the EOW brane. In this setup, the entanglement entropy may be computed as
\begin{align}\label{eq_ae2}
S(A \cup B)= \frac{c}{3} \left( \frac{\sigma}{\ell} + \log \left[ \frac{2 \cosh T}{\epsilon _R} \right] + \log \left[ \frac{2 \sinh X_1}{\epsilon _R} \right]+\log \left[ \frac{2 \ell}{\epsilon _y \sech (\sigma / \ell)} \right] \right).
\end{align}
In this entanglement entropy phase, four distinct reflected entropy phases emerge. They are represented by the dashed lines in \cref{fig_AE2}, and the computations of the reflected entropy are as follows.

\subsubsection*{Reflected Entropy Phase 1}

In this configuration, the subsystems $A$ and $B$ are considered to be of comparable sizes. The field theory analysis of the reflected entropy between $A$ and $B$ follow from the island prescription detailed in \cref{eq_sris}, where the computations proceed in a similar fashion as detailed in the first phase in \cref{sssec_ae1r1}. Consequently, the final reflected entropy using the island prescription may be expressed directly as \cref{eq_ae1r1}.

Similarly from the bulk perspective, the holographic reflected entropy may be computed using the DES prescription, where the first term may be expressed in terms of the EWCS depicted by the blue dashed line in \cref{fig_AE2}. Given the resemblance with the bulk analysis detailed in the first phase in \cref{sssec_ae1r1}, the holographic reflected entropy can thus be obtained directly and it agrees with the field theory result described above.

\subsubsection*{Reflected Entropy Phase 2}\label{sssec_ae2r2}

\textbf{Island prescription:} In this phase the subsystem $A$ is considered to be smaller than the subsystem $B$. This leads to an entanglement structure where the entire entanglement entropy island of $A \cup B$ corresponds to the reflected entropy island of $B$, while the subsystem $A$ has no reflected entropy island on the brane. Applying the island prescription given in \cref{eq_sris}, the effective reflected entropy may be computed using \cref{eq_re}, where the numerator can be expressed as an 8-point correlator of twist fields located at the boundary points $K,J, I, M, N,O$ and the bulk points $I_1$ and $M_1$. In the large $c$ limit, this correlator undergoes a factorization as
\begin{align}\label{eq_sreff_ae2r2}
& \langle \sigma_{g_B}(K) \sigma_{g_A g_B^{-1}}(J) \sigma_{g_A^{-1}}(I) \sigma_{g_A}(I_1) \sigma_{g_A^{-1}}(M_1) \sigma_{g_A}(M) \sigma_{g_A^{-1} g_B}(N) \sigma_{g_B^{-1}}(O) \rangle _{CFT^{\otimes mn}} \notag \\
& \quad \approx \langle \sigma_{g_B}(K) \sigma_{g_B^{-1}}(O) \rangle _{CFT^{\otimes mn}} \langle \sigma_{g_A g_B^{-1}}(J) \sigma_{g_A^{-1}}(I) \sigma_{g_A}(I_1) \rangle _{CFT^{\otimes mn}} \notag \\
& \qquad \qquad \times \langle \sigma_{g_A^{-1}}(M_1) \sigma_{g_A}(M) \sigma_{g_A^{-1} g_B}(N) \rangle _{CFT^{\otimes mn}}.
\end{align}
The 6-point correlator of twist fields $\sigma_{g_m}$ in the denominator factorizes in a similar manner. The effective reflected entropy may then be expressed as
\begin{align}
S^{eff}_R (A:B)=\lim _{m,n \to 1} \frac{1}{1-n} \log & \left( \Omega_{I_1}^{2 h_{g_{A}}} \frac{  \langle \sigma_{g_A g_B^{-1}}(J) \sigma_{g_A^{-1}}(I) \sigma_{g_A}(I_1)\rangle _{CFT^{\otimes mn}}}{\langle \sigma_{g_m^{-1}}(I) \sigma_{g_m}(I_1)\rangle ^n  _{CFT^{\otimes m}}} \right. \notag \\
 & \left. \qquad \times \Omega_{M_1}^{2 h_{g_{A}}} \frac{ \langle \sigma_{g_A}(M) \sigma_{g_A^{-1}}(M_1) \sigma_{g_A^{-1} g_B}(N) \rangle _{CFT^{\otimes mn}}}{ \langle \sigma_{g_m}(M) \sigma_{g_m^{-1}}(M_1)  \rangle ^n  _{CFT^{\otimes m}}} \right),
\end{align}
where in the limit $n \to 1$, the contribution from the 2-point correlator in the numerator cancels out with that in the denominator. Once again, the area term does not contribute to the final reflected entropy due to vanishing reflected entropy island cross section on the brane, resulting in the simplification $S^{gen}_R=S^{eff}_R$. Applying the formula for the 2-point correlator \cref{ssec_eedes} and the 3-point correlator from \cite{Dutta:2019gen}, the final expression for the reflected entropy between $A$ and $B$ may be obtained as
\begin{align}\label{eq_ae2r2}
S_R(A : B) & =\frac{2 c}{3} \log \left[ \frac{\sqrt{((x_2-x_1)^2+(\tau_2-\tau_1)^2)((x_2-x_1)^2+(\tau_2+\tau_1)^2)}}{\epsilon _2 \tau_1} \right]  \notag \\
& =\frac{2c}{3} \log \left[ \frac{2 (\coth X_1-\cosh X_2 \csch X_1)}{\epsilon _R} \right].
\end{align}

\vspace{0.5cm}

\textbf{Bulk prescription:} The holographic reflected entropy may be computed using the DES formula outlined in \cref{eq_srdes}, where the first term can be obtained in terms of the combined contribution to the bulk EWCS represented by the magenta dashed lines in \cref{fig_AE2}. The contributions from the left TFD copy to the bulk EWCS may be evaluated using (check \cref{appA})
\begin{align}\label{eq_ewam2}
E_W=\cosh ^{-1} \left[\sqrt{-\frac{(2 \xi_{21} \xi_{31}+\xi_{32})(2 \xi_{32} \xi_{31}+\xi_{21})}{2 \xi_{31}^{3/2}}} \right],
\end{align}
where we assign $Y_1 \to I$, $Y_2 \to J$ and $Y_3 \to J_1$. Similarly, the contribution from the right TFD copy may be obtained by assigning $Y_1 \to M$, $Y_2 \to N$ and $Y_3 \to M_1$. Combining the two contributions to obtain the bulk EWCS, the first term in \cref{eq_srdes} may be then determined using \cref{eq_hre}. Moreover, the second term in \cref{eq_srdes} does not contribute in this scenario. Thus the final holographic reflected entropy may be obtained, which matches precisely with the field theory result in \cref{eq_ae2r2}.

\subsubsection*{Reflected Entropy Phase 3}

In this phase the subsystem $A$ is assumed to be larger than the subsystem $B$. In this configuration, the entire entanglement entropy island of subsystem $A \cup B$ corresponds to the reflected entropy island of $A$, while $B$ possesses no corresponding reflected entropy island on the brane. The field theory computations for the reflected entropy between $A$ and $B$ is similar to that detailed in the third phase in \cref{sssec_ae1r3}, and the final result may be obtained directly as \cref{eq_ae1r3}.

From the bulk perspective, the computation of the holographic reflected entropy may be expressed in terms of the two contributions to the bulk EWCS, depicted by the orange dashed lines in \cref{fig_AE2}. The bulk EWCS structures in this phase resemble those discussed in the third phase in \cref{sssec_ae1r3}. The holographic reflected entropy can therefore be directly obtained, and it agrees with the field theory computations described above.

\subsubsection*{Reflected Entropy Phase 4}\label{sssec_ae2r4}

\textbf{Island prescription:} In this phase both the subsystems $A$ and $B$ are of comparable sizes and possess reflected entropy islands on the brane. By applying the island prescription detailed in \cref{eq_sris}, the effective reflected entropy may be computed using \cref{eq_re}, where in the numerator we have a 10-point correlator of twist fields positioned at the boundary points $K,J,I,M,N,O$ and the brane points $I_1, J_1, N_1, M_1$. In the large $c$ approximation, it factorizes into product of 2-point correlators as
\begin{align}\label{eq_sreff_ae2r4}
& \langle \sigma_{g_B}(K) \sigma_{g_B^{-1}}(O) \rangle _{CFT^{\otimes mn}} \langle \sigma_{g_A^{-1}}(I) \sigma_{g_A}(I_1) \rangle _{CFT^{\otimes mn}} \langle \sigma_{g_A^{-1}}(M_1) \sigma_{g_A}(M) \rangle _{CFT^{\otimes mn}} \notag \\
& \qquad \times \langle \sigma_{g_A g_B^{-1}}(J) \sigma_{g_A^{-1} g_B}(J_1) \rangle _{CFT^{\otimes mn}} \langle\sigma_{g_A g_B^{-1}}(N_1) \sigma_{g_A^{-1} g_B}(N) \rangle _{CFT^{\otimes mn}}.
\end{align}
The 6-point correlator in the denominator can once again be obtained by taking the $n \to 1$ limit to that in the numerator, and it factorizes in a similar manner. The effective reflected entropy is then expressed as 
\begin{align}
S^{eff}_R (A:B)=\lim _{m,n \to 1} \frac{1}{1-n} \log & \left( \Omega_{J_1}^{2 h_{g_{AB}}} \langle \sigma_{g_A g_B^{-1}}(J) \sigma_{g_A^{-1} g_B}(J_1) \rangle _{CFT^{\otimes mn}} \right. \notag \\ 
& \qquad \times \left. \Omega_{N_1}^{2 h_{g_{AB}}} \langle \sigma_{g_A g_B^{-1}}(N_1) \sigma_{g_A^{-1} g_B}(N) \rangle _{CFT^{\otimes mn}} \right),
\end{align}
where in the $n \to 1$ limit the first three correlators in the numerator cancel out with those in the denominator, effectively simplifying the expression. From symmetry of the configurations, the brane points may be considered to be at $J_1 \equiv (-y,-x_y)$ and $N_1 \equiv (-y,x_y)$, where $y$ is a coordinate along the brane. Using the standard 2-point twist field correlator we obtain
\begin{align}
S^{eff}_R=\frac{2 c}{3} \left( \log \left[ \frac{(y+\tau_2 )^2+(x_2-x_y)^2}{\epsilon_2} \right] + \log \left[ \frac{ \ell}{\epsilon _y y \sech(\sigma_0 / \ell)} \right] \right),
\end{align}
where once again $\epsilon_y$ denotes the UV cut-off on the brane. The contributions due to the area term (the second term in \cref{eq_sris}) must also be taken into account in this phase, and can be directly obtained from \cref{eq_area}. Combining the two terms we arrive at the generalized reflected entropy $S^{gen}_R$ corresponding to this phase, which on extremization with respect to $y$ and $x_y$ give
\begin{align}
y=\tau_2, \qquad x_y= x_2.
\end{align}
Substituting the values of $y$ and $x_y$, the final reflected entropy is then obtained by as
\begin{align}\label{eq_ae2r4}
S_R(A : B) & =\frac{2 c}{3} \left( \frac{\sigma}{\ell} + \log \left[ \frac{2 \tau_2}{\epsilon _2} \right] \right) + \frac{2c}{3} \log \left[ \frac{2 \ell}{\epsilon _y \sech(\sigma / \ell)} \right] \notag \\
& =\frac{2c}{3} \left( \frac{\sigma}{\ell} +  \log \left[ \frac{2 \sinh X_2}{\epsilon _R} \right] + \log \left[ \frac{2 \ell}{\epsilon _y \sech(\sigma / \ell)} \right] \right).
\end{align}

\vspace{0.5cm}

\textbf{Bulk prescription:} The holographic reflected entropy may be computed using the DES formula given in \cref{eq_srdes}. The first term may be determined in terms of the two contributions to the bulk EWCS represented by the red dashed lines in \cref{fig_AE2}, both of which have an endpoint on the brane. The structure of the two contributions have similarities with the two RT surfaces described in the disconnected phase in \cref{ssec_eedes}. Consequently, the points where the two bulk EWCS intersect with the EOW brane can directly be determined as $J_1 \equiv (- x_2,-\tau_2 \tanh (\sigma / \ell),\tau_2 \sech (\sigma / \ell))$ and $N_1 \equiv (x_2,-\tau_2 \tanh (\sigma / \ell),\tau_2 \sech (\sigma / \ell))$. The contribution from the left and right TFD copy to the bulk EWCS may be determined by using \cref{eq_geol} and assigning $Y_1 \to J, Y_2 \to J_1$ and $Y_1 \to N, Y_2 \to N_1$ respectively. Adding the two contributions, the first term in \cref{eq_srdes} is then determined using \cref{eq_hre}.

Additionally, the contribution arising from brane defects can be obtained directly from \cite{Li:2021dmf}, as specified in \cref{eq_area}. Adding the two contributions, the holographic reflected entropy between $A$ and $B$ may finally be obtained, and it matches with the field theory computations described above in \cref{eq_ae2r4}.

\subsection{Entanglement Entropy Phase 3}

\begin{figure}[t]
\centering
\includegraphics[scale=1]{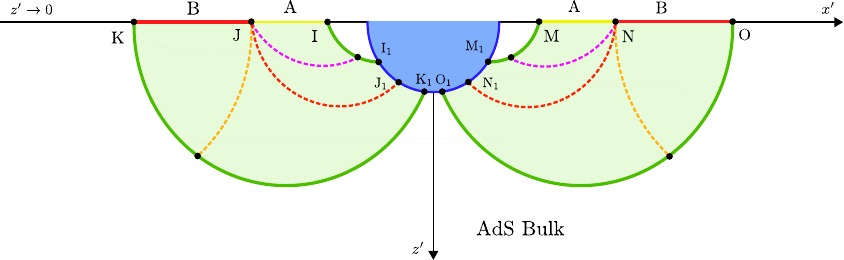}
\caption{EE Phase-3 for two adjacent intervals $A$ and $B$. The RT surfaces corresponding to $A \cup B$ are represented by the green lines, while the dashed lines depict the different configurations of the EWCS.}\label{fig_AE3}
\end{figure}

In this phase, both the subsystems $A$ and $B$ are positioned near the brane. Consequently, the entanglement entropy can be expressed in terms of the RT surfaces $KK_1, OO_1, II_1$ and $MM_1$ (depicted by the green lines in \cref{fig_AE3}), all of which terminate on the brane. The entanglement entropy is thus determined as
\begin{align}\label{eq_ae3}
S(A \cup B)= \frac{c}{3} \left( 2 \frac{\sigma}{\ell} + \log \left[ \frac{2 \sinh X_3}{\epsilon _R} \right] + \log \left[ \frac{2 \sinh X_1}{\epsilon _R} \right]+ 2 \log \left[ \frac{2 \ell}{\epsilon _y \sech (\sigma / \ell)} \right] \right).
\end{align}
In this phase, three distinct reflected entropy configurations emerge, each represented by the dashed lines in \cref{fig_AE3}. We next outline the computations for each phase below.

\subsubsection*{Reflected Entropy Phase 1}

In this phase the subsystems $A$ and $B$ are considered to be similar in size, resulting in both $A$ and $B$ to possess reflected entropy islands on the brane. The computations of the reflected entropy using both the island prescription and the DES prescription is similar to that detailed in the fourth phase in \cref{sssec_ae2r4}, where in the bulk perspective the EWCS is illustrated by the dashed red line in \cref{fig_AE3}. Consequently, the reflected entropy between $A$ and $B$ in this phase may be expressed directly as \cref{eq_ae2r4}.

\subsubsection*{Reflected Entropy Phase 2}

In this phase the subsystem $A$ is assumed to be smaller than the subsystem $B$, leading to an entanglement structure where only $B$ possesses a reflected entropy island on the brane. The effective reflected entropy and bulk EWCS (depicted by the dashed magenta line in \cref{fig_AE3}) corresponding to this phase are once again similar to those described in the second phase in \cref{sssec_ae2r2}. Consequently the reflected entropy computed using both the island and the DES prescription may be directly determined to be \cref{eq_ae2r2}.

\subsubsection*{Reflected Entropy Phase 3}

In this phase the subsystem $A$ is considered to be larger than the subsystem $B$, such that only the subsystem $A$ has a reflected entropy island on the brane. The effective reflected entropy and the bulk EWCS (depicted by the orange dashed lines in \cref{fig_AE3}) corresponding to this scenario are similar to those detailed in the second phase in \cref{sssec_ae2r2} upon substituting $X_1 \to X_3$. As a result, the final reflected entropy determined from both the island and the DES prescription may directly be obtained from \cref{eq_ae2r2} via the above-mentioned substitution as 
\begin{align}\label{eq_ae3r3}
S_R(A : B)=\frac{2c}{3} \log \left[ \frac{2 (\coth X_3-\cosh X_2 \csch X_3)}{\epsilon _R} \right].
\end{align}

\subsection{Entanglement Entropy Phase 4}

\begin{figure}[t]
\centering
\includegraphics[scale=1]{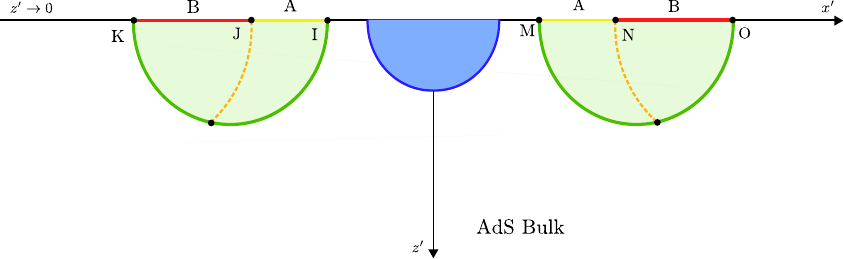}
\caption{EE Phase-4 for two adjacent intervals $A$ and $B$. The RT surfaces corresponding to $A \cup B$ are represented by the green lines, while the dashed lines depict the EWCS.}\label{fig_AE4}
\end{figure}

In this phase the subsystems $A$ and $B$ are considered to be located far away from the brane. The entanglement entropy in this scenario is then determined in terms of the two RT surfaces $KI$ and $MO$, which are denoted by the green lines in \cref{fig_AE4}, and can be expressed as 
\begin{align}\label{eq_ae4}
S(A \cup B)=\frac{2c}{3} \log \left[ \frac{2 \sinh \left( \frac{X_3-X_1}{2} \right) }{\epsilon _R} \right].
\end{align}
In this phase, only a single reflected entropy phase is observed, the details of which are as follows.

\subsubsection*{Reflected Entropy Phase 1}\label{ae4r1}

\textbf{Island prescription:} The reflected entropy for this phase may be computed using the island formula described in \cref{eq_sris}, where the effective term can be evaluated utilizing \cref{eq_re}. In the numerator we have a 6-point correlator of twist fields at the boundary points $K,J,I,M,N$ and $O$, which factorize in the large $c$ approximation as 
\begin{align}
\langle \sigma_{g_B}(K) \sigma_{g_A g_B^{-1}}(J) \sigma_{g_A^{-1}}(I) \rangle _{CFT^{\otimes mn}} \times \langle \sigma_{g_A}(M) \sigma_{g_A^{-1} g_B}(N) \sigma_{g_B^{-1}}(O) \rangle _{CFT^{\otimes mn}}
\end{align}
Subsequently, the 4-point correlator in the denominator is similarly factorized. Since the area term does not contribute in this phase, the final reflected entropy is equivalent to the effective reflected entropy, which can now be expressed as
\begin{align}
S_R(A : B)=S^{eff}_R (A:B)=\lim _{m,n \to 1} \frac{1}{1-n} \log & \left( \frac{ \langle \sigma_{g_B}(K) \sigma_{g_A g_B^{-1}}(J) \sigma_{g_A^{-1}}(I) \rangle _{CFT^{\otimes mn}}}{\langle  \sigma_{g_m}(K) \sigma_{g_m}(I) \rangle ^n  _{CFT^{\otimes m}}} \right. \notag \\
& \qquad \left. \times \frac{ \langle \sigma_{g_A}(M) \sigma_{g_A^{-1} g_B}(N) \sigma_{g_B^{-1}}(O) \rangle _{CFT^{\otimes mn}}}{\langle  \sigma_{g_m}(M) \sigma_{g_m}(O) \rangle ^n  _{CFT^{\otimes m}}} \right).
\end{align}
Utilizing the standard form of the 2-point correlator described in \cref{ssec_eedes} and 3-point correlator from \cite{Dutta:2019gen}, the reflected entropy between $A$ and $B$ may then be computed as
\begin{align}\label{eq_ae4r1}
S_R(A : B) & =\frac{2 c}{3} \log \left[ \frac{2 \sqrt{((x_2-x_1)^2+(\tau_2-\tau_1)^2)}\sqrt{((x_3-x_2)^2+(\tau_3-\tau_2)^2)}}{\epsilon _2 \sqrt{((x_3-x_1)^2+(\tau_3-\tau_1)^2)}} \right] \notag \\
& =\frac{2c}{3} \log \left[ \frac{4 \csch \left( \frac{X_3-X_1}{2} \right) \sinh \left( \frac{X_2-X_1}{2} \right) \sinh \left( \frac{X_3-X_2}{2} \right) }{\epsilon _R} \right] 
\end{align}
where in the final step the transformations in \cref{eq_trans1,eq_trans21,eq_trans22} are applied to describe the result in terms of the near horizon Rindler coordinates.

\vspace{0.5cm}

\textbf{Bulk prescription:} In this phase, the holographic reflected entropy is determined using the DES formula from \cref{eq_srdes}, where the first term can be expressed in terms of the two contributions to the bulk EWCS represented by the orange dashed lines in \cref{fig_AE4}. The contribution from the left TFD copy may be determined using \cref{eq_ewa} with the assignments $Y_1 \to K, Y_2 \to J$ and $Y_3 \to I$, while assigning $Y_1 \to M, Y_2 \to N$ and $Y_3 \to O$ gives the contribution from the right TFD copy. Adding the two contributions, the first term can be evaluated using \cref{eq_hre}. 

Since no reflected entropy islands exist on the brane, the bulk defect matter does not contribute to the total holographic reflected entropy. Consequently, the final result matches the field theory expression given in \cref{eq_ae4r1} upon application of the Brown-Henneaux relation.

\subsection{Page Curves}\label{ssec_pcadjacent}

\begin{figure}[t]
\centering
\begin{subfigure}{0.45\textwidth}
\includegraphics[scale=0.4]{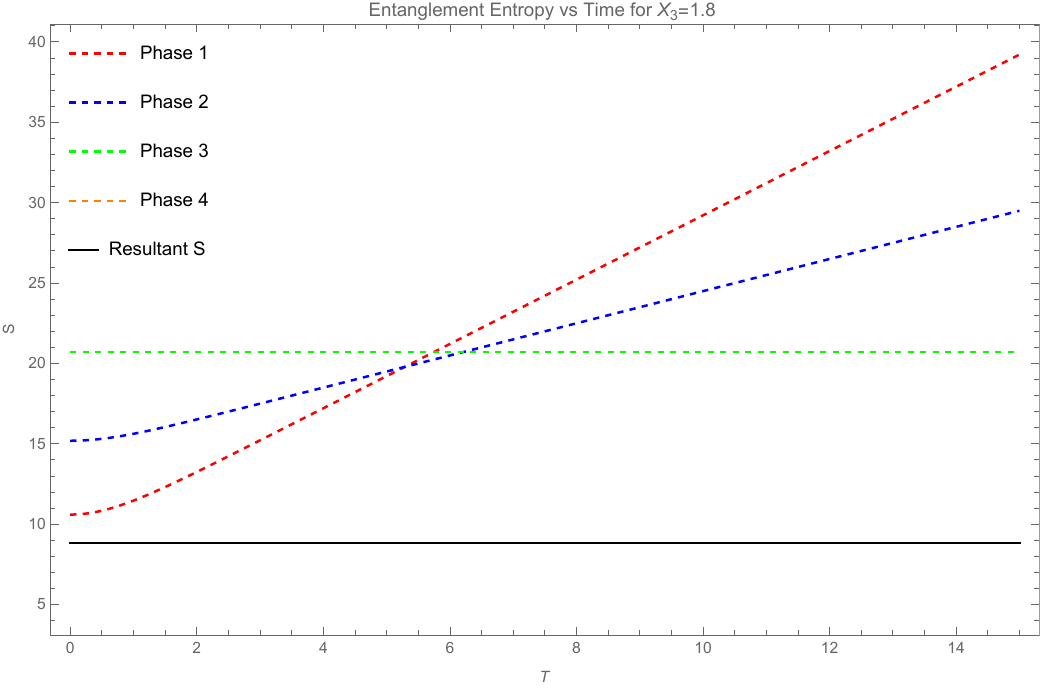}\caption{}
\label{fig_ae1}
\end{subfigure}\hfill
\begin{subfigure}{0.45\textwidth}
\includegraphics[scale=0.4]{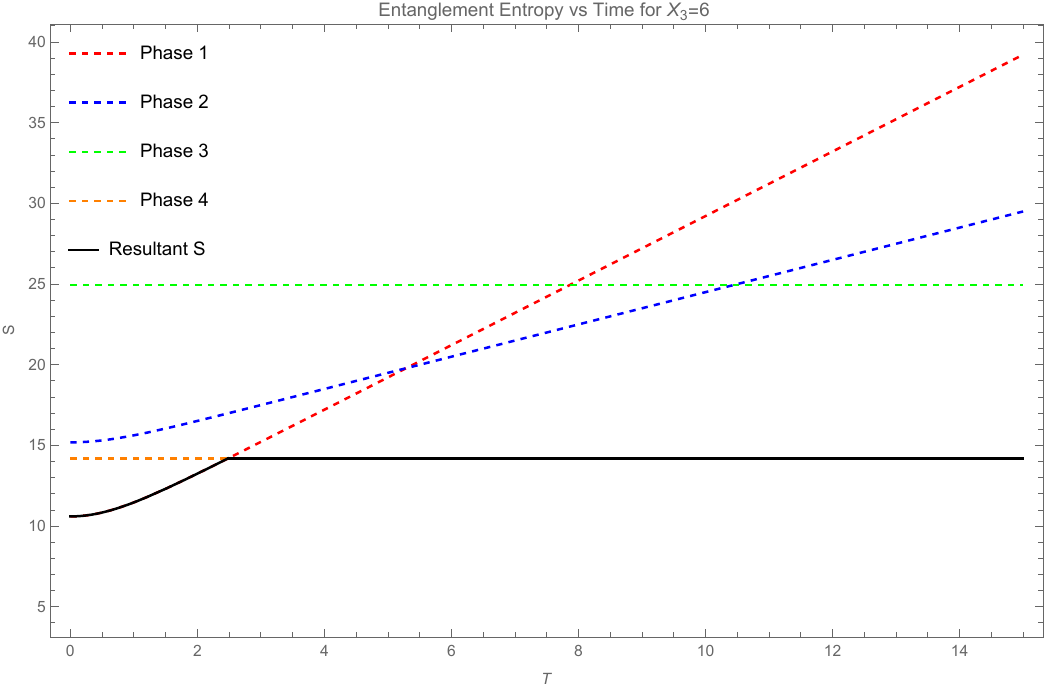}\caption{}
\label{fig_ae2}
\end{subfigure}\hfill
\begin{subfigure}{0.45\textwidth}
\includegraphics[scale=0.4]{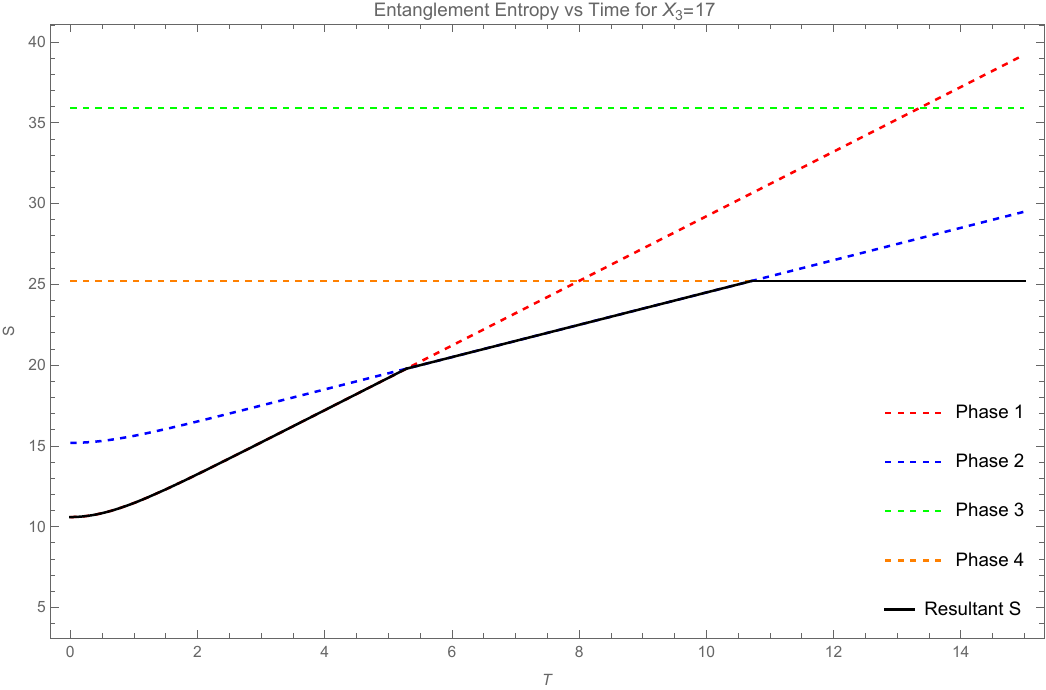}\caption{}
\label{fig_ae3}
\end{subfigure}\hfill
\begin{subfigure}{0.45\textwidth}
\includegraphics[scale=0.4]{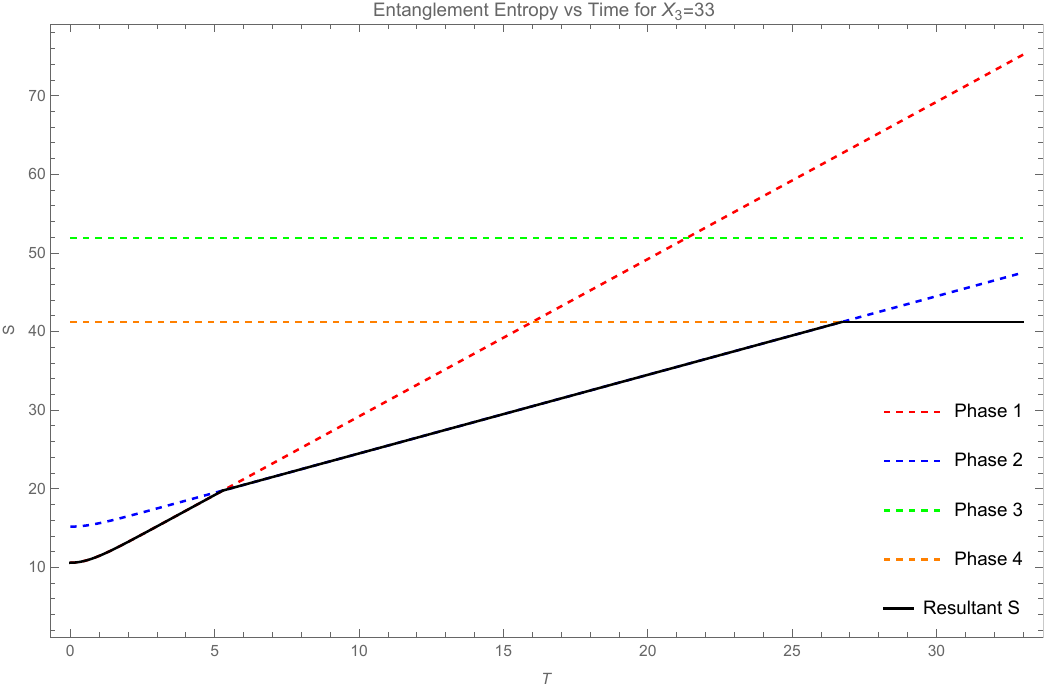}\caption{}
\label{fig_ae4}
\end{subfigure}\hfill
\caption{Entanglement Entropy vs time plots for two adjacent radiation subsystems for different values of $X_3$. The dashed curves represent the time evolution of the different EE phases, while the black solid curve denotes the minimum EE at any given time. We fix $c=3,\ell =1,\sigma_0 =1, X_1 = 1, \epsilon_R=0.01$ and $\epsilon_y=0.1$ to obtain the plots.}
\label{fig_ae}
\end{figure}

In this subsection we plot and analyze the Page curves for the entanglement entropy of the subsystem $A \cup B$ in the left and right radiation bath. Subsequently, a similar analysis is conducted for the reflected entropy between the subsystems $A$ and $B$. Depending on the sizes and locations of the subsystems with respect to the brane, we observe multiple entanglement entropy and analogous reflected entropy Page curves, which exhibit a rich entanglement structure between $A$ and $B$. To systematically analyze the various Page curves, the parameters are fixed at $c=3,\ell =1,\sigma_0 =1, X_1 = 1, \epsilon_R=0.01$, while $X_3$ and $X_2$ are varied across the different scenarios.

A similar analysis can be conducted to explore how the variation of the brane angle $\sigma_0$ influences the time evolution of the entanglement entropy and the reflected entropy. However, as discussed in \cref{ssec_pcdisjoint}, they are excluded from this article as they provide no new understanding of the entanglement structure between the intervals considered.

\subsubsection{Entanglement Entropy Page Curves}

The Page curves for the entanglement entropy of the subsystem $A \cup B$ are illustrated in \cref{fig_ae}. To showcase the complex entanglement structure we consider different configurations specified by different values of $X_3=(1.8,6,17,33)$, while we fix $X_1=1$. The dashed curves represent the time evolution of the different entanglement entropy phases, while the solid black curve represents the dominant phase at any given time. The general expressions for the time of transitions relevant to the plots in \cref{fig_ae} are as follows 
\begin{align}\label{eq_att}
\text{Phase 1 to Phase 4 :} \quad T_{1 \to 4} ^{\text{Adj}} & = \cosh ^{-1} \left[ \sinh \left( \frac{X_3-X_1}{2} \right) \right], \notag \\
\text{Phase 1 to Phase 2 :} \quad T_{1 \to 2} ^{\text{Adj}} & =\cosh ^{-1} \left[ \frac{(1+e^{2 \sigma_0 /\ell}) \ell \sinh X_1}{\epsilon _y} \right], \\
\text{Phase 2 to Phase 4 :} \quad T_{2 \to 4} ^{\text{Adj}} & =\sech ^{-1} \left[ \frac{(1+e^{2 \sigma_0 /\ell}) \ell \sinh X_1}{\epsilon _y \sinh ^2 \left( \frac{X_3-X_1}{2} \right)} \right]. \notag
\end{align}
Note that while the subplots \ref{fig_ae3} and \ref{fig_ae4} exhibit a similar overall patterns (with the minor difference in the transition time $T_{2 \to 4}^{\text{Adj}}$ due to different values of $X_3$), the reflected entropy Page curves between $A$ and $B$ corresponding to each scenario reveal distinct features. These variations highlight the rich entanglement structure between $A$ and $B$.

\subsubsection{Reflected Entropy Page Curves}

\subsubsection*{Case 1: $X_3=1.8, X_2=1.2$}

\begin{figure}[t]
\centering
\begin{subfigure}{0.45\textwidth}
\includegraphics[scale=0.4]{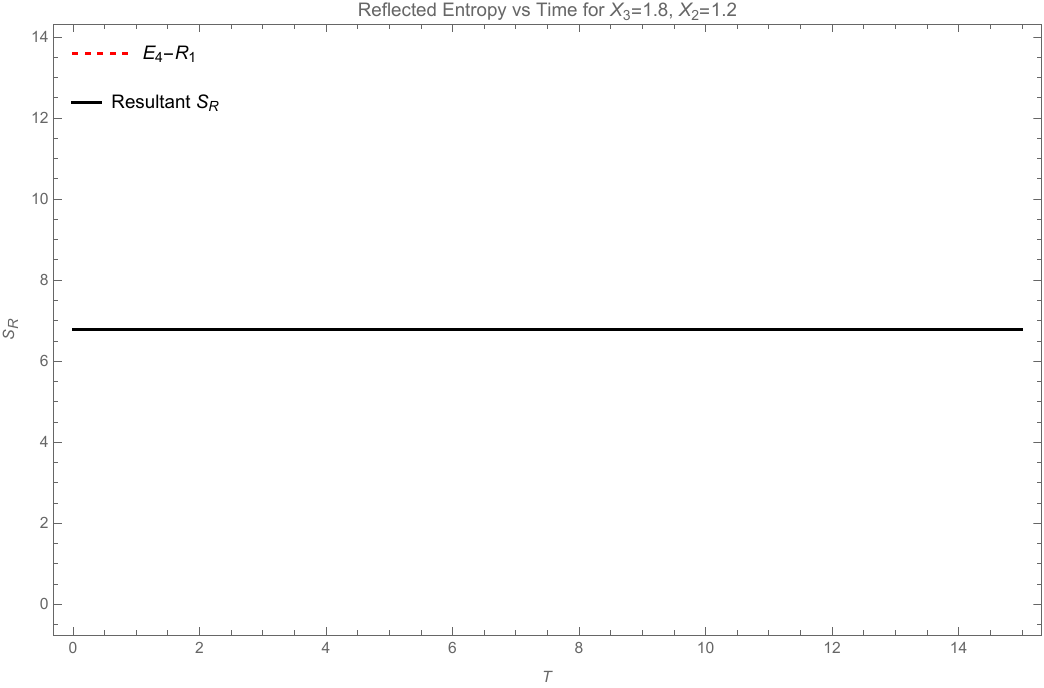}
\caption{}\label{fig_ar1}
\end{subfigure}\hfill
\begin{subfigure}{0.45\textwidth}
\includegraphics[scale=0.4]{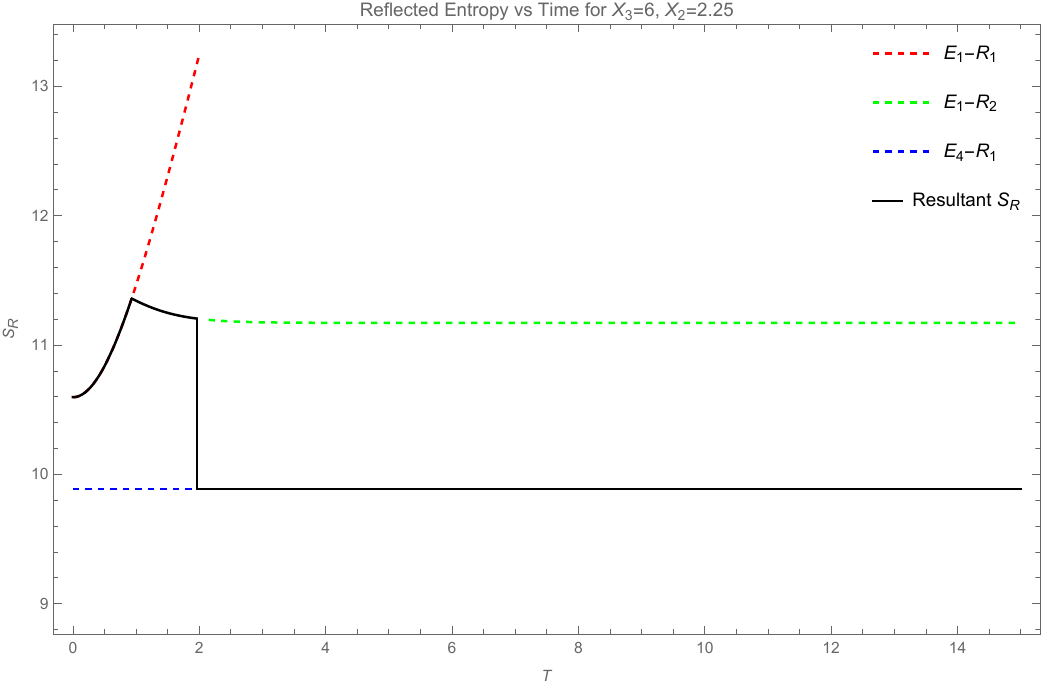}
\caption{}\label{fig_ar2}
\end{subfigure}\hfill
\begin{subfigure}{0.45\textwidth}
\includegraphics[scale=0.4]{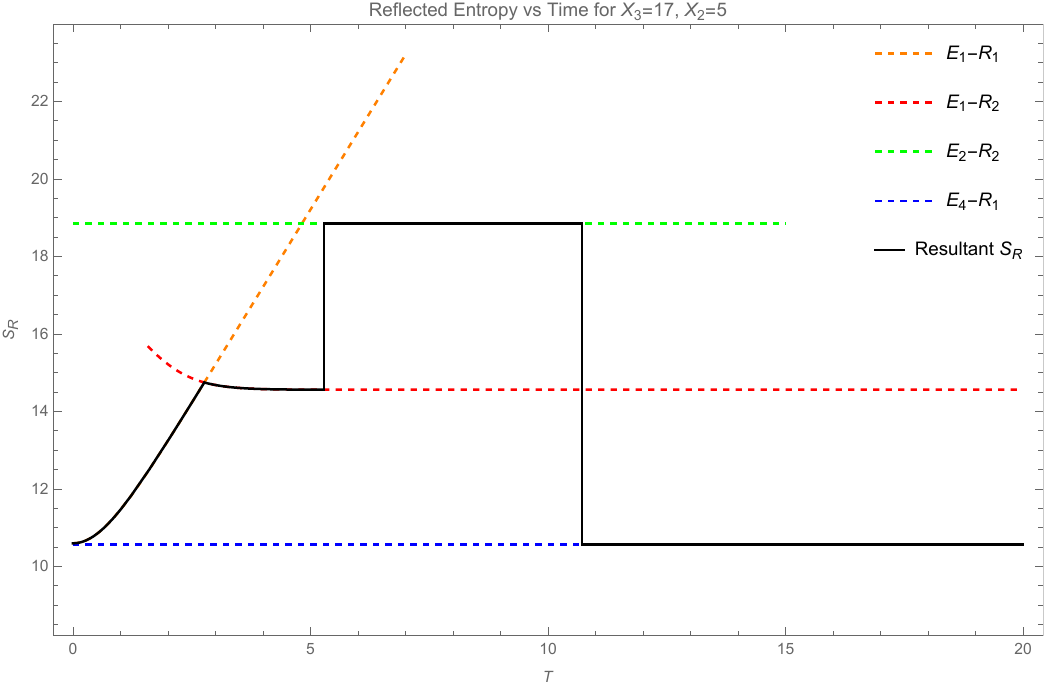}
\caption{}\label{fig_ar3}
\end{subfigure}\hfill
\begin{subfigure}{0.45\textwidth}
\includegraphics[scale=0.4]{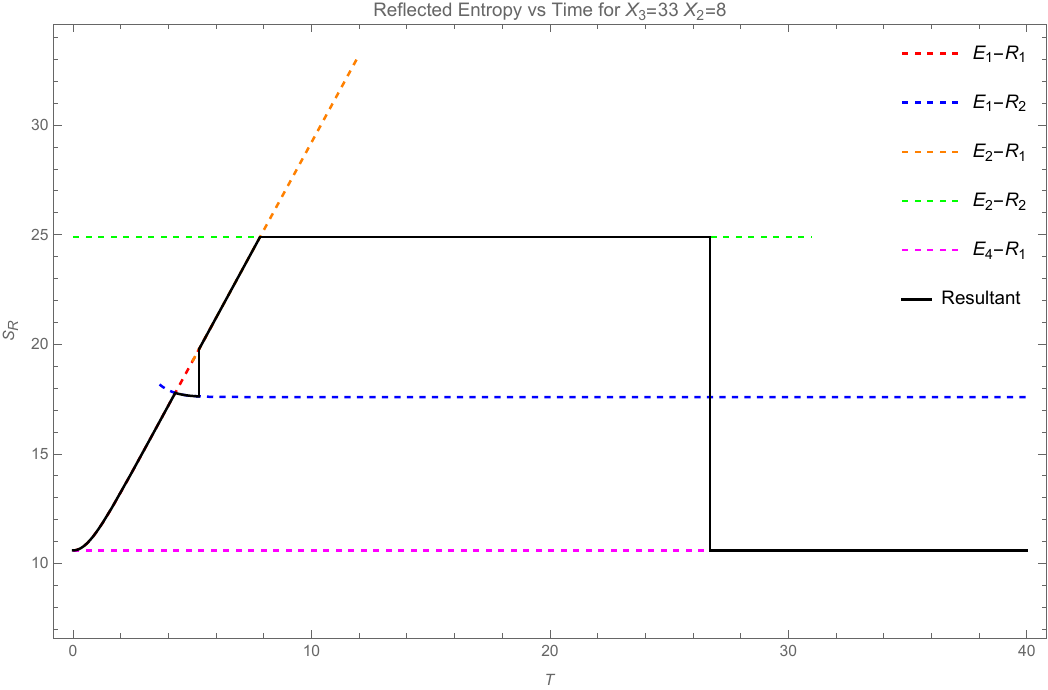}
\caption{}\label{fig_ar4}
\end{subfigure}\hfill
\caption{Analogous reflected entropy Page curves for (a) $X_3=1.8, X_2=1.2$, and (b) $X_3=6, X_2=2.25$, (c) $X_3=17, X_2=5$, and (d) $X_3=33, X_2=8$.}
\end{figure}

In this case both the subsystems $A$ and $B$ are assumed to be very small. The entanglement entropy Page curve corresponding to this configuration is illustrated in \cref{fig_ae1}, where the \hyperref[eq_ae4]{phase-4} of the EE dominates at all time. This EE phase contains a single reflected entropy phase between $A$ and $B$, which also remains constant at all times and is depicted by the dashed curve in \cref{fig_ar1}. The black solid curve denotes the minimum of the reflected entropy, which in this scenario overlaps with the dashed curve.

\subsubsection*{Case 2: $X_3=6, X_2=2.25$}

In this case the subsystems $A$ and $B$ are larger than that in the previous scenario, with $B$ positioned further away from the brane. The entanglement entropy Page curve is shown in \cref{fig_ae2} where we observe a transition between \hyperref[eq_ae1]{phase-1} and \hyperref[eq_ae4]{phase-4} of the entanglement entropy.

The reflected entropy Page curve is illustrated in \cref{fig_ar2}, where the dashed curves represent the time evolution of the different reflected entropy phases involved  and the black solid curve signifies the minimum reflected entropy at any given time. In this configuration, the reflected entropy initially rises sharply and is followed by a gradual decrease during EE \hyperref[eq_ae1]{phase-1}. In EE \hyperref[eq_ae4]{phase-4}, the reflected entropy stabilizes to a lower constant value.

\subsubsection*{Case 3: $X_3=17, X_2=5$}

In this scenario we consider the subsystem $B$ to be much larger than subsystem $A$ and is positioned much further from the brane. \Cref{fig_ae3} represents the EE Page curves corresponding to this scenario, where we observe that the EE initially increases rapidly in \hyperref[eq_ae1]{phase-1}, followed by a gradual rise in \hyperref[eq_ae2]{phase-2}. Finally the EE stabilizes in \hyperref[eq_ae4]{phase-4} to a constant value.

Subsequently the Page curve for the reflected entropy between $A$ and $B$ is shown in \cref{fig_ar3}, where the dashed curves and the solid curve have their usual meaning. As evident from \cref{fig_ar3}, the reflected entropy experiences a rapid increase before entering a decreasing phase in \hyperref[eq_ae1]{phase-1} of the EE. It then increases abruptly to a higher constant value in EE \hyperref[eq_ae2]{phase-2}, followed by a transition to a lower but steady value in \hyperref[eq_ae4]{phase-4} of the EE.

\subsubsection*{Case 4: $X_3=33, X_2=8$}

In this scenario the subsystem $B$ is considered to be very large, and very far from the brane. The Page curve for the entanglement entropy of $A \cup B$ is illustrated in \cref{fig_ae4}. The overall structure of the Page curve is similar to that of \cref{fig_ae3}, with a key difference arising in the transition time $T_{2 \to 4}^{\text{Adj}}$ given in \cref{eq_att} due to difference in $X_3$.

The corresponding Page curve for the reflected entropy between $A$ and $B$ is presented in \cref{fig_ae4}. In the EE \hyperref[eq_ae1]{phase-1} the reflected entropy initially rises rapidly, before transiting to a decreasing phase. As the EE shifts to \hyperref[eq_ae2]{phase-2}, the reflected entropy once again enters an increasing phase, before stabilizing to a high constant value. Finally, in EE \hyperref[eq_ae4]{phase-4}, the reflected entropy undergoes a final transition to a lower, steady value.

\
\section{Summary and Discussions}\label{sec_summary}

To summarize, in this article we have investigated the mixed state entanglement structure for subsystems in radiation baths for $2d$ eternal black holes in a brane world model through the reflected entropy measure in the context of a modified AdS$_3$/BCFT$_2$ framework. The $2d$ eternal black hole on the EOW brane arises from the Euclidean AdS$_3$/BCFT$_2$ scenario via a set of conformal transformations, followed by the analytic continuation of the Euclidean time coordinate. For a mixed state configuration of two disjoint subsystems on a constant time slice in the radiation bath we have demonstrated a rich entanglement structure through the reflected entropy, corresponding to distinct entanglement entropy phases utilizing both the island formulation and the bulk DES prescription. The results from the lower dimensional effective field theory and the bulk computations matched exactly. Subsequently we have analyzed the Page curves for the entanglement entropy and the reflected entropy for various subsystem location and sizes. This analysis was then further extended to configurations involving adjacent subsystems of varying location and sizes and demonstrated exact match between the reflected entropy obtained through the island framework and the bulk DES prescription as earlier. 

A significant aspect of our investigations involve the analysis of the time evolution of the reflected entropy in this model through the corresponding analogs of the Page curve. As illustrated in \cref{ssec_pcdisjoint,ssec_pcadjacent}, the reflected entropy exhibits a complex phase structure which depends on the location of the subsystems, with respect to both the EOW brane and each other. Furthermore the corresponding reflected entropy Page curves demonstrate a behaviour distinct from those described in \cite{Li:2021dmf}. The reflected entropy initially increases with time, reaches a maximum value, before rapidly decreasing to a lower constant value.\footnote{The exact evolution of the reflected entropy varies with the configuration and size of the two subsystems, but the overall trend is similar for all the cases.} The Page curves indicate that the final phase corresponds to the lowest possible reflected entropy. This claim is supported by the corresponding plots in \cite{Basu:2023jtf}, where the reflected entropy evolves in a similar fashion.

Our computations serve as a strong verification of the proposed duality between the island formula and the DES prescription. Note that the results for the reflected entropy of adjacent subsystems may also be obtained from those of the disjoint subsystems in the adjacent limit when their separation is of the order of the field theory cut-off scale. This serves as an additional consistency check for our computations. We should state here that for certain cases the bulk EWCS computations for the DES formula involving adjacent subsystems in the radiation baths are similar to those described in \cite{Shao:2022wrm} where the authors evaluated various phases of the entanglement negativity, which is a distinct mixed state entanglement measure, for two different entanglement entropy phases. However we would like to emphasize here that our computations involve different reflected entropy phases for additional entanglement entropy phases other than those considered in \cite{Shao:2022wrm} for the entanglement negativity.

It is important to note here that in some of the reflected entropy phases, particularly when the subsystems are time-dependent, the bulk EWCS is required to be determined through a complex extremization procedure involving points located on the asymptotic boundary as well as the bulk. This is computationally difficult due to the large number of parameters. To this end, using the techniques described in \cite{Boruch:2020wbe} generalized expressions may be established for the bulk EWCS in terms of the inner products of the position vectors of the above mentioned points which considerably simplify the extremization procedure, the details of which are provided in the Appendices. 

Our investigations lead to several interesting open questions for the future. This includes the study of mixed state entanglement structures in corresponding models involving deformations of the bath CFT$_2$. Furthermore extension of our analysis to different EOW brane configurations for the model under consideration and generalization to higher dimensions are also interesting open issues. Additionally the investigation of the proposed duality between the bulk DES formula and the island prescription in the context of other black hole geometries is a promising future direction. 

Another intriguing extension of this analysis involves multipartite states and the corresponding generalizations of the reflected entropy. In the holographic context, such extensions require multipartite entanglement wedge cross sections, for which several proposals exist in the literature\cite{Umemoto:2018jpc,Bao:2019zqc,Chu:2019etd,Yuan:2024yfg}. Remarkably, these quantities have proven useful as probes \cite{Akers:2019gcv,Hayden:2021gno,Balasubramanian:2024ysu,Bao:2025psl,Mori:2025gqe}, with certain constructions satisfying the stringent requirements for genuine multipartite entanglement measures \cite{Basak:2024uwc,Ahn:2025bdm}. We leave these interesting issues for future consideration.

\section*{Acknowledgement}

We would like to thank Debarshi Basu, Vinayak Raj and Vinay Malvimat for several crucial discussions and clarifications.

\appendix

\section{Geodesics between a fixed point and a minimal surface}\label{appA}

We now derive the expression for the minimal geodesic length (the red curves labelled as $E_W$ in the figures given below) between the point $Y_2$ and a second minimal geodesic with endpoints at $Y_1$ and $Y_2$. In this case, we come across a scenario where all the points are on the asymptotic boundary, and two scenarios where we have one bulk point and two boundary points. The explicit derivations are as follows.

\subsection*{For three boundary points}

\begin{figure}[t]
\centering
\includegraphics[scale=1]{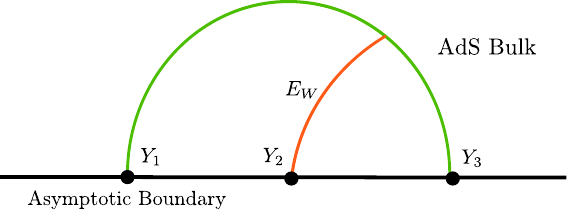}
\caption{EWCS specified by 3 boundary points}\label{fig_3pew}
\end{figure}

In this scenario we consider all the three points $Y_1, Y_2$ and $Y_3$ to be on the asymptotic boundary, as depicted in \cref{fig_3pew}. A spacelike geodesic anchored on the two boundary points $Y_1$ and $Y_3$ may be parametrized by an affine parameter $\lambda$ as follows \cite{Boruch:2020wbe}
\begin{align}\label{eq_geobound}
Y_{31}(\lambda) & =\frac{Y_1 e^{-\lambda}+Y_3 e^{\lambda}}{\sqrt{-2 \xi_{31}}}.
\end{align}
Utilizing \cref{eq_geol}, we may then determine the geodesic length between the point $Y_2$ and an arbitrary point on $Y_{31}(\lambda)$ as
\begin{align}\label{eq_3ewbound}
L(\lambda)=\cosh ^{-1} \left[ - \frac{\xi_{21} e^{-\lambda} + \xi_{32} e^{\lambda}}{\sqrt{-2 \xi_{31}}} \right].
\end{align}
The above equation is extremized at
\begin{align}
\lambda=\frac{1}{2} \log \left[ \frac{\xi_{21}}{\xi_{32}} \right],
\end{align}
which when substituted back into the expression of $L(\lambda)$ gives us the minimal length $E_W$ as
\begin{align}\label{eq_ew3bound}
E_W=\cosh ^{-1} \left[ \sqrt{-\frac{2 \xi_{21} \xi_{32}}{\xi_{31}}} \right].
\end{align}

\subsection*{For two boundary points and a bulk point}

We now consider one of the three points to be in the bulk while the other two points to lie on the asymptotic boundary. In this scenario we encounter two possibilities, which are discussed below. 

\begin{figure}[t]
\centering
\begin{subfigure}{0.45\textwidth}
\includegraphics[scale=1]{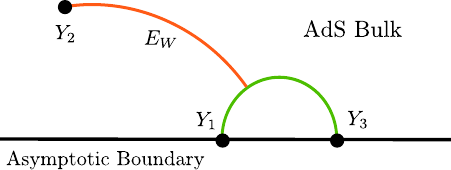}\caption{}
\label{fig_3pewm1}
\end{subfigure}\hfill
\begin{subfigure}{0.45\textwidth}
\includegraphics[scale=1]{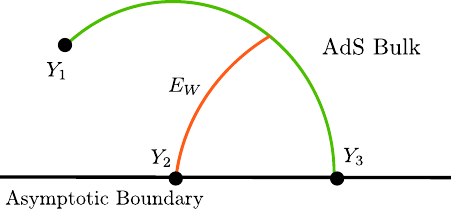}\caption{}
\label{fig_3pewm2}
\end{subfigure}
\caption{EWCS specified by 2 boundary points and 1 bulk point}
\end{figure}

\underline{Case 1}: This scenario corresponds to \cref{fig_3pewm1}, where we assume the point $Y_2$ to be located in the bulk, while $Y_1$ and $Y_3$ are considered to be on the asymptotic boundary. The spacelike geodesic anchored at the points $Y_1$ and $Y_3$ is then parametrized by \cref{eq_geobound}. The geodesic length between an arbitrary point on this spacelike geodesic and the bulk point $Y_2$ may once again be obtained using \cref{eq_geol} as \cref{eq_3ewbound}. Consequently the expression for minimal length $E_W$ in this case may be directly obtained as \cref{eq_ew3bound}.

\vspace{0.5cm}

\underline{Case 2}: In this scenario we assume that $Y_1$ lies in the bulk while $Y_2$ and $Y_3$ are located on the asymptotic boundary, as illustrated in \cref{fig_3pewm2}. The spacelike geodesic $Y_{31}(\lambda)$ in terms of the parameter $\lambda$ may be expressed as \cite{Boruch:2020wbe} 
\begin{align}\label{eq_geobulk}
Y_{31}(\lambda) & =\left( \frac{Y_1}{(-2 \xi_{31})^{1/2}} - \frac{Y_3}{(-2 \xi_{31})^{3/2}} \right) e^{-\lambda} + \left( \frac{Y_3}{(-2 \xi_{31})^{1/2}} - \frac{Y_1}{(-2 \xi_{31})^{3/2}} \right) e^{\lambda}.
\end{align}
The geodesic length between $Y_2$ and an arbitrary point on $Y_{31}(\lambda)$ may then be determined using \cref{eq_geol} as
\begin{align}
L(\lambda)= \cosh ^{-1} \left[ \frac{(2 \xi_{21} \xi_{31}+\xi_{32}) e^{-\lambda}+ (2 \xi_{32} \xi_{31}+\xi_{21}) e^{\lambda}}{(-2 \xi_{31})^{3/2}} \right],
\end{align}
which extremizes at
\begin{align}
\lambda = \frac{1}{2} \log \left[ \frac{2 \xi_{21} \xi_{31}+\xi_{32}}{2 \xi_{32} \xi_{31}+\xi_{21}} \right].
\end{align}
The final expression for the length $E_W$ in this scenario may then be obtained by substituting the value of $\lambda$ into $L(\lambda)$ as
\begin{align}
E_W=\cosh ^{-1} \left[\sqrt{-\frac{(2 \xi_{21} \xi_{31}+\xi_{32})(2 \xi_{32} \xi_{31}+\xi_{21})}{2 \xi_{31}^3}} \right].
\end{align}

\section{Geodesics between two minimal geodesics}\label{appD}

We now derive the expression for the minimal geodesic length $E_W$, described by the red curves in the figures provided below, between two spacelike geodesics $Y_{41}(\lambda)$ and $Y_{32}(\mu)$. In this case we discuss two different scenarios.

\subsection*{For four boundary points}

\begin{figure}[t]
\centering
\includegraphics[scale=1]{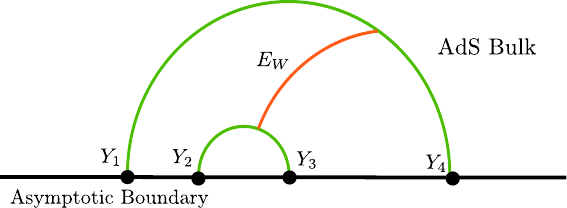}
\caption{EWCS specified by 4 boundary points}\label{fig_4pew}
\end{figure}

In this scenario we assume that the endpoints $Y_1, Y_2, Y_3$ and $Y_4$ of the spacelike geodesics $Y_{41}(\lambda)$ and $Y_{32}(\mu)$ are located on the asymptotic boundary, as depicted in \cref{fig_4pew}. The parametric equation of the two geodesics in terms of affine parameters $\lambda$ and $\mu$ may once again be expressed as
\begin{align}
Y_{41}(\lambda) =\frac{Y_1 e^{-\lambda}+Y_4 e^{\lambda}}{\sqrt{-2 \xi_{41}}}, \qquad \qquad
Y_{32}(\mu) =\frac{Y_2 e^{-\mu}+Y_3 e^{\mu}}{\sqrt{-2 \xi_{32}}}.
\end{align}
The minimal distance between any two arbitrary points on the two geodesics may be obtained using \cref{eq_geol} as
\begin{align}
L(\lambda,\mu)=\cosh ^{-1} \left( - \frac{\xi_{21}e^{-(\lambda+\mu)}+\xi_{31} e^{-\lambda+\mu} + \xi_{42} e^{\lambda-\mu} + \xi_{43} e^{\lambda+\mu}}{2 \sqrt{\xi_{32} \xi_{41}}} \right).
\end{align}
Exxtremizing the above equation over $\lambda$ and $\mu$ we obtain
\begin{align}
\lambda = \frac{1}{4} \log \left[ \frac{\xi_{31} \xi_{21}}{\xi_{42} \xi_{43}} \right] , \qquad \mu = \frac{1}{4} \log \left[ \frac{\xi_{21} \xi_{42}}{\xi_{43} \xi_{31}} \right], 
\end{align}
which when substituted in the expression of $L(\lambda,\mu)$ gives us the minimal length $E_W$ as
\begin{align}
E_W=\cosh ^{-1} \left[ \frac{\sqrt{\xi_{31}\xi_{42}}+\sqrt{\xi_{21}\xi_{43}}}{\sqrt{\xi_{41}\xi_{32}}} \right].
\end{align}

\subsection*{For three boundary points and a bulk point}

\begin{figure}[t]
\centering
\includegraphics[scale=1.2]{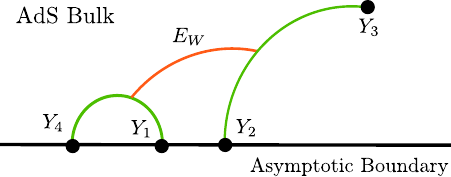}
\caption{EWCS specified by 3 boundary points and a bulk point}\label{fig_4pewm}
\end{figure}

We now assume that the endpoint $Y_2$ of the minimal geodesic $Y_{32}$ to be in the bulk, while the other endpoints $Y_1, Y_3$ and $Y_4$ are located on the asymptotic boundary. This is illustrated in \cref{fig_4pewm}. The parametric expression of $Y_{41}(\lambda)$ thus remains the same as described in the previous case, while that for $Y_{32}(\mu)$ is now expressed as
\begin{align}
Y_{32}(\mu) & =\left( \frac{Y_2}{(-2 \xi_{32})^{1/2}} - \frac{Y_3}{(-2 \xi_{32})^{3/2}} \right) e^{-\mu} + \left( \frac{Y_3}{(-2 \xi_{32})^{1/2}} - \frac{Y_2}{(-2 \xi_{32})^{3/2}} \right) e^{\mu}.
\end{align}
The minimal distance between any two arbitrary points on the geodesics $Y_{41}$ and $Y_{32}$ may then be determined as 
\begin{align}
L(\lambda,\mu)=\cosh ^{-1} & \left[ \frac{   (2 \xi_{21} \xi_{32}+\xi_{31}) e^{-(\lambda+\mu)}   +    (2 \xi_{31} \xi_{32}+\xi_{21}) e^{\mu-\lambda} }    {4 \sqrt{\xi_{41} \xi_{32}^3}} \right. \notag \\
& \qquad + \left. \frac{   (2 \xi_{32} \xi_{42}+\xi_{43}) e^{\lambda-\mu}   +    (2 \xi_{32} \xi_{43}+\xi_{42}) e^{\lambda+\mu} }    {4 \sqrt{\xi_{41} \xi_{32}^3}}  \right].
\end{align}
Extremizing the above expression we obtain
\begin{align*}
\lambda= \frac{1}{4} \log \left[ \frac{(2 \xi_{21} \xi_{32}+\xi_{31}) (2 \xi_{31} \xi_{32}+\xi_{21})}{(2 \xi_{32} \xi_{42}+\xi_{43}) (2 \xi_{32} \xi_{43}+\xi_{42})} \right], \\
\mu= \frac{1}{4} \log \left[ \frac{(2 \xi_{21} \xi_{32}+\xi_{31}) (2 \xi_{32} \xi_{42}+\xi_{43})}{(2 \xi_{31} \xi_{32}+\xi_{21}) (2 \xi_{32} \xi_{43}+\xi_{42})} \right].
\end{align*}
Substituting the values of $\lambda$ and $\mu$ in the expression of $L(\lambda,\mu)$, we finally obtain the minimal length $E_W$ as
\begin{align}
E_W=\cosh ^{-1} \left[ \frac{\sqrt{(2 \xi_{32} \xi_{31}+\xi_{21})(2 \xi_{32} \xi_{42}+\xi_{43})}+\sqrt{(2 \xi_{21} \xi_{32}+\xi_{31})(2 \xi_{32} \xi_{43}+\xi_{42})}}  {2\sqrt{\xi_{41}\xi_{32}^3}} \right].
\end{align}

%%%%%%%%%%%%%%%%%
%%%%%%%%%%%%%%%%%
%%%%%%%%%%%%%%
%%%%%%%%%%%%%%%%%%%

\bibliographystyle{utphys}

\bibliography{ref}

\providecommand{\href}[2]{#2}\begingroup\raggedright\begin{thebibliography}{10}

\bibitem{Page:1993wv}
D.~N. Page, ``{Information in black hole radiation},''
  \href{http://dx.doi.org/10.1103/PhysRevLett.71.3743}{{\em Phys. Rev. Lett.}
  {\bfseries 71} (1993) 3743--3746},
  \href{http://arxiv.org/abs/hep-th/9306083}{{\ttfamily arXiv:hep-th/9306083}}.

\bibitem{Page:1993df}
D.~N. Page, ``{Average entropy of a subsystem},''
  \href{http://dx.doi.org/10.1103/PhysRevLett.71.1291}{{\em Phys. Rev. Lett.}
  {\bfseries 71} (1993) 1291--1294},
  \href{http://arxiv.org/abs/gr-qc/9305007}{{\ttfamily arXiv:gr-qc/9305007}}.

\bibitem{Page:2013dx}
D.~N. Page, ``{Time Dependence of Hawking Radiation Entropy},''
  \href{http://dx.doi.org/10.1088/1475-7516/2013/09/028}{{\em JCAP} {\bfseries
  09} (2013) 028}, \href{http://arxiv.org/abs/1301.4995}{{\ttfamily
  arXiv:1301.4995 [hep-th]}}.

\bibitem{Engelhardt:2014gca}
N.~Engelhardt and A.~C. Wall, ``{Quantum Extremal Surfaces: Holographic
  Entanglement Entropy beyond the Classical Regime},''
  \href{http://dx.doi.org/10.1007/JHEP01(2015)073}{{\em JHEP} {\bfseries 01}
  (2015) 073}, \href{http://arxiv.org/abs/1408.3203}{{\ttfamily arXiv:1408.3203
  [hep-th]}}.

\bibitem{Almheiri:2019hni}
A.~Almheiri, R.~Mahajan, J.~Maldacena, and Y.~Zhao, ``{The Page curve of
  Hawking radiation from semiclassical geometry},''
  \href{http://dx.doi.org/10.1007/JHEP03(2020)149}{{\em JHEP} {\bfseries 03}
  (2020) 149}, \href{http://arxiv.org/abs/1908.10996}{{\ttfamily
  arXiv:1908.10996 [hep-th]}}.

\bibitem{Almheiri:2019psf}
A.~Almheiri, N.~Engelhardt, D.~Marolf, and H.~Maxfield, ``{The entropy of bulk
  quantum fields and the entanglement wedge of an evaporating black hole},''
  \href{http://dx.doi.org/10.1007/JHEP12(2019)063}{{\em JHEP} {\bfseries 12}
  (2019) 063}, \href{http://arxiv.org/abs/1905.08762}{{\ttfamily
  arXiv:1905.08762 [hep-th]}}.

\bibitem{Almheiri:2019qdq}
A.~Almheiri, T.~Hartman, J.~Maldacena, E.~Shaghoulian, and A.~Tajdini,
  ``{Replica Wormholes and the Entropy of Hawking Radiation},''
  \href{http://dx.doi.org/10.1007/JHEP05(2020)013}{{\em JHEP} {\bfseries 05}
  (2020) 013}, \href{http://arxiv.org/abs/1911.12333}{{\ttfamily
  arXiv:1911.12333 [hep-th]}}.

\bibitem{Almheiri:2019psy}
A.~Almheiri, R.~Mahajan, and J.~E. Santos, ``{Entanglement islands in higher
  dimensions},'' \href{http://dx.doi.org/10.21468/SciPostPhys.9.1.001}{{\em
  SciPost Phys.} {\bfseries 9} no.~1, (2020) 001},
  \href{http://arxiv.org/abs/1911.09666}{{\ttfamily arXiv:1911.09666
  [hep-th]}}.

\bibitem{Almheiri:2019yqk}
A.~Almheiri, R.~Mahajan, and J.~Maldacena, ``{Islands outside the horizon},''
  \href{http://arxiv.org/abs/1910.11077}{{\ttfamily arXiv:1910.11077
  [hep-th]}}.

\bibitem{Almheiri:2020cfm}
A.~Almheiri, T.~Hartman, J.~Maldacena, E.~Shaghoulian, and A.~Tajdini, ``{The
  entropy of Hawking radiation},''
  \href{http://arxiv.org/abs/2006.06872}{{\ttfamily arXiv:2006.06872
  [hep-th]}}.

\bibitem{Sully:2020pza}
J.~Sully, M.~V. Raamsdonk, and D.~Wakeham, ``{BCFT entanglement entropy at
  large central charge and the black hole interior},''
  \href{http://dx.doi.org/10.1007/JHEP03(2021)167}{{\em JHEP} {\bfseries 03}
  (2021) 167}, \href{http://arxiv.org/abs/2004.13088}{{\ttfamily
  arXiv:2004.13088 [hep-th]}}.

\bibitem{Rozali:2019day}
M.~Rozali, J.~Sully, M.~Van~Raamsdonk, C.~Waddell, and D.~Wakeham,
  ``{Information radiation in BCFT models of black holes},''
  \href{http://dx.doi.org/10.1007/JHEP05(2020)004}{{\em JHEP} {\bfseries 05}
  (2020) 004}, \href{http://arxiv.org/abs/1910.12836}{{\ttfamily
  arXiv:1910.12836 [hep-th]}}.

\bibitem{Chen:2020uac}
H.~Z. Chen, R.~C. Myers, D.~Neuenfeld, I.~A. Reyes, and J.~Sandor, ``{Quantum
  Extremal Islands Made Easy, Part I: Entanglement on the Brane},''
  \href{http://dx.doi.org/10.1007/JHEP10(2020)166}{{\em JHEP} {\bfseries 10}
  (2020) 166}, \href{http://arxiv.org/abs/2006.04851}{{\ttfamily
  arXiv:2006.04851 [hep-th]}}.

\bibitem{Chen:2020hmv}
H.~Z. Chen, R.~C. Myers, D.~Neuenfeld, I.~A. Reyes, and J.~Sandor, ``{Quantum
  Extremal Islands Made Easy, Part II: Black Holes on the Brane},''
  \href{http://dx.doi.org/10.1007/JHEP12(2020)025}{{\em JHEP} {\bfseries 12}
  (2020) 025}, \href{http://arxiv.org/abs/2010.00018}{{\ttfamily
  arXiv:2010.00018 [hep-th]}}.

\bibitem{Grimaldi:2022suv}
G.~Grimaldi, J.~Hernandez, and R.~C. Myers, ``{Quantum Extremal Islands Made
  Easy, Part IV: Massive Black Holes on the Brane},''
  \href{http://arxiv.org/abs/2202.00679}{{\ttfamily arXiv:2202.00679
  [hep-th]}}.

\bibitem{Suzuki:2022xwv}
K.~Suzuki and T.~Takayanagi, ``{BCFT and Islands in Two Dimensions},''
  \href{http://arxiv.org/abs/2202.08462}{{\ttfamily arXiv:2202.08462
  [hep-th]}}.

\bibitem{Geng:2020qvw}
H.~Geng and A.~Karch, ``{Massive islands},''
  \href{http://dx.doi.org/10.1007/JHEP09(2020)121}{{\em JHEP} {\bfseries 09}
  (2020) 121}, \href{http://arxiv.org/abs/2006.02438}{{\ttfamily
  arXiv:2006.02438 [hep-th]}}.

\bibitem{Geng:2020fxl}
H.~Geng, A.~Karch, C.~Perez-Pardavila, S.~Raju, L.~Randall, M.~Riojas, and
  S.~Shashi, ``{Information Transfer with a Gravitating Bath},''
  \href{http://dx.doi.org/10.21468/SciPostPhys.10.5.103}{{\em SciPost Phys.}
  {\bfseries 10} no.~5, (2021) 103},
  \href{http://arxiv.org/abs/2012.04671}{{\ttfamily arXiv:2012.04671
  [hep-th]}}.

\bibitem{Geng:2021iyq}
H.~Geng, S.~L\"ust, R.~K. Mishra, and D.~Wakeham, ``{Holographic BCFTs and
  Communicating Black Holes},''
  \href{http://dx.doi.org/10.1007/JHEP08(2021)003}{{\em jhep} {\bfseries 08}
  (2021) 003}, \href{http://arxiv.org/abs/2104.07039}{{\ttfamily
  arXiv:2104.07039 [hep-th]}}.

\bibitem{Geng:2021mic}
H.~Geng, A.~Karch, C.~Perez-Pardavila, S.~Raju, L.~Randall, M.~Riojas, and
  S.~Shashi, ``{Entanglement Phase Structure of a Holographic BCFT in a Black
  Hole Background},'' \href{http://arxiv.org/abs/2112.09132}{{\ttfamily
  arXiv:2112.09132 [hep-th]}}.

\bibitem{Geng:2021hlu}
H.~Geng, A.~Karch, C.~Perez-Pardavila, S.~Raju, L.~Randall, M.~Riojas, and
  S.~Shashi, ``{Inconsistency of Islands in Theories with Long-Range
  Gravity},'' \href{http://arxiv.org/abs/2107.03390}{{\ttfamily
  arXiv:2107.03390 [hep-th]}}.

\bibitem{Deng:2020ent}
F.~Deng, J.~Chu, and Y.~Zhou, ``{Defect extremal surface as the holographic
  counterpart of Island formula},''
  \href{http://dx.doi.org/10.1007/JHEP03(2021)008}{{\em JHEP} {\bfseries 03}
  (2021) 008}, \href{http://arxiv.org/abs/2012.07612}{{\ttfamily
  arXiv:2012.07612 [hep-th]}}.

\bibitem{Chu:2021gdb}
J.~Chu, F.~Deng, and Y.~Zhou, ``{Page curve from defect extremal surface and
  island in higher dimensions},''
  \href{http://dx.doi.org/10.1007/JHEP10(2021)149}{{\em JHEP} {\bfseries 10}
  (2021) 149}, \href{http://arxiv.org/abs/2105.09106}{{\ttfamily
  arXiv:2105.09106 [hep-th]}}.

\bibitem{Li:2021dmf}
T.~Li, M.-K. Yuan, and Y.~Zhou, ``{Defect extremal surface for reflected
  entropy},'' \href{http://dx.doi.org/10.1007/JHEP01(2022)018}{{\em JHEP}
  {\bfseries 01} (2022) 018}, \href{http://arxiv.org/abs/2108.08544}{{\ttfamily
  arXiv:2108.08544 [hep-th]}}.

\bibitem{Basu:2022reu}
D.~Basu, H.~Parihar, V.~Raj, and G.~Sengupta, ``{Defect extremal surfaces for
  entanglement negativity},'' \href{http://arxiv.org/abs/2205.07905}{{\ttfamily
  arXiv:2205.07905 [hep-th]}}.

\bibitem{Shao:2022wrm}
Y.~Shao, M.-K. Yuan, and Y.~Zhou, ``{Entanglement negativity and defect
  extremal surface},''
  \href{http://dx.doi.org/10.21468/SciPostPhysCore.7.2.027}{{\em SciPost Phys.
  Core} {\bfseries 7} no.~2, (2024) 027},
  \href{http://arxiv.org/abs/2206.05951}{{\ttfamily arXiv:2206.05951
  [hep-th]}}.

\bibitem{Basu:2024xjq}
D.~Basu, H.~Chourasiya, A.~Dey, and V.~Raj, ``{Bridging Boundaries: $T\bar{T}$,
  Double Holography, and Reflected Entropy},''
  \href{http://arxiv.org/abs/2411.12827}{{\ttfamily arXiv:2411.12827
  [hep-th]}}.

\bibitem{Liu:2023ggg}
Y.~Liu, Q.~Chen, Y.~Ling, C.~Peng, Y.~Tian, and Z.-Y. Xian, ``{Addendum to:
  Entanglement of defect subregions in double holography},''
  \href{http://dx.doi.org/10.1007/JHEP09(2024)194}{{\em JHEP} {\bfseries 09}
  (2024) 194}, \href{http://arxiv.org/abs/2312.08025}{{\ttfamily
  arXiv:2312.08025 [hep-th]}}.

\bibitem{Ling:2021vxe}
Y.~Ling, P.~Liu, Y.~Liu, C.~Niu, Z.-Y. Xian, and C.-Y. Zhang, ``{Reflected
  entropy in double holography},''
  \href{http://dx.doi.org/10.1007/JHEP02(2022)037}{{\em JHEP} {\bfseries 02}
  (2022) 037}, \href{http://arxiv.org/abs/2109.09243}{{\ttfamily
  arXiv:2109.09243 [hep-th]}}.

\bibitem{Akal:2020wfl}
I.~Akal, Y.~Kusuki, T.~Takayanagi, and Z.~Wei, ``{Codimension two holography
  for wedges},'' \href{http://dx.doi.org/10.1103/PhysRevD.102.126007}{{\em
  Phys. Rev. D} {\bfseries 102} no.~12, (2020) 126007},
  \href{http://arxiv.org/abs/2007.06800}{{\ttfamily arXiv:2007.06800
  [hep-th]}}.

\bibitem{Miao:2020oey}
R.-X. Miao, ``{An Exact Construction of Codimension two Holography},''
  \href{http://dx.doi.org/10.1007/JHEP01(2021)150}{{\em JHEP} {\bfseries 01}
  (2021) 150}, \href{http://arxiv.org/abs/2009.06263}{{\ttfamily
  arXiv:2009.06263 [hep-th]}}.

\bibitem{Myers:2024zhb}
R.~C. Myers, S.-M. Ruan, and T.~Ugajin, ``{Double holography of entangled
  universes},'' \href{http://dx.doi.org/10.1007/JHEP07(2024)035}{{\em JHEP}
  {\bfseries 07} (2024) 035}, \href{http://arxiv.org/abs/2403.17483}{{\ttfamily
  arXiv:2403.17483 [hep-th]}}.

\bibitem{Cardy:2004hm}
J.~L. Cardy, ``{Boundary conformal field theory},''
  \href{http://arxiv.org/abs/hep-th/0411189}{{\ttfamily arXiv:hep-th/0411189}}.

\bibitem{Takayanagi:2011zk}
T.~Takayanagi, ``{Holographic Dual of BCFT},''
  \href{http://dx.doi.org/10.1103/PhysRevLett.107.101602}{{\em Phys. Rev.
  Lett.} {\bfseries 107} (2011) 101602},
  \href{http://arxiv.org/abs/1105.5165}{{\ttfamily arXiv:1105.5165 [hep-th]}}.

\bibitem{Fujita:2011fp}
M.~Fujita, T.~Takayanagi, and E.~Tonni, ``{Aspects of AdS/BCFT},''
  \href{http://dx.doi.org/10.1007/JHEP11(2011)043}{{\em JHEP} {\bfseries 11}
  (2011) 043}, \href{http://arxiv.org/abs/1108.5152}{{\ttfamily arXiv:1108.5152
  [hep-th]}}.

\bibitem{Kastikainen:2021ybu}
J.~Kastikainen and S.~Shashi, ``{Structure of holographic BCFT correlators from
  geodesics},'' \href{http://dx.doi.org/10.1103/PhysRevD.105.046007}{{\em Phys.
  Rev. D} {\bfseries 105} no.~4, (2022) 046007},
  \href{http://arxiv.org/abs/2109.00079}{{\ttfamily arXiv:2109.00079
  [hep-th]}}.

\bibitem{Izumi:2022opi}
K.~Izumi, T.~Shiromizu, K.~Suzuki, T.~Takayanagi, and N.~Tanahashi, ``{Brane
  dynamics of holographic BCFTs},''
  \href{http://dx.doi.org/10.1007/JHEP10(2022)050}{{\em JHEP} {\bfseries 10}
  (2022) 050}, \href{http://arxiv.org/abs/2205.15500}{{\ttfamily
  arXiv:2205.15500 [hep-th]}}.

\bibitem{Cavalcanti:2018pta}
A.~G. Cavalcanti, D.~Melnikov, and M.~R.~O. Silva, ``{Studies of Boundary
  Entropy in AdS/BCFT},''
  \href{http://dx.doi.org/10.1088/1361-6382/ab8142}{{\em Class. Quant. Grav.}
  {\bfseries 37} no.~10, (2020) 105009},
  \href{http://arxiv.org/abs/1808.07966}{{\ttfamily arXiv:1808.07966
  [hep-th]}}.

\bibitem{Magan:2014dwa}
J.~M. Mag\'an, D.~Melnikov, and M.~R.~O. Silva, ``{Black Holes in AdS/BCFT and
  Fluid/Gravity Correspondence},''
  \href{http://dx.doi.org/10.1007/JHEP11(2014)069}{{\em JHEP} {\bfseries 11}
  (2014) 069}, \href{http://arxiv.org/abs/1408.2580}{{\ttfamily arXiv:1408.2580
  [hep-th]}}.

\bibitem{Cavalcanti:2020rsp}
A.~G. Cavalcanti and D.~Melnikov, ``{Bubble quenches in the AdS/BCFT model},''
  \href{http://dx.doi.org/10.1103/PhysRevD.103.046022}{{\em Phys. Rev. D}
  {\bfseries 103} no.~4, (2021) 046022},
  \href{http://arxiv.org/abs/2006.04623}{{\ttfamily arXiv:2006.04623
  [hep-th]}}.

\bibitem{Takayanagi:2020njm}
T.~Takayanagi and T.~Uetoko, ``{Chern-Simons Gravity Dual of BCFT},''
  \href{http://dx.doi.org/10.1007/JHEP04(2021)193}{{\em JHEP} {\bfseries 04}
  (2021) 193}, \href{http://arxiv.org/abs/2011.02513}{{\ttfamily
  arXiv:2011.02513 [hep-th]}}.

\bibitem{Vidal:2002zz}
G.~Vidal and R.~F. Werner, ``{Computable measure of entanglement},''
  \href{http://dx.doi.org/10.1103/PhysRevA.65.032314}{{\em Phys. Rev. A}
  {\bfseries 65} (2002) 032314},
  \href{http://arxiv.org/abs/quant-ph/0102117}{{\ttfamily
  arXiv:quant-ph/0102117}}.

\bibitem{Plenio:2005cwa}
M.~B. Plenio, ``{Logarithmic Negativity: A Full Entanglement Monotone That is
  not Convex},'' \href{http://dx.doi.org/10.1103/PhysRevLett.95.090503}{{\em
  Phys. Rev. Lett.} {\bfseries 95} no.~9, (2005) 090503},
  \href{http://arxiv.org/abs/quant-ph/0505071}{{\ttfamily
  arXiv:quant-ph/0505071}}.

\bibitem{Dutta:2019gen}
S.~Dutta and T.~Faulkner, ``{A canonical purification for the entanglement
  wedge cross-section},'' \href{http://dx.doi.org/10.1007/JHEP03(2021)178}{{\em
  JHEP} {\bfseries 03} (2021) 178},
  \href{http://arxiv.org/abs/1905.00577}{{\ttfamily arXiv:1905.00577
  [hep-th]}}.

\bibitem{Akers:2021pvd}
C.~Akers, T.~Faulkner, S.~Lin, and P.~Rath, ``{Reflected entropy in random
  tensor networks},'' \href{http://arxiv.org/abs/2112.09122}{{\ttfamily
  arXiv:2112.09122 [hep-th]}}.

\bibitem{Takayanagi:2017knl}
T.~Takayanagi and K.~Umemoto, ``{Entanglement of purification through
  holographic duality},''
  \href{http://dx.doi.org/10.1038/s41567-018-0075-2}{{\em Nature Phys.}
  {\bfseries 14} no.~6, (2018) 573--577},
  \href{http://arxiv.org/abs/1708.09393}{{\ttfamily arXiv:1708.09393
  [hep-th]}}.

\bibitem{Nguyen:2017yqw}
P.~Nguyen, T.~Devakul, M.~G. Halbasch, M.~P. Zaletel, and B.~Swingle,
  ``{Entanglement of purification: from spin chains to holography},''
  \href{http://dx.doi.org/10.1007/JHEP01(2018)098}{{\em JHEP} {\bfseries 01}
  (2018) 098}, \href{http://arxiv.org/abs/1709.07424}{{\ttfamily
  arXiv:1709.07424 [hep-th]}}.

\bibitem{Wen:2021qgx}
Q.~Wen, ``{Balanced Partial Entanglement and the Entanglement Wedge Cross
  Section},'' \href{http://dx.doi.org/10.1007/JHEP04(2021)301}{{\em JHEP}
  {\bfseries 04} (2021) 301}, \href{http://arxiv.org/abs/2103.00415}{{\ttfamily
  arXiv:2103.00415 [hep-th]}}.

\bibitem{Tamaoka:2018ned}
K.~Tamaoka, ``{Entanglement Wedge Cross Section from the Dual Density
  Matrix},'' \href{http://dx.doi.org/10.1103/PhysRevLett.122.141601}{{\em Phys.
  Rev. Lett.} {\bfseries 122} no.~14, (2019) 141601},
  \href{http://arxiv.org/abs/1809.09109}{{\ttfamily arXiv:1809.09109
  [hep-th]}}.

\bibitem{Li:2020ceg}
T.~Li, J.~Chu, and Y.~Zhou, ``{Reflected Entropy for an Evaporating Black
  Hole},'' \href{http://dx.doi.org/10.1007/JHEP11(2020)155}{{\em JHEP}
  {\bfseries 11} (2020) 155}, \href{http://arxiv.org/abs/2006.10846}{{\ttfamily
  arXiv:2006.10846 [hep-th]}}.

\bibitem{Chandrasekaran:2020qtn}
V.~Chandrasekaran, M.~Miyaji, and P.~Rath, ``{Including contributions from
  entanglement islands to the reflected entropy},''
  \href{http://dx.doi.org/10.1103/PhysRevD.102.086009}{{\em Phys. Rev. D}
  {\bfseries 102} no.~8, (2020) 086009},
  \href{http://arxiv.org/abs/2006.10754}{{\ttfamily arXiv:2006.10754
  [hep-th]}}.

\bibitem{DiFrancesco:639405}
P.~Di~Francesco, P.~Mathieu, and D.~Sénéchal,
  \href{http://dx.doi.org/10.1007/978-1-4612-2256-9}{{\em {Conformal field
  theory}}}.
\newblock Graduate texts in contemporary physics. Springer, New York, NY, 1997.
\newblock \url{https://cds.cern.ch/record/639405}.

\bibitem{Brown:1986nw}
J.~D. Brown and M.~Henneaux, ``{Central Charges in the Canonical Realization of
  Asymptotic Symmetries: An Example from Three-Dimensional Gravity},''
  \href{http://dx.doi.org/10.1007/BF01211590}{{\em Commun. Math. Phys.}
  {\bfseries 104} (1986) 207--226}.

\bibitem{Maldacena:2001kr}
J.~M. Maldacena, ``{Eternal black holes in anti-de Sitter},''
  \href{http://dx.doi.org/10.1088/1126-6708/2003/04/021}{{\em JHEP} {\bfseries
  04} (2003) 021}, \href{http://arxiv.org/abs/hep-th/0106112}{{\ttfamily
  arXiv:hep-th/0106112}}.

\bibitem{Penington:2019npb}
G.~Penington, ``{Entanglement Wedge Reconstruction and the Information
  Paradox},'' \href{http://dx.doi.org/10.1007/JHEP09(2020)002}{{\em JHEP}
  {\bfseries 09} (2020) 002}, \href{http://arxiv.org/abs/1905.08255}{{\ttfamily
  arXiv:1905.08255 [hep-th]}}.

\bibitem{Akers:2022max}
C.~Akers, T.~Faulkner, S.~Lin, and P.~Rath, ``{The Page Curve for Reflected
  Entropy},'' \href{http://arxiv.org/abs/2201.11730}{{\ttfamily
  arXiv:2201.11730 [hep-th]}}.

\bibitem{Kusuki:2019evw}
Y.~Kusuki and K.~Tamaoka, ``{Entanglement Wedge Cross Section from CFT:
  Dynamics of Local Operator Quench},''
  \href{http://dx.doi.org/10.1007/JHEP02(2020)017}{{\em JHEP} {\bfseries 02}
  (2020) 017}, \href{http://arxiv.org/abs/1909.06790}{{\ttfamily
  arXiv:1909.06790 [hep-th]}}.

\bibitem{Fitzpatrick:2014vua}
A.~L. Fitzpatrick, J.~Kaplan, and M.~T. Walters, ``{Universality of
  Long-Distance AdS Physics from the CFT Bootstrap},''
  \href{http://dx.doi.org/10.1007/JHEP08(2014)145}{{\em JHEP} {\bfseries 08}
  (2014) 145}, \href{http://arxiv.org/abs/1403.6829}{{\ttfamily arXiv:1403.6829
  [hep-th]}}.

\bibitem{Banerjee:2016qca}
P.~Banerjee, S.~Datta, and R.~Sinha, ``{Higher-point conformal blocks and
  entanglement entropy in heavy states},''
  \href{http://dx.doi.org/10.1007/JHEP05(2016)127}{{\em JHEP} {\bfseries 05}
  (2016) 127}, \href{http://arxiv.org/abs/1601.06794}{{\ttfamily
  arXiv:1601.06794 [hep-th]}}.

\bibitem{Basu:2023jtf}
D.~Basu, H.~Chourasiya, V.~Raj, and G.~Sengupta, ``{Reflected entropy in a BCFT
  on a black hole background},''
  \href{http://dx.doi.org/10.1007/JHEP05(2024)054}{{\em JHEP} {\bfseries 05}
  (2024) 054}, \href{http://arxiv.org/abs/2311.17023}{{\ttfamily
  arXiv:2311.17023 [hep-th]}}.

\bibitem{Boruch:2020wbe}
J.~Boruch, ``{Entanglement wedge cross-section in shock wave geometries},''
  \href{http://dx.doi.org/10.1007/JHEP07(2020)208}{{\em JHEP} {\bfseries 07}
  (2020) 208}, \href{http://arxiv.org/abs/2006.10625}{{\ttfamily
  arXiv:2006.10625 [hep-th]}}.

\bibitem{Umemoto:2018jpc}
K.~Umemoto and Y.~Zhou, ``{Entanglement of Purification for Multipartite States
  and its Holographic Dual},''
  \href{http://dx.doi.org/10.1007/JHEP10(2018)152}{{\em JHEP} {\bfseries 10}
  (2018) 152}, \href{http://arxiv.org/abs/1805.02625}{{\ttfamily
  arXiv:1805.02625 [hep-th]}}.

\bibitem{Bao:2019zqc}
N.~Bao and N.~Cheng, ``{Multipartite Reflected Entropy},''
  \href{http://dx.doi.org/10.1007/JHEP10(2019)102}{{\em JHEP} {\bfseries 10}
  (2019) 102}, \href{http://arxiv.org/abs/1909.03154}{{\ttfamily
  arXiv:1909.03154 [hep-th]}}.

\bibitem{Chu:2019etd}
J.~Chu, R.~Qi, and Y.~Zhou, ``{Generalizations of Reflected Entropy and the
  Holographic Dual},'' \href{http://dx.doi.org/10.1007/JHEP03(2020)151}{{\em
  JHEP} {\bfseries 03} (2020) 151},
  \href{http://arxiv.org/abs/1909.10456}{{\ttfamily arXiv:1909.10456
  [hep-th]}}.

\bibitem{Yuan:2024yfg}
M.-K. Yuan, M.~Li, and Y.~Zhou, ``{Reflected Multientropy and Its Holographic
  Dual},'' \href{http://dx.doi.org/10.1103/76vs-rxcs}{{\em Phys. Rev. Lett.}
  {\bfseries 135} no.~9, (2025) 091604},
  \href{http://arxiv.org/abs/2410.08546}{{\ttfamily arXiv:2410.08546
  [hep-th]}}.

\bibitem{Akers:2019gcv}
C.~Akers and P.~Rath, ``{Entanglement Wedge Cross Sections Require Tripartite
  Entanglement},'' \href{http://dx.doi.org/10.1007/JHEP04(2020)208}{{\em JHEP}
  {\bfseries 04} (2020) 208}, \href{http://arxiv.org/abs/1911.07852}{{\ttfamily
  arXiv:1911.07852 [hep-th]}}.

\bibitem{Hayden:2021gno}
P.~Hayden, O.~Parrikar, and J.~Sorce, ``{The Markov gap for geometric reflected
  entropy},'' \href{http://dx.doi.org/10.1007/JHEP10(2021)047}{{\em JHEP}
  {\bfseries 10} (2021) 047}, \href{http://arxiv.org/abs/2107.00009}{{\ttfamily
  arXiv:2107.00009 [hep-th]}}.

\bibitem{Balasubramanian:2024ysu}
V.~Balasubramanian, M.~J. Kang, C.~Murdia, and S.~F. Ross, ``{Signals of
  multiparty entanglement and holography},''
  \href{http://dx.doi.org/10.1007/JHEP06(2025)068}{{\em JHEP} {\bfseries 06}
  (2025) 068}, \href{http://arxiv.org/abs/2411.03422}{{\ttfamily
  arXiv:2411.03422 [hep-th]}}.

\bibitem{Bao:2025psl}
N.~Bao, K.~Furuya, and J.~Naskar, ``{Tripartite Entanglement Signal from
  Multipartite Entanglement of Purification},''
  \href{http://arxiv.org/abs/2509.08209}{{\ttfamily arXiv:2509.08209
  [hep-th]}}.

\bibitem{Mori:2025gqe}
T.~Mori, ``{Quantum correlation beyond entanglement: Holographic discord and
  multipartite generalizations},''
  \href{http://arxiv.org/abs/2506.02131}{{\ttfamily arXiv:2506.02131
  [hep-th]}}.

\bibitem{Basak:2024uwc}
J.~K. Basak, V.~Malvimat, and J.~Yoon, ``{A New Genuine Multipartite
  Entanglement Measure: from Qubits to Multiboundary Wormholes},''
  \href{http://arxiv.org/abs/2411.11961}{{\ttfamily arXiv:2411.11961
  [hep-th]}}.

\bibitem{Ahn:2025bdm}
B.~Ahn, J.~K. Basak, K.-Y. Kim, G.~B. Koo, V.~Malvimat, and J.~Yoon, ``{Probing
  the Hierarchy of Genuine Multipartite Entanglement with Generalized Latent
  Entropy},'' \href{http://arxiv.org/abs/2510.19922}{{\ttfamily
  arXiv:2510.19922 [hep-th]}}.

\end{thebibliography}\endgroup

\end{document}